\documentclass[a4paper, 12pt]{report}

\usepackage{fontspec}
\usepackage{xeCJK}
\usepackage{geometry}
\usepackage[singlespacing]{setspace} % line spacing
\usepackage{indentfirst} % indented first paragraph
\usepackage{multicol} % multi-column
\usepackage{xeCJK}

\usepackage{tocloft}
\usepackage{nameref}
\usepackage{setspace}

\usepackage{url}
\usepackage[colorlinks,allcolors=black]{hyperref}
\usepackage{caption}
\usepackage{capt-of}
\usepackage{graphicx}
\usepackage{adjustbox}
\usepackage{float}
\usepackage{subfig}
\usepackage{wrapfig}

\usepackage{amsfonts,amsthm,amsmath,amssymb}
\usepackage{mathrsfs}
\usepackage{bbm}
\usepackage{bm}

\usepackage{cite} % citation standard, must load after package hyperref 
\usepackage{cleveref}
\usepackage{siunitx}

\usepackage{optidef} % Optimization form
\usepackage{listings} % code listing

\usepackage{xcolor} % Review rebuttals

\title{
\vspace{-1cm}
\Large{Doctoral Thesis \\
\vspace{0.3cm}
博士論文 \\
\vspace{2cm} 
}

\vspace{0.5cm} 
\LARGE{Geometry and Mechanics of Multistable Origami Blocks\\
\vspace{0.3cm} 
}
\Large{（多安定折紙ブロックの幾何学と力学）\\}
\vspace{2cm}

\normalsize{Advisor: Prof. Tomohiro Tachi \\
\vspace{1cm}
Department of Architecture \\
\vspace{0.2cm}
Graduate School of Engineering \\
\vspace{0.2cm} 
The University of Tokyo \\
\vspace{0.2cm} 
37-237047 \\
\vspace{0.2cm} 
}

\Large{
\author{\textbf{LEE \quad MUNKYUN}\\ イ \quad ムンギュン \\ 李 \quad 文均}
}
\date{}
}

\begin{document}
	\maketitle
    \clearpage
	\pagenumbering{roman}
    
    \newpage
    \thispagestyle{empty}
    \mbox{}
    \newpage

    \addcontentsline{toc}{chapter}{Abstract}
    \chapter*{Abstract}

Origami, which transforms flat sheets into complex three-dimensional shapes through folding patterns, has inspired the emergence of deployable systems in architecture and civil realms.
Most existing origami-inspired deployable systems are based on rigid or curved-crease origami types.
While these types provide smooth deployability and high design freedom, they inherently lack shape stability and require additional supports to maintain their deployed shapes.
These lead to a fundamental trade-off between deployability and shape stability, which remains a major challenge for large-scale origami systems.

Multistable origami, in contrast, achieves energy stability across multiple configurations by allowing elastic deformation of panels during deployment.
This unique characteristic enables it to maintain stable shapes even under external loads and to exhibit snap-through transitions between them.
These properties allow multistable origami to achieve both shape stability and deployability, offering high potential for self-supporting deployable systems in architectural applications.

However, realizing large-scale and structurally stable systems using a single origami with a single sheet faces many practical constraints.
To overcome these limitations, origami block assembly has emerged as an effective approach to form global systems.
This approach enables flexibility in global geometry and mechanical behaviors while offering reconfigurability.
However, existing origami block assemblies often suffer from geometric mismatches and insufficient self-supporting capacity, indicating the need for modules that inherently combine deployability and stability.
These challenges indicate that the complementary potential of multistable origami and block assemblies can provide a promising solution.

This study aims to address the challenges of applying deployable origami to large-scale architectural systems by leveraging the potential of multistable origami as modular building blocks.
From a geometric standpoint, we explore design methods for stable configurations of multistable origami blocks that can align and interlock with each other.
From a mechanical standpoint, we explore stiffness-controllable design methods that ensure self-supporting and load-bearing capabilities through geometric parameters.

To achieve these goals, we introduce three design practices of multistable origami blocks, named \emph{Q-Bellows}, \emph{T-Toroid}, and \emph{CC-Block}, based on over-constraining design approaches applied to rigid or curved-crease origami geometries.
Each multistable origami block is designed to achieve specific goals: stiffness control in Q-Bellows, and stable configuration design in T-Toroid and CC-Block.
The mechanics of the designed blocks are investigated through simulations and desktop-scale prototype experiments to clarify the relationship and trends between geometric parameters and mechanical behaviors.
By assembling these blocks, we demonstrate global systems capable of programming deployment sequences, partial deployments, and forms.
Finally, large-scale prototype fabrications validate the feasibility of the proposed modular system, showcasing its scalability, reconfigurability, and self-supporting capability, and highlighting its potential for deployable architectural applications such as sequentially deployable structures, curved shells, and curved wall systems.

    \clearpage

    \newpage
    \thispagestyle{empty}
    \mbox{}
    \newpage
    
    \chapter*{Acknowledgment}
Standing at the end of this long doctoral journey, I feel deeply moved as I look back and realize how many brilliant moments have filled my life along the way.
I would like to take this opportunity to express my heartfelt gratitude to those who have supported me throughout this journey.

First and foremost, I would like to express my deepest appreciation to my mentor and advisor, \emph{Professor Tomohiro Tachi}.
I have always been inspired by his remarkable creativity and sharp intuition, and it has been a great honor to work with him over the past five years.
The countless discussions and the process of creating new ideas together have been truly invaluable experiences that have shaped both my academic and personal growth.
Under his guidance, I was able to achieve meaningful research outcomes, receive generous funding support, and gain a wide range of international experiences.
Meeting Professor Tachi has been a turning point in my life, and I am sincerely grateful for all the positive changes he has brought into my journey.

I would also like to thank the members of my doctoral examination committee, \emph{Prof. Kenichi Kawaguchi, Prof. Joseph M. Gattas, Prof. Yasushi Ikeda, and Prof. Sachiko Ishida}, for their time, effort, and insightful comments.
In particular, I am deeply grateful to Prof. Kenichi Kawaguchi for his kind advice and thoughtful discussions each semester, which consistently guided my research in the right direction.

I am also sincerely grateful to my co-authors, \emph{Yuki Miyajima, Dr. Kiumars Sharifmoghaddam, Dr. Mahmoud Abu-Saleem, Dr. Felix Delinger, Prof. Joseph M. Gattas, Prof. Martin Killian, Prof. Christian Müller, and Prof. Georg Nawratil}, for their invaluable collaboration.

Especially, with \emph{Prof. Joseph M. Gattas}, I had the privilege of working closely through annual research exchanges between Australia and Japan.
Our discussions led to many fruitful outcomes, and I deeply resonated with his creative ideas and clear research direction.
The time I spent in Brisbane each year remains among the happiest memories of my life.

I am also grateful to \emph{Dr. Kiumars Sharifmoghaddam}, whose earlier research inspired my own and whose insightful discussions helped refine and further develop our work.
I especially appreciate his patience and open-mindedness throughout our collaboration.

I am thankful to \emph{Prof. Glaucio H. Paulino} and \emph{Prof. Sigrid Adriaenssens} at Princeton University, \emph{Prof. Ivan Izmestiev} at TU Wien, and \emph{Prof. Jinkyu Yang} at Seoul National University for their warm hospitality, valuable discussions during my visits, and encouragement to share my work.

I also appreciate the support and advice provided by \emph{Kawakami Sangyo} and \emph{NIKKEN Sekkei} during my research activities.

I am grateful to the various funding programs that supported my studies from my undergraduate to doctoral degree.
I sincerely thank the Korea-Japan Joint Government Scholarship Program (KJSP) for allowing me the opportunity to study in Japan.
I also acknowledge the financial support from the WINGS-CFS and DFS Fellowships and JSPS DC2 KAKENHI (Grant No. 24KJ0648).
I thank the Samsung Electronics DS Global Talent Scholarship Program for supporting my future career.

Finally, I would like to thank those who supported me in my daily life.
My warm thanks go to \emph{Rinki Imada, Seri Nishimoto, Yiwei Zhang}, and all of our lab members for their help and friendship.
I would also like to acknowledge my long-time companions, \emph{Junoh Kim and Sunghyun Park}, from the Kansai-UTokyo Korean community.
Above all, I am deeply grateful to \emph{my loving parents}, who have always supported and cared for me throughout every step of this journey.

% \begin{flushright}
% \emph{Toward the next chapter...}

% \textbf{Munkyun Lee}
% \end{flushright}

    \clearpage
    
    \newpage
    \thispagestyle{empty}
    \mbox{}
    \newpage
    
    \setlength{\cftbeforetoctitleskip}{12pt}
    \setlength{\cftaftertoctitleskip}{12pt}
 	\tableofcontents
    \clearpage

    \newpage
    \thispagestyle{empty}
    \mbox{}
    \newpage
    
    \pagenumbering{arabic}
	\chapter{Introduction}
\label{chap:intro}

\section{Background}
\subsection{Origami}
\label{Intro_background}

\begin{figure}[tbhp]
  \centering
  \includegraphics[page=1,width=\linewidth]{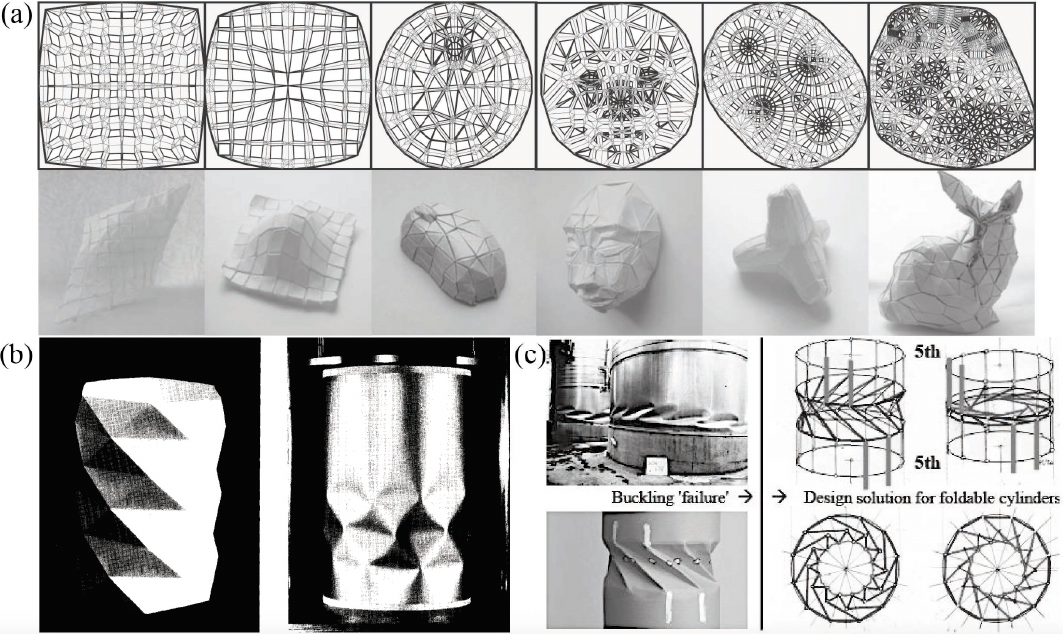}
   \caption{(a) Origami patterns for freeforms, (\emph{Figures from \cite{tachi2009origamizing} in grayscale}). (b) Yoshimura pattern origami (left) and the compression loading buckling mode of cylindrical shell (right), (\emph{Figures from \cite{miura1969proposition} in grayscale}). (c) Torsional loading buckling mode of cylindrical shell (left top) and Kresling pattern origami (left bottom and right), (\emph{Figures from \cite{kresling2020fifth} in grayscale}).}
  \label{fig:Intro_Background}
\end{figure}

\emph{Origami}, which can turn a flat sheet into complex three-dimensional shapes by following a folding pattern, has recently gained significant attention as an innovative design approach across art, mathematics, science, and engineering. 
This ability to transform from 2 to 3 dimensions makes origami valuable not only as an artistic expression but also as a powerful tool for complex design applications. 
Lang~\cite{lang1996computational}, Demaine~\cite{demaine2007geometric}, and Tachi~\cite{tachi2009generalization} have mathematically established the relationship between geometric crease patterns and three-dimensional forms, leading to the development of various origami design software tools~\cite{tachi2009origamizing, tachi2009simulation, ghassaei2018fast}, which greatly expand the design flexibility of origami systems (see Figure~\ref{fig:Intro_Background} (a)).

Also, origami is closely related to the mechanical behavior of thin panels. 
This connection has long been studied, with well-known examples including the Yoshimura pattern observed under axial compression of thin cylindrical shells (Figure~\ref{fig:Intro_Background} (b))~\cite{yoshimura1955mechanism, miura1969proposition}, and the Kresling pattern generated under torsional loading (Figure~\ref{fig:Intro_Background} (c))~\cite{donnell1934stability, kresling2008natural, kresling2020fifth}. 
Both emerge through local buckling, and other follow-up studies have shown that these patterns are affected by boundary conditions and loading conditions~\cite{guest1996folding3, lee2019elastic, suh2021new}. 
Such findings highlight the possibility of inverse design for controlling structural performance, thereby broadening the engineering applications of origami.

These developments and understanding of origami have inspired deployable systems in various fields, such as space structures~\cite{koryo1985method}, robotics~\cite{wu2024modular, lee2021high}, electronic devices~\cite{song2014origami, tang2014origami}, and metamaterials~\cite{yasuda2017origami, zhao2025modular}. 
In particular, advantages such as flat-foldability, stackable storage, scalability, and lightweight have also attracted researchers and engineers in the civil and architectural realms, leading to the emergence of origami-inspired deployable structures for architectural use.

\subsection{Origami-inspired Deployable Architectural Systems}
\label{Intro_ArchSys}

We broadly categorize origami into three types, \emph{Rigid Origami, Curved-Crease Origami, and Multistable Origami}, based on the panel deformation during deployment, and then investigate the architectural applications in each type.

\paragraph{Rigid Origami}
\label{Intro_ArchSys_Rigid}

\begin{figure}[tbhp]
  \centering
  \includegraphics[page=2,width=\linewidth]{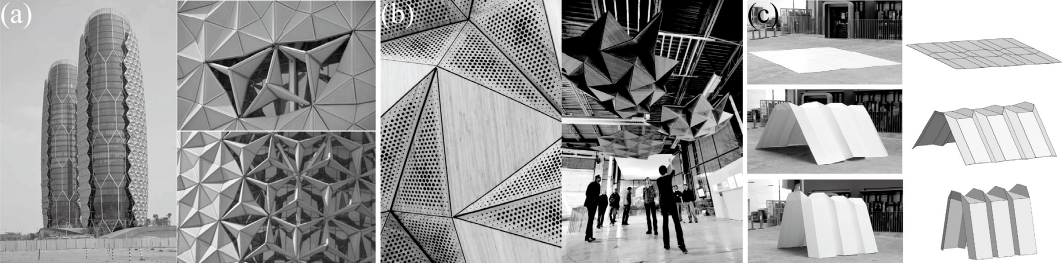}
   \caption{Architectural applications of rigid origami: (a) building facade (\emph{Figures from~\cite{attia2017evaluation} in grayscale}), (b) acoustic ceiling system (\emph{Figures from~\cite{thun2012soundspheres} in grayscale}), and (c) emergency shelter (\emph{Figures from~\cite{lee2016geometric} in grayscale}).}
  \label{fig:Intro_RigidOri}
\end{figure}

\emph{rigid origami} (also called rigid-foldable origami) is the most well-known type, which is a transformable structure consisting of rigid panels connected by rotational joints.
Rigid origami can deploy from the folded state to the deployed state without distortion of the panels~\cite{tachi2009simulation}.
This property makes it possible to construct structures from thick panels and to exploit any of the continuous families of folded configurations~\cite{tachi2016rigid, hull2017double, lang2018review}.
In addition, rigid origami offers advantages such as smooth deployment, a high design flexibility (almost freeform)~\cite{tachi2009origamizing}, and tunable mechanical behavior depending on the choice and combination of crease patterns~\cite{filipov2015origami}.
These features have inspired a wide range of deployable architectural systems, including 
building facades (Figure~\ref{fig:Intro_RigidOri} (a))~\cite{attia2017evaluation, pesenti2018exploration, Web_kiefer}, 
acoustic ceiling systems (Figure~\ref{fig:Intro_RigidOri} (b))~\cite{thun2012soundspheres}, 
spatial structures~\cite{Web_xile, chudoba2015oricreate, ando2020lightweight}, 
bridges~\cite{Web_rolling, maleczek2023bridging, zhu2024large}, 
and emergency shelters (Figure~\ref{fig:Intro_RigidOri} (c))~\cite{lee2016geometric, gattas2016design, Web_cardborigami}.

However, the very feature of smooth deployability can also present challenges in architectural applications.
Maintaining or locking a structure at a specific deployed state often requires additional engineering constraints.
Moreover, more complex geometric patterns, increased self-weight, and stress concentration at the hinges can make it difficult to sustain structural stability, which is the trade-off design challenge in applying rigid origami at larger architectural scales.

\paragraph{Curved-Crease Origami}
\label{Intro_ArchSys_Curved}

\begin{figure}[tbhp]
  \centering
  \includegraphics[page=3,width=\linewidth]{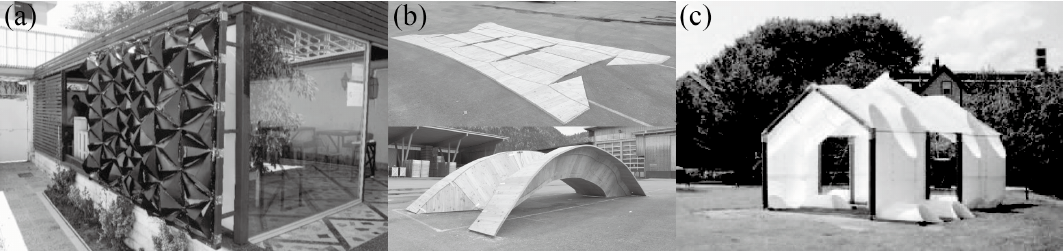}
   \caption{Architectural applications of curved-crease origami: (a) building facade (\emph{Figure from~\cite{korner2018arch} in grayscale}), (b) bridge system (\emph{Figure from~\cite{maleczek2020large} in grayscale}), and (c) pavilion (\emph{Figure from~\cite{fayyad2023bending} in grayscale}).}
  \label{fig:Intro_CurvedOri}
\end{figure}

\emph{Curved-crease origami}, originating from its use in art and sculpture~\cite{demaine2011curved}, with the high aesthetic appeal, has led to the widespread adoption across various contemporary design and engineering domains. 
This consists of curved creases and bendable panels that generate smooth developable surface curvature within elastic deformation during deployment.
The configurations of folding and bending of thin materials can create not only aesthetically appealing geometries but also simultaneously enhance structural performance, including bending stiffness~\cite{mirzajanzadeh2025reprogrammable}, buckling resistance~\cite{lee2019elastic}, and stress stiffening~\cite{lee2021compliant, zhou2024tunable}.
These features have inspired a wide range of deployable architectural systems, including 
column design~\cite{lalvani2020form, koschitz2016designing},
beams~\cite{lee2016folded},
building facade (Figure~\ref{fig:Intro_CurvedOri} (a))~\cite{korner2018arch}, 
bridges (Figure~\ref{fig:Intro_CurvedOri} (b))~\cite{nagy2015conceptual, maleczek2020large},
spatial structures and pavilions (Figure~\ref{fig:Intro_CurvedOri} (c))~\cite{bhooshan2015curve, koschitz2019curved, martin2019developable, siefert2020programming, scheder2022curved, fayyad2023bending, bernhard2023multi}.

This offers a rapid and reversible construction method for curved surfaces, but it also imposes limitations on the design space of curved geometries and restricts panel deformation to the elastic range.
Because curved-crease origami inherently stores bending stress within its panels, additional fixing mechanisms, such as rib stiffeners or edge beams, are often required to ensure structural stability and to reduce stress concentrations~\cite{bhooshan2015curve, koschitz2019curved}.
Consequently, the design process involves a fundamental trade-off between shape stability and deployability, which remains an essential challenge for large-scale applications.

\paragraph{Multistable Origami}
\label{Intro_ArchSys_Multi}

\begin{figure}[tbhp]
  \centering
  \includegraphics[page=4,width=\linewidth]{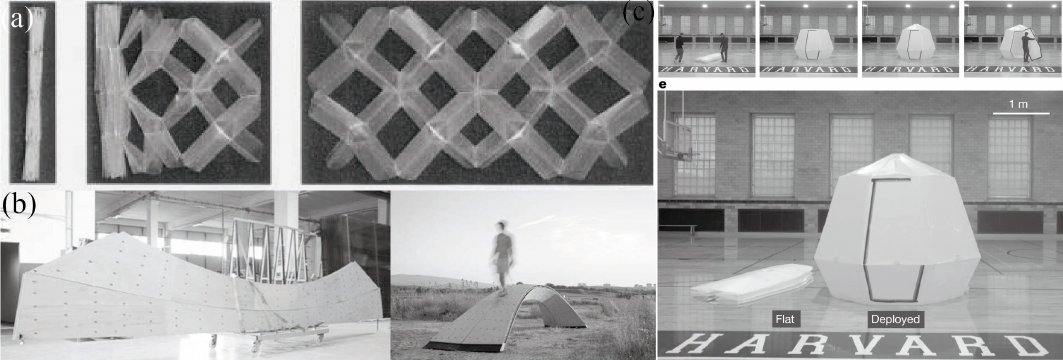}
   \caption{Architectural application concepts of multistable origami. Concepts of spatial structures with (a) deployable solids(named as ~\emph{Ebara Tetra}), (\emph{Figures from~\cite{ebara2003deployable} in grayscale}), (b) multistable curved origami bridge (beam) (\emph{Figures from~\cite{rihaczek2022timbr} in grayscale}), and (c) emergency shelter (\emph{Figures from~\cite{melancon2021multistable} in grayscale}).}
  \label{fig:Intro_MultiOri}
\end{figure}

Multistable origami achieves energy stability across multiple configurations through elastic deformation of the panel during deployment. 
This characteristic enables multistable origami to remain stable in certain deployed states, even under external loads, and exhibit snap-through transitions with non-linear mechanical behavior between them.
Multistable origami can be categorized into two types based on where elastic energy is primarily stored: \emph{Over-constraint type and Mode-bifurcation type}.

The over-constrained type can be designed by closing or assembling specific types of rigid or curved origami units into an over-constrained system, as exemplified by the Kresling cylindrical origami (see Figure~\ref{fig:Intro_Background} (c))~\cite{jianguo2015bistable, yasuda2017origami, zang2024kresling} and loop-closed Miura origami~\cite{kamrava2019origami}.
Kinematically, it has zero degrees of freedom, but folding or deployment becomes possible by allowing panel deformation.
Thus, this type primarily stores its strain energy in the panels during deployment.

The mode-bifurcation type follows rigid or curved-crease origami kinematics, where a kinematic bifurcation causes the configuration to snap into a lower-energy mode during deployment, thereby inducing multistability.
Multistability and stable configurations of this type are governed by the initial conditions of the creases, including rest angles, pre-stress, and stiffness in the creases.
The representative examples are the Waterbomb origami~\cite{hanna2014waterbomb} and the Hypar origami~\cite{filipov2018mechanical, liu2019invariant}.

These unique features of multistable origami, which combine shape stability and deployability with nonlinear mechanical behavior, are referred to as \emph{Deployable Solid}~\cite{ebara2003deployable}.
In this study, we focus on the over-constrained type to ensure self-supporting and load-bearing capacity, rather than the mode-bifurcation type, which mainly relies on kinematics.
By enabling self-supporting deployable systems, they highlight the high potential of multistable origami for large-scale and architectural applications (Figure~\ref{fig:Intro_MultiOri} (a)).
Building on this potential, several concepts have been proposed, including reconfigurable spatial structures (Figure~\ref{fig:Intro_MultiOri} (a))~\cite{ebara2003deployable, hanna2014modeling, hoberman2019construction , liu2019invariant, melancon2021multistable}, bridges (Figure~\ref{fig:Intro_MultiOri} (b))~\cite{rihaczek2022timbr, maleczek2024fold}, and emergency shelters (Figure~\ref{fig:Intro_MultiOri} (c))~\cite{melancon2021multistable}.

However, compared to other origami types, large-scale applications of multistable origami remain relatively limited, and most examples are confined to single-unit systems.
These limitations stem from the still-limited understanding of their geometric design principles and structural performance, particularly in stable-configuration design and nonlinear stiffness control.
Moreover, as the scale increases, single-unit systems face compounded geometric, mechanical, and manufacturing constraints, which restrict the practical advantages of multistable origami.
This design complexity naturally motivates modularization strategies, leading to the need for a deeper geometric and mechanical understanding of multistable origami to enable the development of systematic and reconfigurable modular systems for large-scale applications.

\subsection{Origami Block Assembly}

While some types of origami can provide wide design freedom in both geometry and mechanical behavior, realizing complex forms and structurally stable systems on a large scale with a single origami unit faces practical constraints, such as material size, fab machine size, manufacturing precision, part replacement, reconfiguration, and cost efficiency.
Furthermore, those cutting patterns become even more complex when panel thickness is considered, and geometric interference and play around crease regions add additional design and fabrication challenges.
Even when such systems are successfully fabricated, the resulting structures usually exhibit low structural stiffness, making it difficult to achieve sufficient structural stability at large scales.
Therefore, the block assembly (modularization) becomes an essential strategy for scaling origami-inspired systems to architectural and engineering applications.

Among various modularization methods, the common but effective approaches is \emph{block assembly}, in which multiple modular foldable or pre-folded components are fabricated independently and assembled to form larger systems, as shown in Figures~\ref{fig:Intro_RigidOri} (a), (b), \ref{fig:Intro_CurvedOri} (a), and \ref{fig:Intro_MultiOri} (a).
It can be realized in several ways, such as fixing individual origami blocks onto a supporting frame, connecting them at edges or points through mechanical joints, or interlocking and tessellating their faces.
These block assembly approaches can offer a practical direction toward overcoming the challenges of scaling origami systems while providing greater flexibility in both global form design and mechanical behavior control, including expandable motion ranges~\cite{wang2025bloom}, programmable mechanical behaviors~\cite{filipov2015origami, yasuda2017origami, zhao2025modular, mirzajanzadeh2025reprogrammable}, reconfigurable morphologies~\cite{wang2024reconfigurable}, and large-scale architectural contexts~\cite{bhooshan2015curve, maleczek2016curved, korner2018arch}.

However, block assembly for large-scale applications presents several challenges.
The reliance on numerous modules and connections makes the process time-consuming, while discontinuous surfaces often cause geometric mismatches that interfere with deployment.
Moreover, each block typically lacks self-supporting or load-bearing capacity, even if the origami-like deployable structures are worse, usually requiring additional frames or supports.

These challenges demonstrate the complementary potential of multistable origami, which inherently can offer both shape stability and deployability.
Therefore, the integration of multistable origami with a block assembly approach opens up new possibilities for modular building systems that could overcome the current challenges associated with large-scale origami applications.

% Multistable Deployable Spatial Structure System
% origami~\cite{ebara2003deployable, melancon2021multistable}\\
% kirigami~\cite{chen2021bistable,toyoka2022programming}\\
% scissors~\cite{arnouts2019computational, arnouts2020multi, ren2022umbrella}\\

\section{Objective}
\label{Intro_Obj}
This study aims to address the challenges of applying deployable origami to large-scale architectural systems by leveraging the potential of multistable origami as modular building blocks.
These challenges can be viewed from two perspectives.

From a geometric standpoint, the discontinuity of assembled shapes can be addressed by designing multistable origami blocks whose geometries of stable configuration align and interlock.
These geometries should be designed with respect to the global deployment motion and the assembled configuration.
By achieving such geometric matching, the global system allows reconfigurability.

From a mechanical standpoint, ensuring the self-supporting and load-bearing capacity of multistable origami blocks requires control of their nonlinear stiffness.
Through this control, global systems can exhibit programmable deployment sequences and partial deployments, thereby achieving structural reconfigurability.

\begin{figure}[tbhp]
  \centering
  \includegraphics[page=5,width=\linewidth]{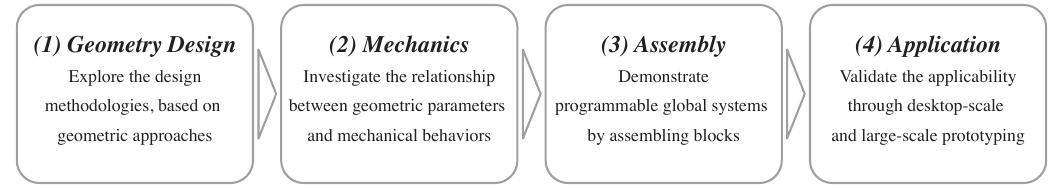}
   \caption{Overview of research objectives and the investigation workflow.}
  \label{fig:workflow}
\end{figure}

To achieve these goals, we propose and investigate systems in the following four research objectives (Figure~\ref{fig:workflow}):
(1) We explore the design methodologies of multistable origami blocks tailored to specific goals, based on geometric design approaches, and (2) investigate the relationship between geometric parameters and mechanical behaviors through simulations and experiments. 
(3) By assembling designed blocks, we demonstrate programmable global systems capable of controlling overall geometry or mechanical properties. 
Finally, (4) the proposed concepts are validated through the assembly of desktop- and large-scale prototypes, showcasing the feasibility of using multistable origami as a foundation for deployable, self-supporting, scalable, and reconfigurable large-scale system applications.

To demonstrate the validity of the concept of multistable origami block systems, this study proposes three novel systems in Chapters \ref{chap:qbellows}, \ref{chap:ttoroid}, and \ref{chap:ccblock}; each system addresses the aspects mentioned above (1)-(4).

\begin{itemize}
    \item In Chapter~\ref{chap:qbellows}, we propose the stiffness-controllable design method for multistable origami blocks by optimizing the geometric incompatibility between rigid origami and linkage. Furthermore, based on the linkage structure, we define the prescribed deployment motion and achieve the stable configurations of the blocks through optimization. The mechanical behaviors of the designed blocks are estimated using the equivalent strain derived from the incompatibility metric and validated through bar-and-hinge simulations and experiments. Finally, by fabricating serial chain assemblies, we demonstrate that both the deployment sequence and partial deploying behavior of global systems can be programmed.

    \item In Chapter~\ref{chap:ttoroid}, we propose the stable-configuration design method for multistable origami blocks with toroidal topology by using the loop-closing method. The mechanical behaviors of the designed blocks are investigated through finite element analyses across various geometric parameters and loading conditions, revealing the trends of parameter dependence and indicating the stiffness-controllability. By assembling these blocks based on the Koebe polyhedra and edge-offset meshes, we showcase that the global systems of three-dimensional curved surfaces can be programmable. Furthermore, we showcase scalability and feasibility by fabricating a large-scale prototype.

    \item In Chapter~\ref{chap:ccblock}, we propose a stable-configuration design method for multistable origami blocks with cylindrical topology by extending the loop-closing method to curved-crease origami. The mechanical behaviors of the designed blocks are investigated through finite element analyses across various geometric parameters and loading conditions, revealing the trends of parameter dependence and indicating the stiffness-controllability. We demonstrate that various curved wall systems can be programmable by integrating blocks and assembly design processes using the generative shape grammar. Furthermore, we showcase scalability and feasibility by fabricating a large-scale prototype.
    
\end{itemize}

%multistable moduleを使うことで、deployable, self-supportingな構造が作れる、
% "mechanical" block assemblyにすることでやreconfiguruable systems/programmable deploymentが作れる
%deployable self-supportingにするためには、nonlinearなstiffness controlが課題になる 
% (3: optimization / geometric incompatibility, 4, 5, FEMによるevaluation)
% "geometry" reconfigurableにするためには、互いにフィットする幾何学をかんがえる必要がある。(3: linkage, bit-operation 4: 多面体, 5: interlocking curved fold)

% 3. stiffness control, prescribed motion, programmable deployment
% 4. curved shell assembly, self-supporting
% 5. reconfigurable system (block assembly), self-supporting, (challenges in curved folding)

\section{Contents of Thesis}
This thesis consists of 6 chapters.
\begin{itemize}
	\item Chapter~\ref{chap:intro} discusses the introduction and outline of this thesis.
    \item Chapter~\ref{chap:methodology} discusses the fundamental methodologies and the workflows that underpin the following chapters.
    \item Chapter~\ref{chap:qbellows} discusses the multistable quadrilateral boundary origami blocks (\emph{Q-bellows}).
    \item Chapter~\ref{chap:ttoroid} discusses the multistable polyhedral origami blocks (\emph{T-toroid}).
    \item Chapter~\ref{chap:ccblock} discusses the multistable curved-crease origami blocks (\emph{CC-block}).
    \item Chapter~\ref{chap:conclusion} concludes this thesis.
\end{itemize}
	\chapter{Methodology}
\label{chap:methodology}

This chapter introduces the fundamental methodologies for designing, analyzing, assembling, and prototyping multistable origami.
We first describe two rigid-origami-based design approaches that underpin multistable origami design.
Next, we describe the characteristics of multistability and present the numerical simulation methods used for structural analysis, along with the modeling frameworks.
Then, we introduce two assembly methods for programming the global systems.
Finally, we discuss material selections and fabrication methods for both desktop and large-scale prototypes.
The flow of this chapter establishes the workflow for the later chapters~\ref{chap:qbellows},~\ref{chap:ttoroid}, and~\ref{chap:ccblock}.

\section{Geometric Design Methods}
\label{sec:2geometry}

In this study, we focus on the design of \emph{over-constrained types} of multistable origami (introduced in Section~\ref{Intro_ArchSys_Multi}).
This type can offer advantages for building-block applications, as panels primarily contribute to the load-bearing capacity in the deployed state.
We broadly categorize two design approaches for the existing multistable origami design method based on rigid origami overconstraining, which facilitate the design of tunable stiffness and stable shapes, respectively.

\subsection{Geometrical Incompatibility Design Method}
\label{method:incomp}

\begin{figure}[tbhp]
  \centering
  \includegraphics[page=1,width=\linewidth]{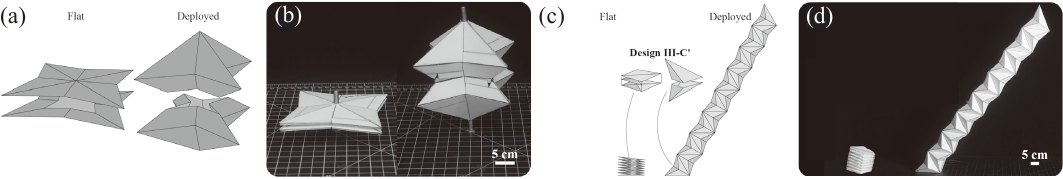}
   \caption{Examples of the geometric incompatibility design method for multistable origami: (a) flat and deployed states of a rigid origami module, and (b) a multistable origami obtained by combining two modules; (c) flat and deployed states of another rigid origami module, and (d) a multistable deployable origami boom assembled from multiple modules. (\emph{Figures (a--d) are adopted from~\cite{melancon2021multistable} in grayscale}.) }
  \label{fig:incompatibilitymelanconn}
\end{figure}

The \emph{geometric incompatibility method} was introduced in~\cite{tachi2017capping, hoberman2019construction, melancon2021multistable}.
This method is based on assembling multiple one-degree-of-freedom(DOF) rigid origami modules by interlocking their open ends in specific deployed configurations (see Figure~\ref{fig:incompatibilitymelanconn} (a) and (c)).
With this approach, the modules are intentionally designed to align only at certain deployed states while remaining misaligned in between.
When these modules are assembled, the system becomes over-constrained, and the system reaches a geometrically stable state at these specific matching configurations (see Figure~\ref{fig:incompatibilitymelanconn} (b) and (d)).
However, during the deployment between matching configurations, it induces geometric incompatibility, leading to panel deformation.
Melançon et al.~\cite{melancon2021multistable} quantified the strain energy arising from this incompatibility and demonstrated that, by adjusting the rigid origami design parameters, the various multistable origami shapes can be designed.
In Chapter~\ref{chap:qbellows}, we use this methodology for designing multistable origami with tunable stiffness.

\subsection{Loop-closing Design Method}
\label{method:loop}

\begin{figure}[tbhp]
  \centering
  \includegraphics[page=2,width=\linewidth]{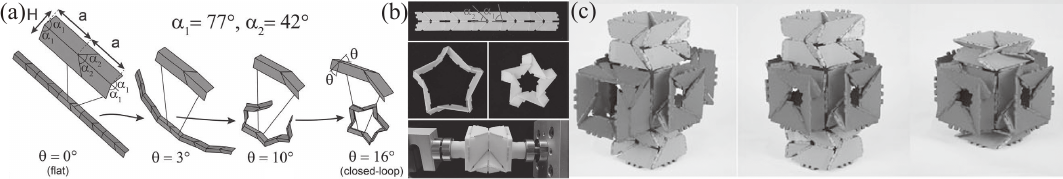}
   \caption{Examples of the loop-closing design method for multistable origami: (a) Loop-closing design using a rigid origami strip, and (b) its physical prototype; (c) a metamaterial application obtained by assembling modules, demonstrating multiple stable configurations with anisotropic stiffness. (\emph{Figures (a--c) are adopted from~\cite{kamrava2019origami} in grayscale.}) }
  \label{fig:loopclosingkamrava}
\end{figure}

The basic concept of this method is that a one-degree-of-freedom rigid origami strip generates curvature during folding, and its two ends coincide at multiple specific configurations.
By connecting both ends, the system becomes over-constrained, and it reaches a stable configuration at certain deployed states, while between these states, panel deformation occurs, resulting in an energy barrier.
A representative example of this design is the Kresling origami and its tubular form~\cite{kresling1994folded,jianguo2015bistable,yasuda2017origami,kresling2020fifth}.

In~\cite{kamrava2019origami}, this approach was termed the \emph{closed-loop}, where a strip of Miura origami was loop-closed to design multistable origami (see Figure~\ref{fig:loopclosingkamrava} (a) and (b)).
Furthermore, they demonstrated that the stiffness could be tuned by adjusting the geometry of the Miura pattern and can design metamaterials with anisotropic stiffness by assembling designed multistable origami (see Figure~\ref{fig:loopclosingkamrava} (c)).

In this study, we named this design method the \emph{loop-closing method}.
In Chapter~\ref{chap:ttoroid}, we employ this design method to construct multistable origami blocks in the form of polygonal frustums.
In Chapter~\ref{chap:ccblock}, we extend this approach to curved origami, resulting in multistable curved-crease origami blocks.

\section{Mechanics Analysis Methods}
\label{sec:2simulations}
In this section, we explain the mechanical characteristics of the multistable structure that serve as the analysis metrics for later analysis. 
We then present the two numerical simulation methods employed to analyze these characteristics, together with the corresponding modeling frameworks.

\subsection{Mechanics of Multistability}
\label{subsec:2multistability}

Multistable origami exhibits multiple stable configurations along a specific parameter vector space representing its deployment. 
In this study, we idealize the origami structure as being composed of thin, linear-elastic panels with no in-plane deformation, connected by zero-stiffness hinges. 
Therefore, while the panel deformation consists of continuous infinitesimal deformations within the elastic range, the origami structure itself undergoes large deformations during deployment. 
This characteristic gives rise to the geometric nonlinearity that governs the mechanical behavior of multistable origami. 
Throughout the deployment process, the deformation of these thin panels induces strain energy, and the resulting energy plot reveals multiple energy valleys.

\begin{wrapfigure}{l}{0.45\textwidth}
    \centering
    \includegraphics[keepaspectratio,width=0.85\linewidth, page=9]{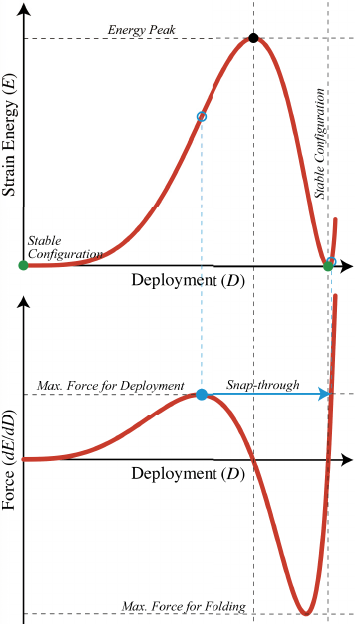}
    \caption{Strain energy (top) and force (bottom) plots of Multistable (bistable) structure.}
    \label{fig:multistability}
\end{wrapfigure}

A fundamental example of multistability is bistability. 
As illustrated in the Figure~\ref{fig:multistability}-top, a bistable origami structure has two energy valleys (green points) along the deployment coordinate (stability depends on the presence of an energy valley, not on having zero energy). 
At the energy peak, the slope of the energy plot also becomes zero, as in energy valleys; however, this configuration is actually unstable, as a tiny perturbation leads the structure away from this state.

The derivative of the strain energy with respect to the deployment parameter corresponds to the force (load) required to deploy the structure, known as the \emph{equilibrium path} (Figure~\ref{fig:multistability}-bottom). 
Therefore, the equilibrium path crosses zero at both stable configurations and the energy peak, where the slope is zero in the energy plot. 
From this force plot, the maximum force required for transition (deployment) from the first stable configuration to the next, and the maximum force(absolute value) needed for the reverse transition (folding) can be identified.
These required forces for deployment and folding can be either symmetric or asymmetric, and this difference in behavior is also a key metric for the block application of multistable origami.

A characteristic feature of a multistable structure is snap-through behavior (blue arrow in Figure~\ref{fig:multistability}-bottom). 
When a certain force (maximum deployment force) is reached, the structure suddenly transitions to a lower-energy configuration while the applied force remains unchanged.
Under force-controlling conditions, i.e., gradually increasing load, this snap-through prevents the full equilibrium path from being captured, making it difficult to understand the overall mechanical behavior. 
Therefore, in both our simulations and experiments, we adopt the forced displacement load condition, gradually increasing the deployment parameter to obtain the full equilibrium path.

\subsection{Bar-and-Hinge Analysis}
\label{subsec:merlin}

\begin{figure}[tbhp]
  \centering
  \includegraphics[page=3,width=\linewidth]{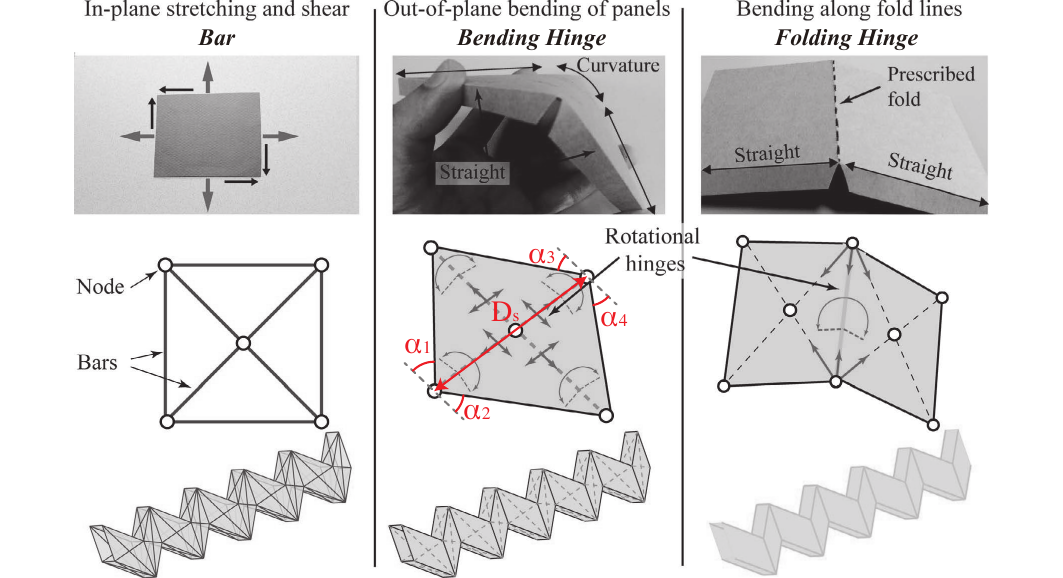}
   \caption{Bar-and-Hinge configurations. (\emph{Figures are adopted from~\cite{filipov2017bar} in grayscale.}) }
  \label{fig:barhingemodeling}
\end{figure}

The \emph{bar-and-hinge model}~\cite{filipov2017bar, Liu2017Mer1EN, liu2018highly} is a simplified yet highly efficient structural analysis method for origami systems, based on reduced-order modeling.
Here, we only adopt the N5B8 model (see Figure~\ref{fig:barhingemodeling}-left), which has better accuracy in capturing the mechanical behavior of origami structures than another reduced-order model, such as N4B5.
The bar-and-hinge model consists of three mechanical behavior components (see Figure~\ref{fig:barhingemodeling}; for details, refer to~\cite{filipov2017bar}):

\paragraph{In-plane stretching and shearing of panels}
This behavior is represented using bar elements (see Figure~\ref{fig:barhingemodeling}-left).
The stiffness of each bar ($K_{bar}~[N/mm]$) can be calculated from its length ($L_{bar}~[mm]$) and cross-sectional area of the bar ($A_{bar}~[mm^2]$) (Eq.~\ref{eq:barstiffness}).
Here, the $A_{bar}$ is defined as a uniform area proposed in~\cite{liu2018highly}, calculated from the polygon panel area ($S~[mm^2]$), the polygon edge length ($L_i~[mm]$), and the panel thickness ($t~[mm]$) (Eq.~\ref{eq:bararea}).
$E~[N/mm^2]$ and $\nu$ represent Young's modulus and Poisson's ratio, respectively.

\begin{align}
\label{eq:barstiffness}
    K_{bar} = E \frac{A_{bar}}{L_{bar}}
\end{align}
\begin{align}
\label{eq:bararea}
    A_{bar} = \frac{2St}{(1-\nu)\Sigma_iL_i}
\end{align}

\paragraph{Out-of-plane bending of panels}
This behavior is represented by the rotation of hinges located along the diagonals of the panels (see Figure~\ref{fig:barhingemodeling}-mid).
The stiffness of the bending hinges ($K_{panel}~[N \cdot mm]$) is computed from the panel thickness ($t~[mm]$), the bending modulus ($B~[N \cdot mm]$, Eq.~\ref{eq:bendingmodulus}), the short diagonal length ($D_s~[mm]$) of the panel, and the corner geometry of the short diagonal ($\alpha_i~[rad]$) (Eq.~\ref{eq:Kpanel}).

\begin{align}
\label{eq:bendingmodulus}
     B = \frac{Et^3}{12(1-\nu^2)}
\end{align}

\begin{align}
\label{eq:Kpanel}
    K_{panel} = \left(0.55-0.42\frac{\Sigma\alpha}{\pi}\right) B \left(\frac{D_s}{t}\right)^{\frac{1}{3}}
\end{align}

\paragraph{Bending of creases}
Similarly, this behavior is modeled by rotational hinges.
The stiffness of a folding hinge ($K_{hinge}~[N \cdot mm]$) is treated as a parameter in this study and is computed from the bending modulus ($B~[N \cdot mm]$), the crease length ($L_{crease}~[mm]$), and a length scale factor ($f~[mm]$) (Eq.~\ref{eq:Khinge}).
To prevent unrealistically large hinge stiffness, $K_{max}~[N \cdot mm]$ is introduced for the upper bound, and the crease stiffness ($K_{crease}~[N \cdot mm]$) is then determined by combining these two quantities as expressed in Eq.~\ref{eq:Kcrease}.
As noted in~\cite{filipov2017bar}, however, $K_{max}$ only acts as an upper bound and has negligible influence on the results of the analysis, meaning that the $K_{hinge}$ is the dominant value of $K_{crease}$; therefore, $K_{max}$ is omitted in this study.

\begin{align}
\label{eq:Khinge}
    K_{hinge} = \frac{L_{crease}}{f} B
\end{align}
\begin{align}
\label{eq:Kcrease}
    K_{crease} = \left(\frac{1}{K_{hinge}} + \frac{1}{K_{max}}\right)^{-1}
\end{align}

In this study, the bar-and-hinge model simulations were carried out using \emph{MERLIN2}~\cite{liu2018highly}, a \emph{MATLAB}~\cite{matlab}-based software.
We focus solely on the geometric and elastic behavior; therefore, the hinge stiffness is minimized (approaching zero) by assigning an extremely large length scale factor ($f~[mm]$).
It should also be noted that Young's modulus ($E$) functions as a scaling parameter for comparing relative stiffness, rather than representing an actual material property.

\subsection{Finite Element Analysis}
\label{subsec:FEA}

\begin{figure}[tbhp]
  \centering
  \includegraphics[page=4,width=\linewidth]{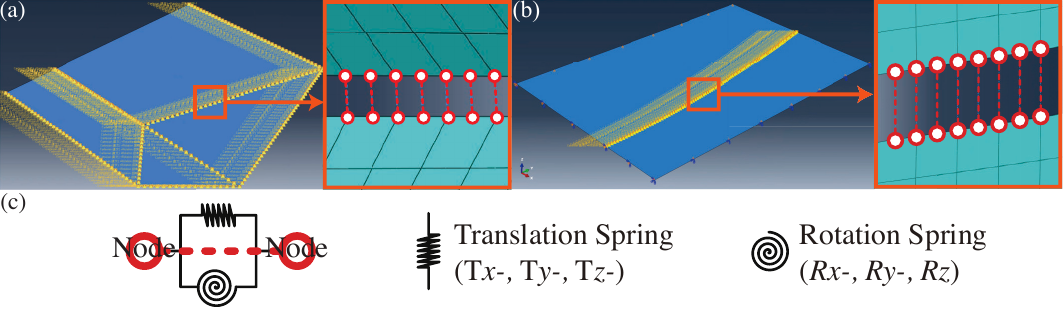}
   \caption{(a) Linear- and (b) Curved-crease modeling for the finite element analysis(FEA). (c) The types of connectors used in the FEA software interface.}
  \label{fig:feamodeling}
\end{figure}

To capture detailed panel deformation and obtain accurate results, we employ finite element analysis (FEA) using \emph{ABAQUS (Dassault Systèmes)}~\cite{abaqus}.
To stably find the equilibrium path of multistable origami, we used an implicit solver to compute the equilibrium solution at each step. 
Additionally, we performed a quasi-static analysis to determine the equilibrium path governed by geometric nonlinearity by minimizing the effects of inertia
The content covered in this subsection is partially included in publications~\cite{lee2025multistable, snappingtoroids-2025}.

\paragraph{Panel Modeling}
To model the thin panels (faces) of the origami structures, we used shell elements within the FEA software.
Shell elements are more suitable than solid elements for thin panel structures, as they avoid thickness-direction meshing, thereby reducing the total number of elements and improving computational efficiency.
Each panel was discretized into quadrilateral meshes, and the S8R element type, an 8-node (4 corner nodes and 4 midpoint nodes), second-order quadrilateral shell element, was employed.
This element formulation offers improved bending response accuracy and improved computational efficiency by reducing integration, while mitigating overstiffness errors.

\paragraph{Crease Modeling}
Figure~\ref{fig:feamodeling} shows the (a) linear crease and (b) curved crease modeling methods of an origami structure in the FEA software interface.
Adjacent edges (creases) were divided into equal numbers of nodes to ensure node connectivity.
To simulate the folding behavior in the FEA software, we developed a new hinge modeling method, based on the pin-joint array hinge proposed in~\cite{lee2024design}.
We implemented pin joints by connecting mesh nodes at identical locations using connector elements. 
Each connector type was defined with \emph{Cartesian-type} for the translational spring in $x$, $y$, and $z$-direction and \emph{Rotation-type} for the rotational spring in $Rx$, $Ry$, and $Rz$-direction.
By assigning equal stiffness to the $x$, $y$, and $z$ translational directions, and likewise to the $R_x$, $R_y$, and $R_z$ rotational directions, we ensured that each pin-joint maintains the intended translational and rotational stiffness regardless of its position in the global coordinate system.
This approach differs from previous methods~\cite{lahiri2024improving, almessabi2024reprogramming, zhou2024tunable}, which rely on local coordinate systems at each node connection.
In contrast, our method achieves hinge behavior using only a global coordinate, simplifying the modeling process and reducing computation cost.

In this study, we focus only on the geometric and elastic behavior of the structure. 
Therefore, the complex physical phenomena from panels and creases, such as plasticity, viscoelasticity, damping, and contact, are not considered.
As the default setting, we assign only linear stiffness values to each pin joint, with a translational stiffness of ``\SI{1.0e6}{N/mm} / edge nodes count'' and a very low rotational stiffness of ``\SI{0.001}{N\cdot mm/ rad} / edge nodes count''.
Detailed codes for the crease modeling are provided in Appendix~\ref{chap:AppendixA}.

\section{Programmable Assembly Methods}
\label{sec:2assemblies}

In this section, we introduce two fundamental assembly methods: a bottom-up method, where the global system emerges from local assembly logic, and a top-down method, where the local block is derived by subdividing a prescribed global system. 
These two methods establish the foundation for programming the global geometry and mechanical properties of multistable origami block assemblies.

\subsection{Bottom-up Method: Logic-guided Assembly}

\begin{figure}[tbhp]
  \centering
  \includegraphics[page=4,width=\linewidth]{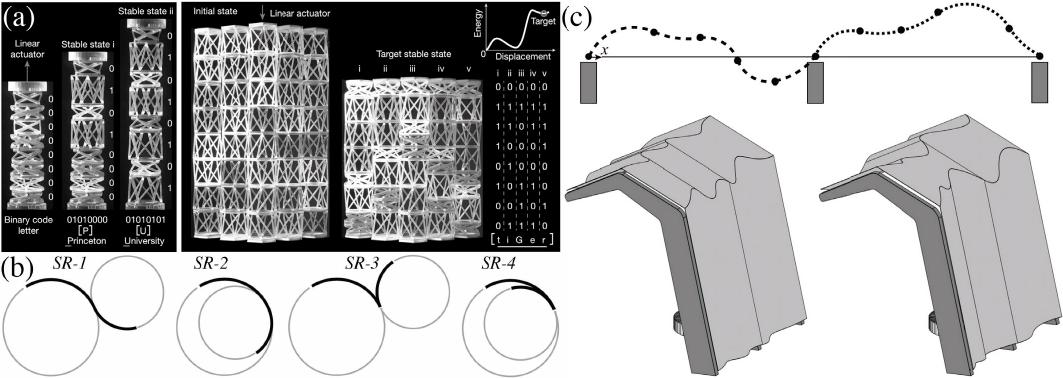}
   \caption{(a) Binary mechanical memory system formed based on the Kresling origami-inspired structures. (\emph{Figures are adopted from ~\cite{zhao2025modular} in grayscale}.) (b) Shape rules (SR) of arc segment combinations, and (c) curved origami claddings generated through the shape grammar based on these shape rules. (\emph{Figures (b) and (c) are adopted from ~\cite{gattas2018generative}}.) }
  \label{fig:logicassembly}
\end{figure}

The bottom-up method constructs global systems by combining blocks according to the logic sequences.
In this logic-guided assembly, blocks are assembled according to predefined assembly rules, and the ordered sequence of these rules serves as the logic input that programs the mechanical behavior or global shape of the system.
The representative logic-guided assembly methods include binary mechanical memory (Figure~\ref{fig:logicassembly} (a))~\cite{yasuda2017origami, yasuda2021mechanical, zhao2025modular}, shape grammar (Figure~\ref{fig:logicassembly} (b))~\cite{stiny1980introduction, stiny1982spatial, gattas2018generative, yu2021rethinking}, graph grammar~\cite{klavins2004graph, klavins2007programmable}, discrete aggregation~\cite{rossi2018voxels} and L-systems~\cite{nouri2023development}.
This method enables the global system to be reconfigurable, including modular assembly, part replacement, deployment sequence control, and global shape control, suitable for our multistable origami modular building system.

In this study, we adopted two of the simplest and most intuitive methods, binary mechanical memory and shape grammar, to achieve the reconfigurability of our modular systems in Chapters~\ref{chap:qbellows} and \ref{chap:ccblock}, respectively.

Binary mechanical memory uses the discrete stable states of bistable blocks to represent physical bits, 0 and 1. 
By arranging these blocks in series, previous studies~\cite{yasuda2017origami, zhao2025modular} have encoded simple logical operations directly into the global system, allowing the snapping order of blocks to function as mechanical memory, as shown in Figure~\ref{fig:logicassembly} (a). 
By extending this binary memory as an assembly method, we can get reconfigurability of the global system, such as programmable deployment sequences, partial deployment, and hysteresis control.

Shape grammar is a geometry-based rule system in which the assembly logic is encoded through shape rules~\cite{stiny1980introduction} that specify how modules connect, such as tangent, stacking, rotating, or interlocking (Figure~\ref{fig:logicassembly} (b)).
By iteratively applying these rules, the global system is generated and programmed, allowing complex assemblies to emerge from simple local shape rules (Figure~\ref{fig:logicassembly} (c)).
By adjusting the rule sequences, intuitive reprogramming, partial replacement, and reconfigurability of the global systems can be achieved.

\subsection{Top-down Method: Mesh-guided Assembly}

\begin{figure}[tbhp]
  \centering
  \includegraphics[page=5,width=\linewidth]{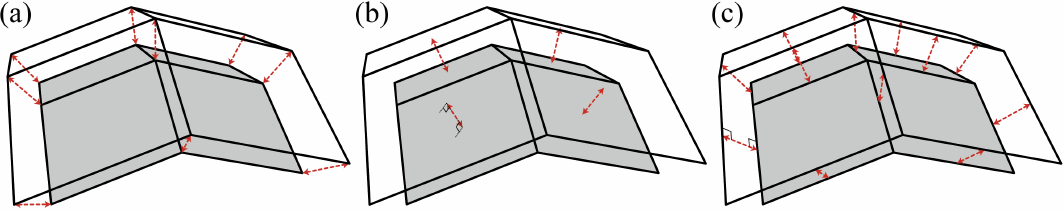}
   \caption{The concept illustrations of (a) vertex-offset, (b) face-offset, and (c) edge-offset meshes.}
  \label{fig:meshassembly}
\end{figure}

The top-down method derives block assemblies from a prescribed global shape by first discretizing the surface into a mesh and then designing blocks that meet the local geometric constraints, such as outermost edges and dihedral angles, defined by that mesh. 
In this study, we limit the meshing mechanism to planar meshes, specifically, \emph{planar quad (PQ) meshes}, to directly correspond to an individual block with defined mesh edges and dihedral angles, ensuring that the assembled global system reproduces the target discretized surface~\cite{bobenko2006minimal, bobenko2017discrete, pottmann2007geometry, pottmann2010edge, liu2006geometric, sauer1970differenzengeometrie, bobenko2008surfaces, pottmann2008focal}.

In discrete differential geometry, PQ meshes are usually modeled based on the conjugate nets (Q-nets) on curved surfaces~\cite{sauer1970differenzengeometrie}. 
By aligning these Q-nets with principal curvature directions, they enable multilayers with constant offset.
These offset capabilities can be valuable for architectural geometry, such as constructing roofs, facades, and shells from panels or beams with uniform thickness~\cite{liu2006geometric, pottmann2007geometry, pottmann2008focal, pottmann2010edge}. 
Depending on the targeted offset property, PQ meshes can be categorized into three types: vertex, face, and edge offset meshes.

\emph{Vertex-offset mesh} (Figure~\ref{fig:meshassembly} (a)) originates from a circular mesh, where every face has a circumcircle and the opposite angles of each face sum to $\pi$.
Circular meshes are Möbius-invariant (conformal) and parallel to meshes whose vertices lie on the unit sphere.
These properties ensure that each vertex can be shifted by a constant distance along each vertex normal, thereby forming consistent vertex-offset families~\cite{bobenko2006minimal, bobenko2008surfaces, pottmann2008focal}.

\emph{Face-offset mesh} (Figure~\ref{fig:meshassembly} (b)) originates from a conical mesh, characterized by the property that all faces around a vertex are tangent to a common cone, meaning that the face normals intersect at a single point.
Geometrically, the sum of opposite angles between edges around a vertex is equal.
Conical meshes are Laguerre-invariant(face-tangential) and parallel to meshes whose faces are tangent to the unit sphere.
These conditions ensure that each face can be shifted by a constant distance along its face normal, forming face-offset families~\cite{liu2006geometric, pottmann2008focal}.

\emph{Edge-offset mesh} (Figure~\ref{fig:meshassembly} (c)) is defined by a parallel-mesh relation, where corresponding edges of two meshes remain strictly parallel.
It is Laguerre-invariant(edge-tangential), and its Gauss map is a Koebe polyhedron, meaning that all edges of the associated mesh are tangent to the unit sphere, producing an incircle for each face that generates an orthogonal circle pattern~\cite{bobenko2006minimal, bobenko2008surfaces}.
This property provides the Christoffel duality, yielding a dual mesh that forms a discrete minimal surface while preserving the edge-parallel directions~\cite{bobenko2017discrete, tovsiae2019design}.
These conditions ensure that each edge can be shifted by a constant distance while preserving edge-parallelism (Combescure transformation~\cite{pirahmad2025area}), forming edge-offset families~\cite{pottmann2007geometry, pottmann2010edge}.

In Chapter~\ref{chap:ttoroid}, we employed the edge-offset mesh method to assemble complex global geometries, taking into account the geometric features of our multistable origami block.

\section{Physical Prototype Methods}
\label{sec:2fabrications}

In this section, we describe the material selections and fabrication processes for constructing multistable origami prototypes at both desktop and large scales.

\subsection{Material Selection}
The prototype fabrication of multistable origami blocks requires panel materials that are thin, lightweight, and capable of undergoing repeated elastic deformation without plastic damage. 
Because the snapping behavior depends on the strain energy from panel bending, the material must deform within the elastic range during snap-through. 
At the same time, to remain stable in deployed configurations, the panels need sufficient in-plane and bending stiffness to support their own weight or external loads. 
This requirement introduces a trade-off between the flexibility and load-bearing of panel material.
Furthermore, repeated deployment–folding cycles also cause stress concentrations in the local areas of the panel, making durability an important consideration for reliable usage.

For these reasons, we selected polyester- and polypropylene-based materials for both desktop and large-scale prototypes.
These materials offer an effective balance between the flexibility required for snapping and the stiffness needed to maintain deployed configurations.
They also provide sufficient durability to withstand repeated use, and are easy to cut, fabricate, and process during prototype construction.

The crease regions (hinges) undergo highly localized and repeated bending(folding); they are required to be durable against fatigue. 
To approximate the ideal origami assumption, where hinge stiffness and energy dissipation (e.g., friction or play) are negligible, the hinge material should possess significantly lower bending stiffness than the panels, and the crease width should be kept minimal.
Additionally, the fabrication methods should be able to accommodate not only straight creases but also curved creases.

For these reasons, we selected membrane-type materials, such as non-woven fabrics, adhesive tapes, and a monolayer polypropylene sheet (for a large-scale case), as a hinge material to realize straight and curved creases. 
These membrane hinge enables smooth, continuous bending along any crease geometry with low stiffness and minimal play, thereby preserving the ideal assumptions of origami kinematics and mechanics more than other hinge structures, such as mechanical hinges and compliant hinges~\cite{lee2024design}. 
Additionally, they provide sufficient durability against folding fatigue while remaining easy to cut, adhesive, and replace during prototype fabrication.

\subsection{Desktop-scale Fabrication}

For desktop-scale prototypes, we propose two fabrication methods that utilize polyester and polypropylene (PP)-based elastic panels and membrane-based low-stiffness hinges, enabling ideal snapping behavior in multistable origami.

\subsubsection{Method 1: Polyester-Laminated Film Panel}
\label{subsubsec:deskfabmethod1}

\begin{figure}[tbhp]
  \centering
  \includegraphics[page=5,width=\linewidth]{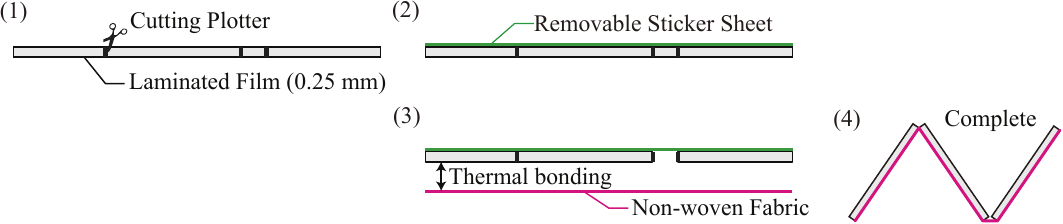}
   \caption{The process illustration of desktop-scale prototype fabrication with PE-laminated film and non-woven fabric.}
  \label{fig:desktopfabmethod1}
% \end{figure}
\vspace{0.5cm}
% \begin{figure}[tbhp]
  \centering
  \includegraphics[page=1,width=\linewidth]{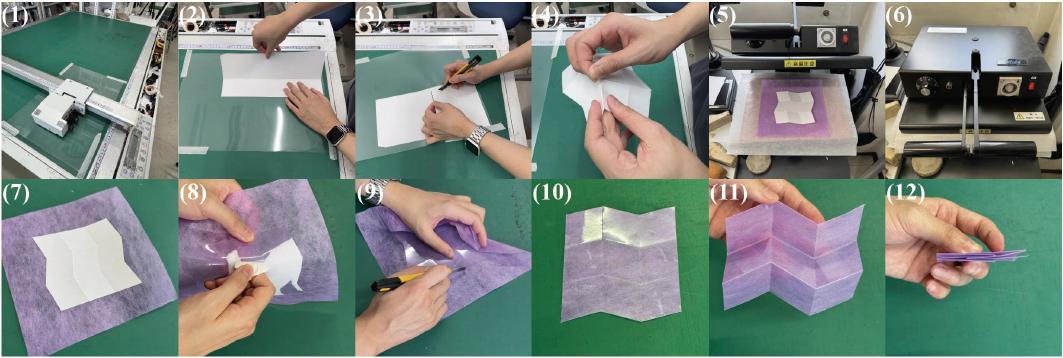}
   \caption{The picture of the actual fabrication process with the polyester-laminated film panel. (1) Cut the laminated film sheet, (2) attach the removable sticker to the sheet, (3, 4) cut off and detach the unnecessary parts, and bridges between panels, (5) place the cut sheet and non-woven fabric into the iron press, (6) bake it together (thermal bonding) for $10-20$ seconds at $120-140^\circ$, (7) put out and cooling, (8) peel off the sticker, (9) cut off the unnecessary parts of fabric, (10--12) complete and fold it.}
  \label{fig:desktopfabmethod1-1}
\end{figure}

We used \SI{0.25}{mm}-thick heat-adhesive polyester-based laminate film as rigid panels and non-woven fabric (also polyester) to create a foldable hinge with low rotational stiffness.
The fabrication process is as follows (see Figure~\ref{fig:desktopfabmethod1}):
\begin{enumerate}
    \item The cutting pattern is cut into a laminated film using a cutting plotter (or laser cutter), using a single-line cut on mountain folds and a \SI{0.7}{mm} wide cut on valley folds.
    \item Attach a removable sticker sheet to the cut laminate film. 
    \item Remove the part that doesn't become the panel of the module, and thermally bond with non-woven fabric using an iron.
    \item Remove the sticker sheet and complete other processes, such as trimming off the unnecessary parts, and assembling the parts using double-sided tape.
\end{enumerate}
Figure~\ref{fig:desktopfabmethod1-1} shows the detailed actual fabrication process of the Miura pattern example.

\subsubsection{Method 2: PP-Coated Paper Panel}
\label{subsubsec:deskfabmethod2}

\begin{figure}[tbhp]
  \centering
  \includegraphics[page=7,width=\linewidth]{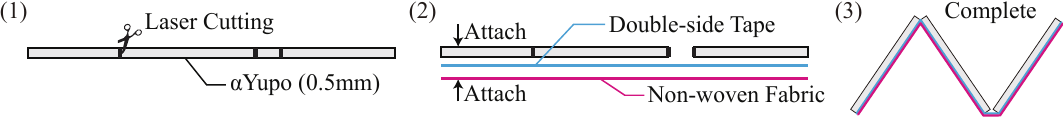}
   \caption{The process illustration of desktop-scale prototype fabrication with PP-coated paper, double-sided tape, and non-woven fabric.}
  \label{fig:desktopfabmethod2}
% \end{figure}
\vspace{0.5cm}
% \begin{figure}[tbhp]
  \centering
  \includegraphics[page=2,width=\linewidth]{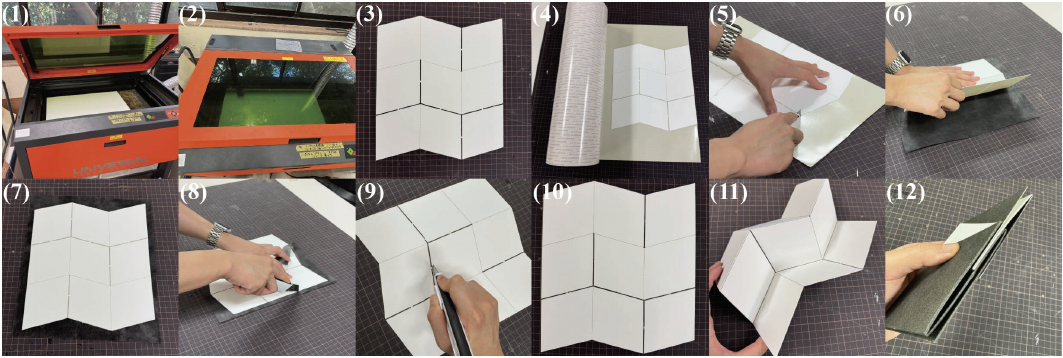}
   \caption{The picture of the actual fabrication process with the PP-coated paper panel. (1, 2) Put the Yupo sheet into the laser cutting machine and operate, (3) take out the cut pattern and clear the unnecessary parts, (4) attach the pattern to the double-sided tape sheet, (5) remove unnecessary parts of tape, (6) attach the pattern on the non-woven fabric, (7) press it to be well bonded, (8) remove unnecessary parts of non-woven fabric, (9) remove the bridges between panels, (10--12) complete and fold it.}
  \label{fig:desktopfabmethod2-1}
\end{figure}

We used a \SI{0.5}{mm}-thick polypropylene-coated paper (\emph{ALPHAYUPO}~\cite{yupo}, referred to here as the Yupo sheet) as the panel material, and we attached non-woven fabric to the entire side face using a double-sided tape sheet. 
The fabric and double-sided tape function as membrane hinges, having relatively low hinge stiffness compared to the panel material. 
The fabrication process is as follows (see Figure~\ref{fig:desktopfabmethod2}):
\begin{enumerate}
    \item The pattern is cut into the Yupo sheet using the laser cutting machine, using a single-line cut on mountain folds and a \SI{1.4}{mm} wide cut on valley folds;
    \item The double-sided tape sheet is attached to one whole side along with the non-woven fabric membrane.
    \item Complete other processes, such as trimming off the unnecessary parts, and assembling the parts using double-sided tape.
\end{enumerate}
Figure~\ref{fig:desktopfabmethod2-1} shows the detailed actual fabrication process of the Miura pattern example.

\subsection{Large-scale Fabrication}
\label{subsec:largefab}

\begin{figure}[tbhp]
  \centering
  \includegraphics[page=3,width=\linewidth]{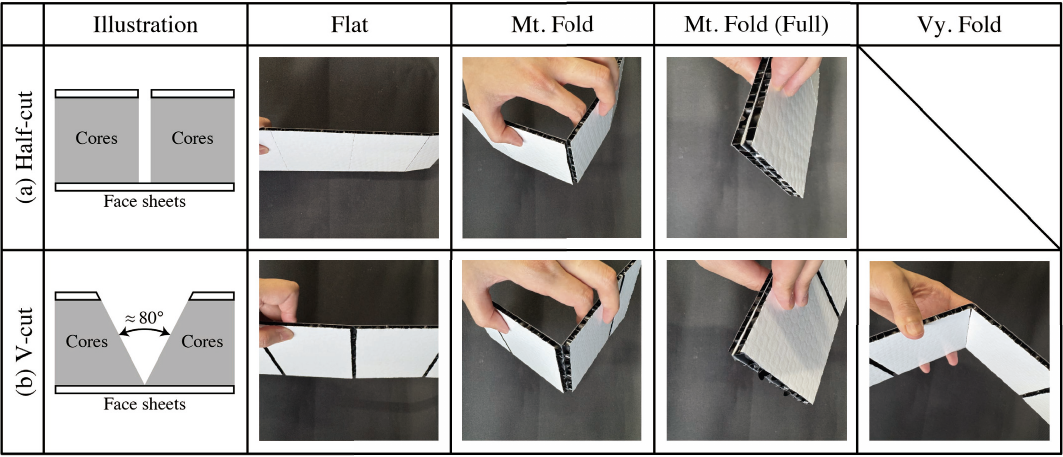}
   \caption{(a) Half-cut and (b) V-cut hinges applied to the Plapearl panel, and folding motions.}
  \label{fig:plapearl}
\end{figure}

For large-scale fabrication, we employed a polypropylene bubble-core sandwich sheet (\emph{Plapearl} \cite{plapearl}) as the panel material, selected for its lightweight and elastic properties.
Especially, in this study, we used four types of \emph{Plapearl} for large-scale fabrications$^1$: \emph{PCPPZ-050, PDPPZ-080, PDPPZ-100, and PGPPZ-300}.
Figure~\ref{fig:plapearl} shows the cutting methods and folding behavior of a \SI{5}{mm}-thick \emph{PDPPZ-100} Plapearl panel$^1$.
Panels were cut using a CNC cutter, and creases were fabricated either as half-cuts or V-cuts.
In both cutting methods, the bottom sheet layer is left intact, allowing the hinge motion to be realized through the plastic deformation of the remaining PP sheet.

Both cutting methods allow complete mountain folds.
However, for valley folds, half-cuts become blocked, while V-cuts can fold up to approximately $100^\circ$ before being blocked.
Therefore, the half-cut method is suitable for relatively simple origami patterns dominated by mountain folds, or when a smooth panel surface without visible crease lines is preferred in the deployed state.
The blocking behavior of valley folds can also provide structural benefits, such as enhanced load-bearing capacity of the panels.

In contrast, the V-cut method is more suitable for complex origami patterns that involve both mountain and valley folds, or when the design requires reducing the panel stiffness.

For connecting panel parts, we employed membrane hinges made of double-sided tape or duct tape, depending on the structural requirements.
These connection strategies are discussed in more detail in each subsection of the large-scale fabrication section.

\subsubsection{Minimum Performance Criteria}
For the large-scale prototypes, we define a set of minimum performance criteria to evaluate whether the fabricated multistable origami blocks can function as basic architectural furniture elements, such as stools, partitions, or beds.
Note that these criteria are not intended to demonstrate full structural compliance, but rather to assess the practical feasibility of multistable origami blocks at large scales.

First, the prototype must be sufficiently lightweight to be handled by a single person, and we set a block weight of less than \SI{5}{kg}.
Second, the block should be deployable and foldable by a single person and exhibit sufficient multistability to remain stable in its deployed state without external supports, ensuring reliable shape fixation during use.
Third, the block must be self-standing and capable of supporting its own weight at least, corresponding to a partition-level structural performance.
Furthermore, if possible, it would be useful to support the weight of a single adult (here, we set \SI{70}{kg}) for use as a stool or bed-level structural performance.

By satisfying these minimum criteria, we demonstrate that the proposed multistable origami blocks can go beyond the conceptual level and achieve real-world architectural and furniture-scale applications, while maintaining the advantages of lightweight, compactness, deployability, and reconfigurability.

\section{Workflow}

The subsequent chapters follow a workflow that builds upon the methodologies introduced in this chapter. 
The workflow for each chapter can be summarized as follows (also, refer to Figure~\ref{fig:workflow}):

\begin{itemize}
    \item First, we design the geometry of multistable origami blocks based on the methods introduced in Section~\ref{sec:2geometry}.
    \item Second, we analyze the mechanical behaviors of the designed blocks across different parameter variations using the simulation methods described in Section~\ref{sec:2simulations}. 
    To validate these results and capture actual behaviors, we fabricate desktop-scale prototypes following the processes outlined in Section~\ref{sec:2fabrications}. 
    \item Third, considering geometric and mechanical features of designed blocks, we employed assembly methods in Section~\ref{sec:2assemblies} and demonstrate global systems.
    \item Finally, each chapter concludes with the fabrication of desktop and large-scale prototypes (Section~\ref{sec:2fabrications}), highlighting the feasibility of our blocks in architectural use.
\end{itemize}

	\chapter{Q-Bellows: Multistable Quadrilateral Boundary Origami Block}
\label{chap:qbellows}

\emph{\textbf{Author's note:}
The content covered in this chapter is partially included in publications~\cite{miyajima2022designing,lee2024designing}.
% It should be noted that the contents in Sections~\ref{sec:Qgeodesign}--\ref{sec:design scheme} of this chapter are attributable to the contributions of ``\emph{Yuki Miyajima}'' and are described in detail in his master degree thesis~\cite{mastermiyajima}.
}

\section{Introduction}

\begin{figure}[thbp]
  \centering
  \includegraphics[page=1,width=\linewidth]{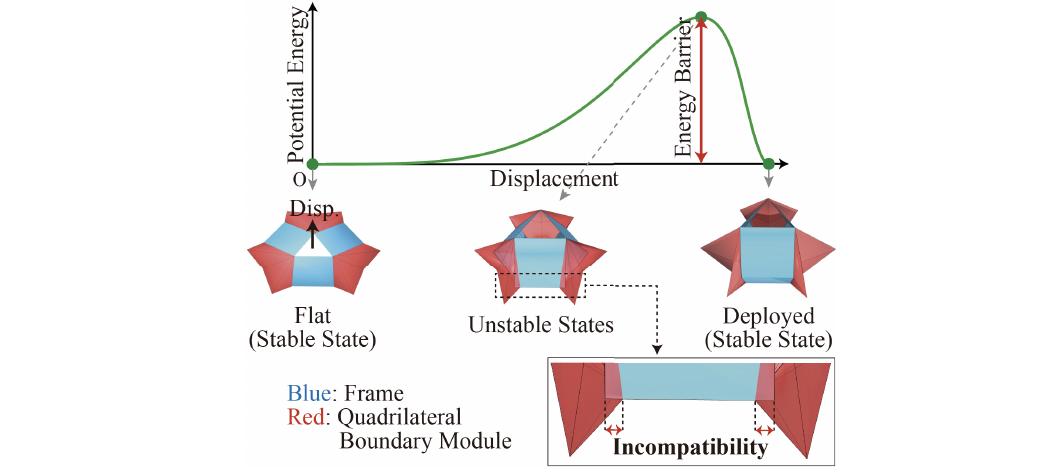}
  \caption{Plot of potential energy and displacement in a bistable structure of our design. Blue indicates the frame with quadrilateral holes, and red indicates the quadrilateral boundary module. The incompatibility between the frame and the modules creates a potential energy barrier.}
  \label{fig:schematic_graph}
\end{figure}

Multistable origami has been explored as a building block for mechanical devices~\cite{ishida2017design, chang2020kirigami}, metamaterials~\cite{yasuda2017origami, liu2022triclinic, zhao2025modular}, and large-scale deployable architectural systems that can remain stable in the deployed state~\cite{hoberman2019construction, melancon2021multistable}.
However, most existing design approaches of multistable origami rely on assembling well-known origami patterns as building blocks.
Because these patterns transform in predefined manners, they inherently limit the range of achievable motions, stable configurations, and mechanical properties.

To tackle these issues, we present the design and analysis of multistable origami blocks with prescribed motion and tunable stiffness.
We create multistable origami blocks by combining the linkage structure and rigid origami modules, while intentionally introducing geometric incompatibility between them to achieve snapping behavior  (see Figure~\ref{fig:schematic_graph}).
We (1) prescribe the overall motion by using linkage structures with quadrilateral holes (called frame) and designing \emph{quadrilateral boundary modules} that fit into the frame, and
(2) introduce controlled geometric incompatibility by providing the representation model of the motion of the module and tuning the snapping stiffness through optimization-based approaches, 
(3) validate the designed blocks through bar-and-hinge simulations, experiments, and 
(4) demonstrate the potential application of mechanical memory systems via serial chain assembly and propose a prediction method for snapping in such assemblies.
These objectives are addressed in Sections~\ref{sec:Qgeodesign},~\ref{sec:design scheme},~\ref{sec:validation}, and~\ref{sec:application}.

\section{Geometry Design}
\label{sec:Qgeodesign}

\subsection{Frame}
\label{sec:linkage}

\begin{figure}[tbhp]
  \centering
  \includegraphics[page=2,width=0.75\linewidth]{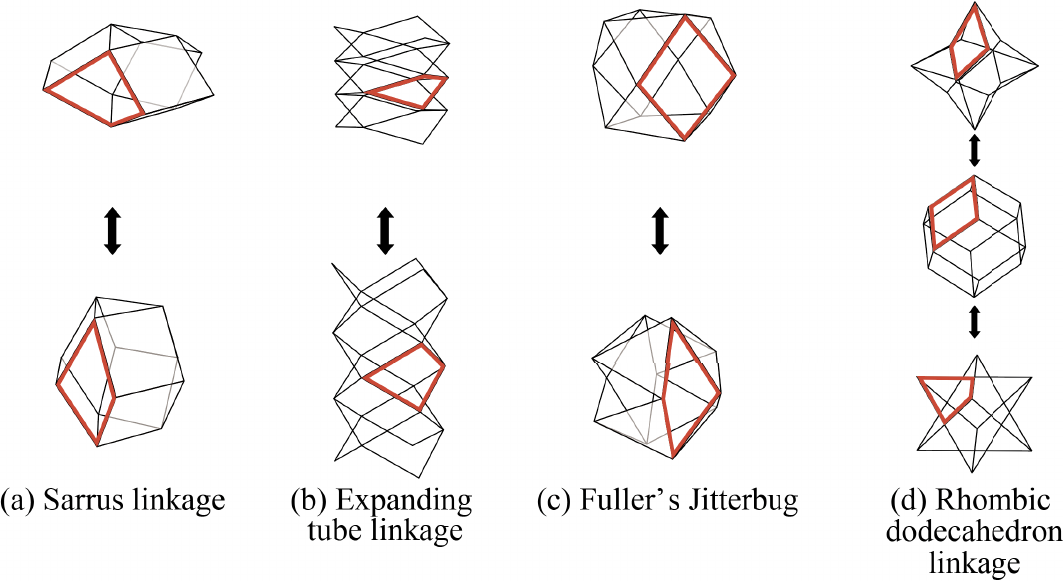}
   \caption{Examples of linkages with quadrilateral frames. Red lines show the quadrilateral frame.}
  \label{fig:linkage_example}
\end{figure}

As the \emph{frame} structure for prescribing the deployment motion, we focus on the Sarrus linkage (Figure~\ref{fig:linkage_example} (a)), which generates linear motion.
By filling its quadrilateral holes with appropriate origami modules, the linkage becomes over-constrained and forms a flat-foldable closed tube, establishing a foundation for designing multistable structures with tunable stiffness and prescribed deployment motion.

Although this study uses the Sarrus linkage as a case study, the same design approach can also be applied to other linkage structures, such as the expanding tube linkage (Figure~\ref{fig:linkage_example} (b)), Fuller's Jitterbug (Figure~\ref{fig:linkage_example} (c)), and the rhombic dodecahedron linkage (Figure~\ref{fig:linkage_example} (d)).

\subsection{Quadrilateral Boundary Modules}
\label{sec:quadrilateralmodule}

\begin{figure}[tbhp]
  \centering
  \includegraphics[page=3,width=0.75\linewidth]{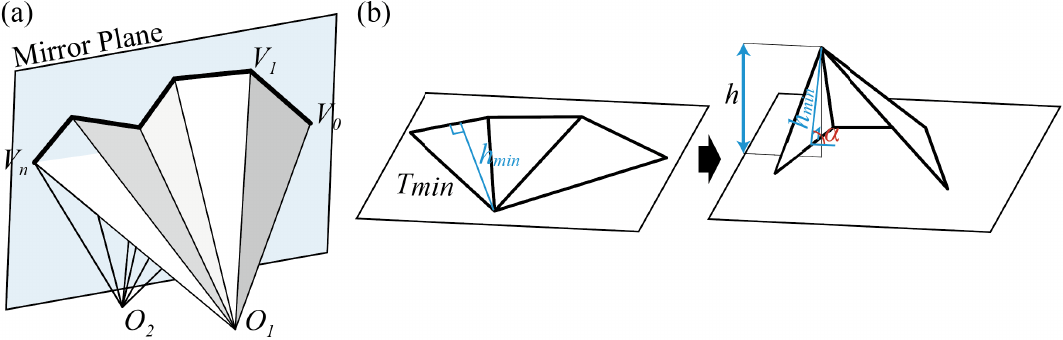}
   \caption{(a) Mirror-symmetric biconical module. (b) Kinematics of a mirror-symmetric biconical module.}
  \label{fig:biconical_module}
\end{figure}

% \begin{figure}[tbhp]
%   \centering
%   \includegraphics[page=4,width=0.75\linewidth]{figure/Chapter3.pdf}
%    \caption{Mode bifurcation by the flip of triangles. There are two triangles with minimum height, resulting in $4$ modes.}
%   \label{fig:flip}
% \end{figure}

A \emph{quadrilateral boundary module} is defined as a polyhedral surface homeomorphic to a disk with a quadrilateral boundary.
When triangulated, such a module forms a $1$-DOF rigid origami mechanism, where the DOF is computed as the number of boundary edges minus three~\cite{tachi2009simulation}.
The mechanism of the quadrilateral boundary module origami with developability has been studied in~\cite{demaine2017folded}.
Here, we extend the use of quadrilateral boundary modules beyond developability constraints and introduce an improved parameter space that provides a reasonable incompatibility metric for tuning snapping stiffness.

\subsubsection{Mirror-symmetric Biconical Modules}
We propose a family of quadrilateral boundary modules called \emph{mirror-symmetric biconical modules}, which is a family that contains the four-panel modules in~\cite{melancon2021multistable} and some of the developable modules in~\cite{demaine2017folded}.

In Figure~\ref{fig:biconical_module}, we give a simple open polyline $V_0V_1...V_n$, called \emph{profile polyline}, that lies on a plane, \emph{mirror plane}, and two points $O_1, O_2$, \emph{pivots}, symmetric about the plane.
A \emph{mirror-symmetric biconical module} is a quadrilateral boundary module that is a union of two generalized polyhedral cones formed by extruding a profile polyline to each of the pivots.
A mirror-symmetric biconical module consists of $n$ pairs of triangular panels and has a sequence of $n-1$ quadrivalent interior vertices along the profile polyline, which produce the 1-DOF mechanism.

We describe the kinematics by finding the profile polyline staying on the mirror plane while the pivots move away from or closer to each other symmetrically about the mirror plane.
The initial completely flatly folded state is given when $O_1$ and $O_2$ are on the mirror plane.
As $O_1$ moves away by distance $h$ from the plane, each triangle rotates about the edge of the profile polyline.
The profile polyline continuously transforms to keep the lifted triangles connected.

The specific motion can be given as follows.
Refer to Figure~\ref{fig:biconical_module} (b). 
Let $T_{min}$ be the triangle with the shortest height $h_{min}$, i.e., the distance from the pivot to the edge on the profile polyline.
Then, $h = h_{min}\sin\alpha > 0$ increases as $\alpha >0$ increases, where $\alpha$ denotes the rotated angle measured between $T_{min}$ and the mirror plane.
Each triangle $T_i$ with distance $h_i$ is then rotated by $\arcsin{\frac{h_{min}}{h_{i}}\sin\alpha}$ to form a 1-DOF motion.
The distance $d$ takes its maximum when $\alpha = 90^\circ$; we call the corresponding folded state the \emph{fully opened state}.

% The entire deformation behavior of a module from the flat state to the fully opened state is unique.
% However, after the module hits the fully opened state, the mechanism bifurcates into multiple modes by the \emph{flipping} of the panels (Figure~\ref{fig:flip}).
% Because $d = h_{min}\sin{\alpha} = h_{min}\sin{\left(180^\circ - \alpha\right)}$, $T_{min}$ at $\alpha = 90^\circ$ can either fold in $(90^\circ,180^\circ)$ (flips) or folds back in $(0,90^\circ)$ with the same pivot positions $d$.
% Each triangle $T_i$ with a larger height $h_i > h_{min}$ cannot flip because it cannot take the rotation angle of $90^\circ$.
% If there are multiple triangles $T_{min}$ with minimum height $h_{min}$, then each of $T_{min}$ can either flip or not, creating $2^{\textrm{\# of } T_{min}}$ different modes.
% Although the mode transition by flipping is an interesting phenomenon by itself, it is necessary to prevent it in our design, as it can cause unpredictable transitions of states.

\subsection{Representation Model}
\label{sec:representation}

\begin{figure}[tbhp]
  \centering
  \includegraphics[page=5,width=\linewidth]{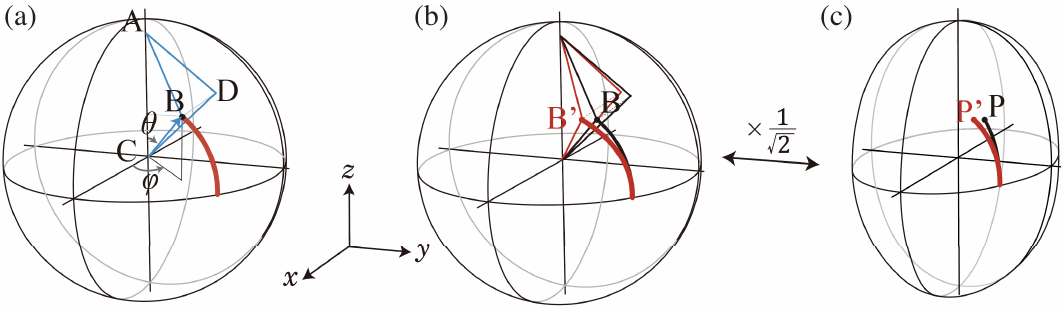}
   \caption{Representation curve. (a) The deformation behavior of a quadrilateral boundary $ABCD$ on a sphere. The black curve represents the locus of point $B$ when the quadrilateral boundary is deformed. (b) locus of points $B$ and $B'$ of two quadrilateral boundary modules. (c) Representation curves, i.e., the locus of $P$ and $P'$ on the ellipsoid (the unit sphere scaled by $1/\sqrt{2}$ in the $y$-axis direction.}
  \label{fig:representation}
\end{figure}

\subsubsection{Representation} 
\label{sec:rep}
% The quadrilateral boundary modeled as a four-bar linkage has 2 DOF, while the rigid origami module inside the quadrilateral boundary is a 1-DOF structure.
% Hence, the deformation of the quadrilateral boundary module can be represented as a 1-dimensional path (configuration space) in a 2-dimensional coordinate space (parameter space).
% In ~\cite{demaine2017folded}, the parameterization of the linkage is to use diagonal lengths; however, the use of diagonal lengths results in a singularity at the flat state, meaning that the infinitesimal (first-order) motion of the frame can be achieved without a first-order length change of diagonals.
% This is problematic when we use the parameter space to measure the incompatibility between different motions.
% Therefore, we provide a different representation such that the distances between configurations can be interpreted as incompatibility metrics with physical meaning.
The quadrilateral boundary modeled as a four-bar linkage has 2 DOF, while the rigid origami module inside it is a 1-DOF structure.
Thus, its deformation can be represented as a 1-dimensional path in a 2-dimensional parameter space.
In Demaine and Ku~\cite{demaine2017folded}, diagonal lengths were used for parameterization, but this leads to a singularity at the flat state, where first-order motions occur without changes in diagonal lengths.
To address this, we propose an alternative representation that allows distances between configurations to serve as incompatibility metrics with physical meaning.

Figure~\ref{fig:representation} (a-c) shows the overview of our parameterization. 
We first align the frame $ABCD$ such that $C$ is the origin, $A$ is on the $z$ axis, $BD$ is parallel to the $x$ axis, and then scale such that each edge length is $1$.
Because the edge lengths are equal, every non-degenerate configuration is mirror symmetric with respect to the $yz$ plane and a plane parallel to the $xy$ plane passing through $B$ and $D$.
Then, the entire motion is represented by the motion of point $B$ on a unit sphere, which can be expressed using the coordinates $(\phi,\theta)$ on a sphere, where $\phi \in [-90^\circ, 90^\circ]$ is the azimuthal angle and $\theta \in [0,90^\circ]$ is the polar angle.

To provide the metrics related to the strain, we further tweak the parameters.
We define point $P$ as the nonuniform scaling of $B$ by $1/\sqrt{2}$ in the $y$-coordinate direction, as the reason for scaling is explained in the next section.
We call $P$ the \emph{representation point} of the boundary configuration, and the set of $P$ corresponding to the folding motion the \emph{representation curve} of the quadrilateral boundary module.
The representation curve is the configuration space that lies on the ellipsoidal parameter space (see Figure~\ref{fig:representation} (c)).

\subsubsection{Strain Evaluation from Incompatibility}

\begin{figure}[tbhp]
  \centering
  \includegraphics[page=6,width=\linewidth]{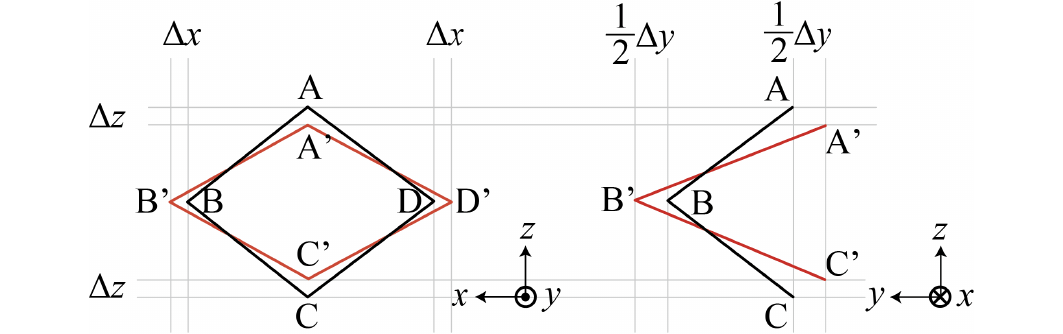}
   \caption{Quadrilateral boundary $ABCD$ of the module obtained as rigid origami without any strain (black lines) and the quadrilateral boundary $A'B'C'D'$ of the module with forced deformation (red lines)}
  \label{fig:evaluation_preparation}
\end{figure}

% Note that the distortion of the module in actual material is realized by the combination of backlash at the crease and the deformation of panels.

% In our study, we use an approximated evaluation of elastic strain by assuming that the deformation is concentrated at the vertices.
% More specifically, we consider rigid origami models of frames and modules connected with zero-length springs attached at the vertices.
% Under the assumption mentioned above, we demonstrate that the Euclidean distances between configurations on the ellipsoidal parameter space can be used to measure elastic strain. 

In practice, module distortion arises from crease backlash and panel deformation.
We approximate elastic strain by assuming deformation is concentrated at the vertices, modeling frames and modules as rigid origami connected by zero-length springs at the vertices.
Under this assumption, Euclidean distances in the ellipsoidal parameter space provide a measure of elastic strain.

Consider the quadrilateral boundary $ABCD$ of the module obtained as rigid origami without any strain, and the target quadrilateral boundary $A'B'C'D'$ after distortion.
Now, we assume that there exists a zero-length spring with spring constant $k$ between the corresponding two vertices $AA', BB', CC'$, and $DD'$ (Figure~\ref{fig:evaluation_preparation}).
We estimate the strain energy of the distorted module by the elastic energy $E$, \emph{equivalent strain energy}, stored in the zero-length spring.

By symmetry, the two frames are aligned to match their orientations ($x$,$y$,$z$ axes) and the center of mass when the potential energy is minimized (Figure~\ref{fig:evaluation_preparation}).
Now we let $\overrightarrow{CB}$ and $\overrightarrow{C'B'}$ be $(x,y,z)$ and $(x',y',z')$.
The difference $(\Delta x, \Delta y, \Delta z) := (x,y,z)-(x',y',z')$ represents the displacement for each edge.
The displacement at the vertices are: $(0,\frac{|\Delta y|}{2} , |\Delta z|)$ at $A$ and $C$ and $(|\Delta x|,\frac{|\Delta y|}{2}, 0)$ at $B$ and $D$.
Therefore, the total energy of the four springs at $A$, $B$, $C$, and $D$ is
\begin{align}
E & =  \frac{1}{2}k(AA'^2+BB'^2+CC'^2+DD'^2)\\
& = \frac{1}{2}k \left(
2\left\|
\left( 0,\frac{\Delta y}{2} , \Delta z \right)
\right\|^2 
+ 2\left\|
\left( \Delta x,\frac{\Delta y}{2}, 0 \right) 
\right\|^2
\right)\\
& =  k\left(\Delta x^2+\frac{1}{2}\Delta y^2+\Delta z^2\right).
\end{align}

Now we can check that the equivalent strain energy is proportional to the squared distance between two representation points defined in Section~\ref{sec:rep}.
Let $P=(x,\frac{1}{\sqrt{2}}y,z)$ and $P'=(x',\frac{1}{\sqrt{2}}y',z')$ represent the representation points of configuration $ABCD$ and $A'B'C'D'$ respectively (Figure~\ref{fig:representation} (c))).
Then the Euclidean distance $d = PP'$ between them is

\begin{eqnarray}
d^2 = \Delta x^2 + \frac{1}{2}\Delta y^2 + \Delta z^2 \propto E.
\end{eqnarray}

We now define the equivalent strain, $\epsilon_q$, as the misalignment($[mm]$) between the ideal and actual quadrilateral boundaries, normalized by the unit edge length($=1[mm]$) of the quadrilateral boundary, as follows:

\begin{eqnarray}
\epsilon_q =d \propto \sqrt{E}.
\end{eqnarray}

Using this representation, multistability is characterized by the distances between the representation curves of the target quadrilateral boundary and the boundary modules.
Stable states occur at the points where these distances take minimal values, and if the curves intersect at two or more points, the module has zero-strain stable states.
The energy barrier is evaluated as the maximum distance between the curves between stable states, defined as the \emph{maximum equivalent strain} $\epsilon_{qMax}$.

\section{Design Scheme}
\label{sec:design scheme}

\begin{figure}[tbhp]
  \centering
  \includegraphics[page=7,width=\linewidth]{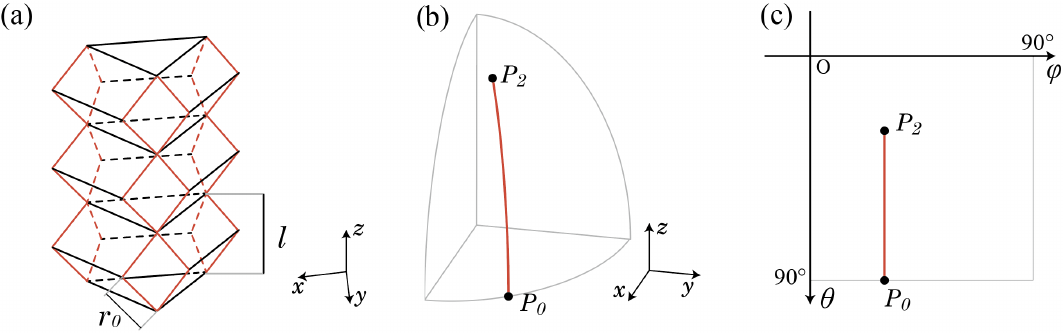}
   \caption{(a) Frame of an extending cylinder with Sarrus linkage. (b) Representation curve of the frame drawn on an ellipsoid. (c) Representation curve of the frame on the $\phi\theta$ plane.}
  \label{fig:quadframe}
\end{figure}

We now propose a design scheme for multistable origami using the representation model.
As an example case, we designed a linearly extending multistable structure by filling the quadrilateral frames of a Sarrus linkage with mirror-symmetric biconical modules.
The same procedure can also be applied to other linkage types introduced in Section~\ref{sec:linkage}.
We let the length of one edge of the quadrilateral frame be $r_0 = 10$, and the height of the Sarrus linkage be $l$.
Since the fully extended state is a singular state of the original linkage, we avoid the singular state by setting the range in $30^\circ\leq\theta\leq90^\circ$, so $l$ moves in $0\leq l \leq 20\cos30^\circ\approx17.3$ (Figure~\ref{fig:quadframe}).
Here, $\theta$ is the polar angle of $B$ of the frame, which is equivalent to the rotation angle of a panel of the Sarrus linkage.
Therefore, we consider solving the motion between the flat state $P_0(\phi,\theta)=(30^\circ,90^\circ)$ and the deployed state $P_2(\phi,\theta)=(30^\circ,30^\circ)$.

\subsection{Optimization}
To solve this design problem, we propose the following optimization scheme.
First, we use a mirror-symmetric biconical module with $6$-vertex polyline and use its parameters as design variables.
Next, we optimized for a module whose representation curve closely matches the target curve.
This yields a structure that approximates the desired mechanism with minimal strain, referred to as a \emph{quasi-mechanism} structure.
Then, using this quasi-mechanism as the reference, we create bistable structures with tuned stiffness of snapping by designing a module with a representation curve that intersects the target curve at specific points but passes through a point of maximum specified distance from the target curve.
Using the \emph{Rhino/Grasshopper}, we design the kinematics of modules and conduct the optimization using the evolutionary algorithm of Galapagos~\cite{rutten2013galapagos}.
The variables and objective functions for both steps are explained below.

\subsubsection{Variables}

\begin{figure}[tbhp]
  \centering
  \includegraphics[page=8,width=\linewidth]{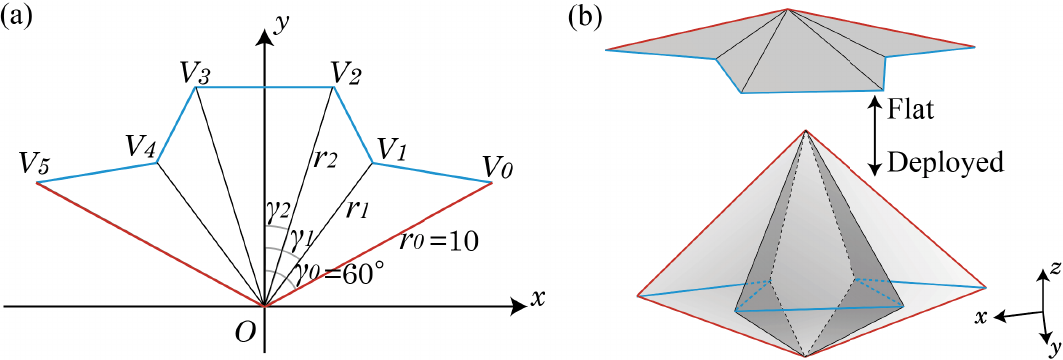}
   \caption{Design variables.}
  \label{fig:variables}
\end{figure}

We use the mirror-symmetric biconical module of six vertices, parameterized at the flat state on the $xy$ plane (Figure~\ref{fig:variables}).
We place the pivots on the origin and consider a further symmetry about the $y$ axis and use the polar coordinates of two interior vertices $(r_1,\gamma_1)$ and $(r_2,\gamma_2)$ as design variables.
Here, the boundary vertex is fixed such that $r_0 = 10$ and $\gamma_0=60^\circ$ have a valid configuration at the flat state.
Also, to prevent self-intersections of the polylines, we set the constraint $0^\circ\leq\gamma_2\leq\gamma_1\leq60^\circ$.

\begin{figure}[tbhp]
  \centering
  \includegraphics[page=9,width=\linewidth]{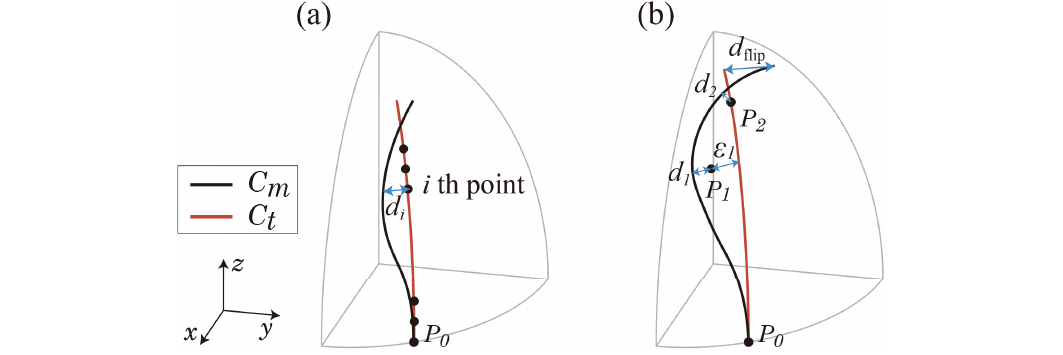}
   \caption{(a) Evaluation of the modules for quasi-mechanism structure. (b) Evaluation of the modules for the bistable structure.}
  \label{fig:idea_of_optimizing}
\end{figure}

\subsubsection{Objective Function: Quasi-mechanism}
To design a quasi-mechanism structure, we want to find a module with its representation curve $C_m(\gamma_1,\gamma_2,r_1,r_2)$ that matches as closely as possible to the target frame's representation curve $C_t$ between $P_0$ and $P_2$.
See Figure~\ref{fig:idea_of_optimizing}  Left. 
To evaluate the mean distances between the target curve $C_t$ and the current $C_m$, we first sample $n+1$ points along $C_t$ between $P_0$ and $P_2$ by dividing it into $n$ equal parts.
For each of the sampled points $i = 0, \dots, n$ on $C_t$, we find the shortest distance $d_i$ to $C_m$.
Then, we minimize the objective function $G$, which is given as the average of the squares of $d_i$.

% \begin{eqnarray}
% G(\gamma_1,\gamma_2,r_1,r_2)=\frac{\sum_{i=0}^{n}{d_i^2}}{n+1}.
% \end{eqnarray}

\begin{mini}
{}{G(\gamma_1,\gamma_2,r_1,r_2)=\frac{\sum_{i=0}^{n}{d_i^2}}{n+1}}{}{}
\addConstraint{0^\circ \le \gamma_2 \le \gamma_1 \le 60^\circ}{}{}
\addConstraint{r_1 > 0,\; r_2 > 0,\; n \in \mathbb{N}}{}{}
\end{mini}

\subsubsection{Objective Function: Bistable mechanism}
To design a bistable structure, we let $C_m(\gamma_1,\gamma_2,r_1,r_2)$ pass through two stable points $P_0,P_2$ on $C_t$ and a point $P_1$ not on $C_t$ between the two points (Figure~\ref{fig:idea_of_optimizing} Right).
The distance from $P_1$ to the $C_t$ is the estimated equivalent strain causing the potential energy barrier between the stable points and is denoted by $\epsilon_1$.

Here, we found that $P_1$ cannot be arbitrarily chosen.
The heuristics for choosing $P_1$ are to first refer to the shape of the representation curve of the quasi-mechanism module and find the farthest point with maximum equivalent strain.
This is around $P_{max} (28.1^\circ,50.8^\circ)$, corresponding to $\epsilon_{qMax} = 1.98\%$ in our case.
Then, we set $P_1$ to be reasonably close to this point but further away from the $C_t$ to achieve the prescribed snapping stiffness.

We minimize $d_1$ and $d_2$, which are the shortest distances from $P_1$ and $P_2$ to $C_m$.
Specifically, we minimize the objective function $H$ given as follows

% \begin{eqnarray}
% H(\gamma_1,\gamma_2,r_1,r_2)=\frac{d_1^2+d_2^2}{2}+f(\epsilon_1, d_{flip})
% \end{eqnarray}

\begin{mini}
{}{H(\gamma_1,\gamma_2,r_1,r_2)=\frac{d_1^2+d_2^2}{2}+f(\epsilon_1, d_{flip})}{}{}
\addConstraint{0^\circ \le \gamma_2 \le \gamma_1 \le 60^\circ}{}{}
\addConstraint{r_1 > 0,\; r_2 > 0}{}{}
\end{mini}

where $f(\epsilon_1, d_{flip})$ is a penalty function to prevent the flipping.
Here, $d_{flip}$ is the distance from the fully opened state on $C_m$ to the frame's representation curve in $0 \leq \theta  \leq 90^\circ$ (full-range version of $C_t$, which is restricted between $P_0$ and $P_2$).
To avoid the flipping while not losing the fitness, we design the penalty function to minimize $d_1/\epsilon_1 < 1$, $d_2/\epsilon_1 <1$, and $\epsilon_1 / d_{flip} < 1$ at the same time.
In our case, we used $f(\epsilon_1, d_{flip})=\frac{\epsilon_1^3}{d_{flip}}$.

\subsection{Optimization Results}

\begin{figure}[tbhp]
  \centering
  \includegraphics[page=10,width=0.8\linewidth]{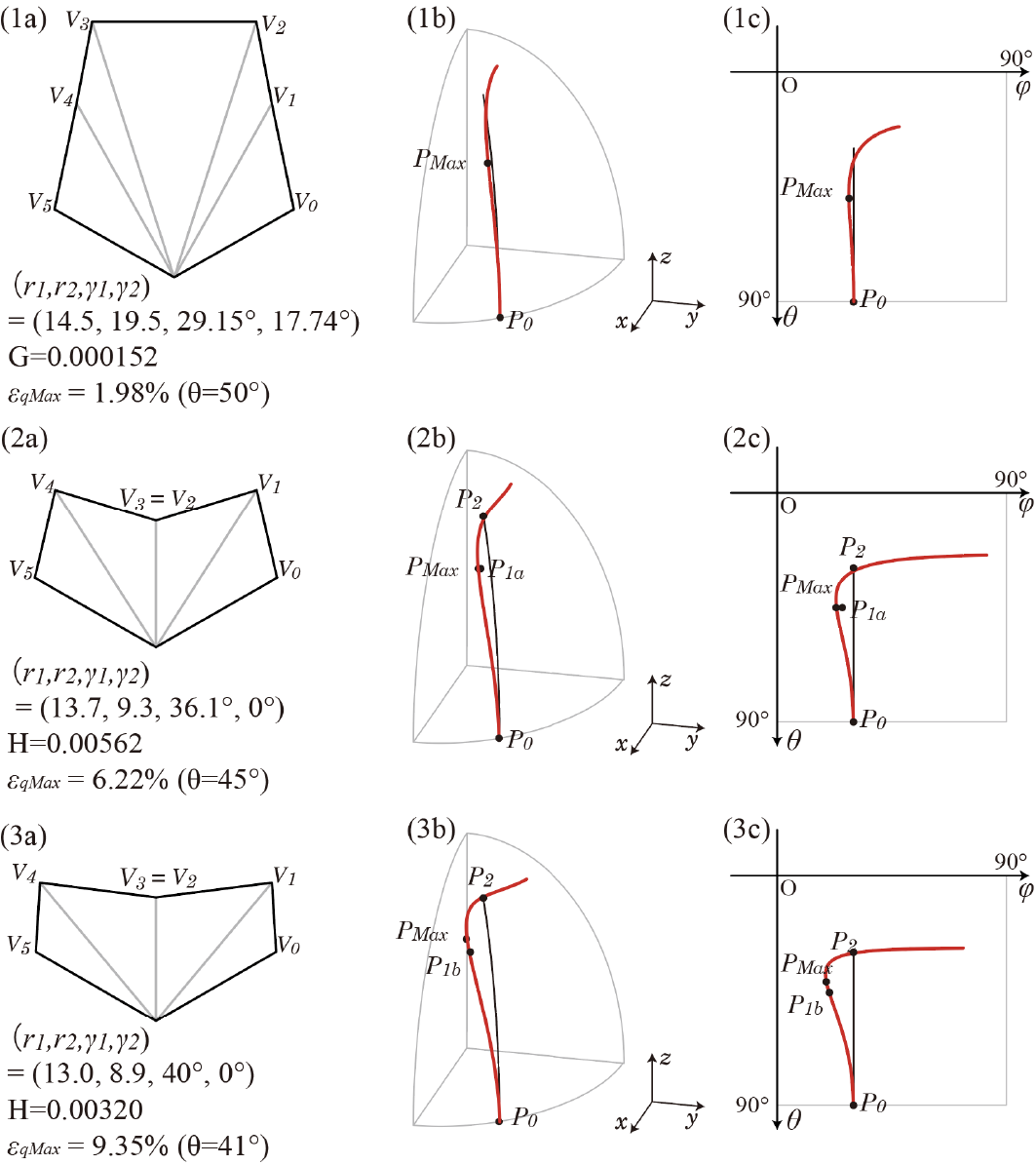}
   \caption{Results of the optimization: The top row is a module for a quasi-mechanism structure, and the middle and the bottom rows are blocks $A$ and $B$ for bistable structures. (1a), (2a), (3a) Mirror-symmetric biconical module obtained by optimization. $G, H$ denote the evaluated values, and $S_{q\mathrm{Max}}$ denotes the maximum strain and $\theta$ when it is taken. (1b), (2b), (3b) Representation curves of the target frame (red curves) and the module (black curves) on an ellipsoidal sphere. (1c), (2c), (3c) The representation curves are drawn on the $\phi\theta$ plane.}
  \label{fig:result}
\end{figure}

% \begin{figure}[tbhp]
%   \centering
%   \includegraphics[page=11,width=\linewidth]{figure/Chapter3.pdf}
%    \caption{Obtained structure with designed modules. Top: structures applied to a single Sarrus linkage module. Bottom: structures applied to a series of Sarrus linkage modules. (1) Quasi-mechanism structure, (2) multistable structure with smaller strain (module $A$), and (3) multistable structure with larger strain (module $B$).}
%   \label{fig:embedded}
% \end{figure}

The result of the optimization is shown in Figures~\ref{fig:result}.
For a quasi-mechanism structure, we obtained a module where $V_0, V_1, V_2$ are almost in a straight line (Figure~\ref{fig:result} (1a)), thus it is approximately realizable only with a four-vertex polyline.
The representation curve almost follows that of the frame, where the maximum equivalent strain $\epsilon_{qMax}$ of $1.98\%$ is obtained at $P_{max} (28.1^\circ,50.8^\circ)$.
The objective function $G$ converged to $0.000152$, which indicates the average (specifically, root mean square) equivalent strain of $1.23\%$.
As already explained, $P_{max}$ is then used as the reference for determining $P_{1a}$ and $P_{1b}$.

For bistable structures, we computed module $A$ with a smaller strain and module $B$ with a larger strain.
(A) For module $A$, we set $P_1$ to be $P_{1a} (25^\circ,45^\circ)$ which corresponds to the target maximal equivalent strain of $\epsilon_{1a} = 4.74\%$.
This resulted in the maximum equivalent strain for module $A$ is $\epsilon_{qMax}=6.22\%$ at $P_{max} = (23.5^\circ,45.5^\circ)$ (Figure~\ref{fig:result} (2a)).
(B) For module $B$, $P_{1b} (20^\circ,45^\circ)$, we set $P_1$ to be $P_{1b} (20^\circ,45^\circ)$, which corresponds to the target maximal strain of $\epsilon_{1b} = 9.38\%$, respectively, as shown in (Figure~\ref{fig:result} (3a)).
This resulted in the maximum equivalent strain for module $B$ being $\epsilon_{qMax}=9.35\%$ at $P_{max} = (19.4^\circ,41.8^\circ)$.
Both modules A and B have converged to $\gamma_2=0$, so they have a five-vertex polyline (See Figure~\ref{fig:result} (2a) and (3a)).
The representation curve of the obtained modules almost passes through given $P_1$ and $P_2$ (Figure~\ref{fig:result} (2b) (3b) (2c) (3c)); the distances were $d_{1a} = 0.015, d_{2a} = 0.0094$ for module $A$ and $d_{1b} = 0.0023, d_{2b} = 0.00032$ for module $B$.

Finally, the objective function of modules $A$ and $B$ converged to $H=5.62\times10^{-5}$ and $H=3.20\times10^{-5}$ with $d_{flip}= 0.261$ and $d_{flip} = 0.258$, respectively. 
The representation curve for module $A$ satisfies
$d_1/\epsilon_1 \approx 0.31 < 1$, $d_2/\epsilon_1 \approx 0.20 < 1$, and $\epsilon_1 / d_{flip} \approx 0.18 < 1$ as desired.
Similarly, the representation curve for module $B$ satisfies
$d_1/\epsilon_1 \approx 0.024 < 1$, $d_2/\epsilon_1 \approx 0.0034 < 1$, and $\epsilon_1 / d_{flip} \approx 0.36 < 1$ as desired.

By fitting the obtained modules within the quadrilateral boundary of the frame (refer to Figure~\ref{fig:Fabrication1} (a) in the next section), the multistable origami blocks, termed \emph{Q-bellows} (Quadrilateral-boundary bellows), are completed.

\section{Mechanics}
\label{sec:validation}
\subsection{Physical Prototypes}
\label{sec:fabrication}

\begin{figure}[tbhp]
  \centering
  \includegraphics[page=12,width=0.8\linewidth]{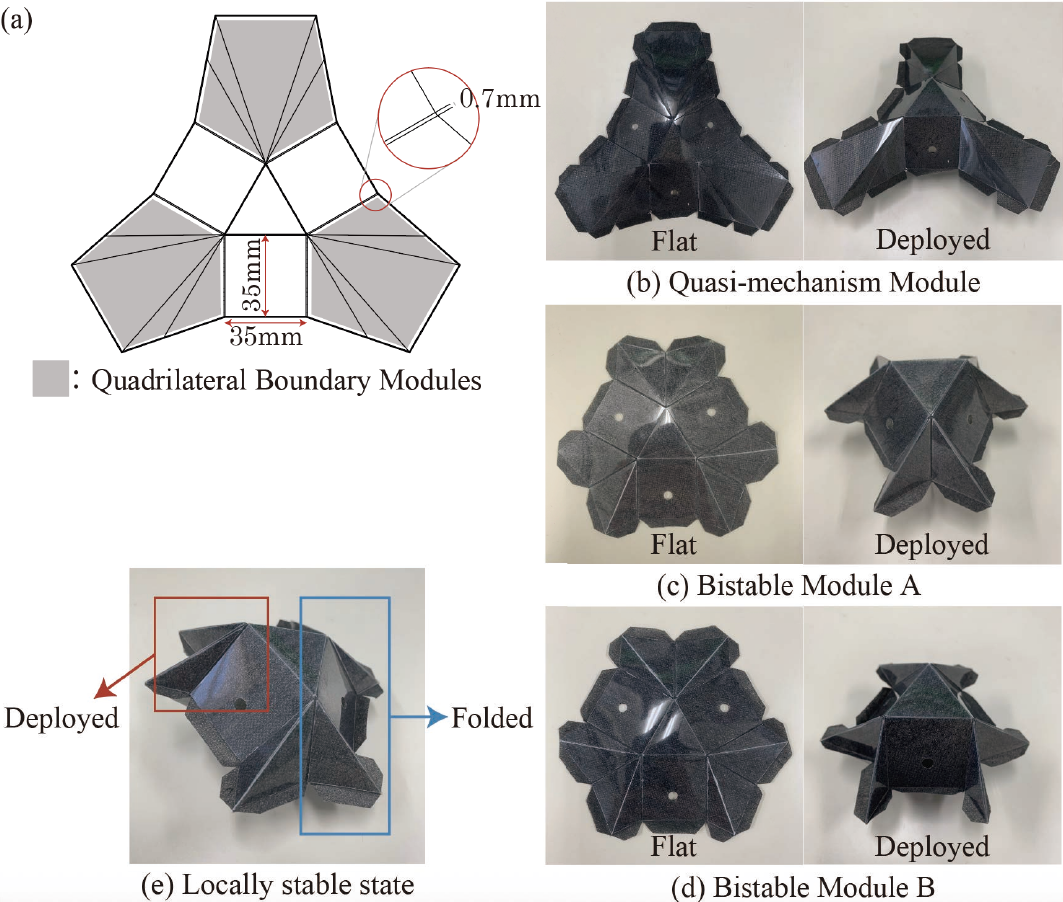}
   \caption{(a) The cutting pattern of a quasi-mechanism module. (b) Flat and deployed state of the physical model of the quasi-mechanism module, (c) bistable module $A$, and (d) bistable module $B$. (e) The locally stable state of the bistable module $B$.}
  \label{fig:Fabrication1}
\end{figure}

We fabricate physical prototypes of Q-bellows of the quasi-mechanism module and bistable modules $A$ and $B$ with a unit length($r_0$) of \SI{35}{mm} as in Figure~\ref{fig:Fabrication1} (a-d) using the fabrication method 1 in section~\ref{subsubsec:deskfabmethod1}.
From hands-on interaction with the physical models, we experienced snapping between the flat position and the fully deployed state for all three blocks.
We also found that when we apply an unbalanced force to bistable modules $A$ and $B$, the structure may be locally stable with one module deployed and two modules folded.
This unintended multistability effect is due to the limited stiffness of the quadrangular panels in the Sarrus linkage (Figure~\ref{fig:Fabrication1} (e)).

\subsection{Snapping Analysis}

Here, we verify the bistability of the designed Q-bellows blocks through experiments and bar-and-hinge analysis.
Then, by comparing these analysis results with our estimated behaviors from equivalent strain energy, we validate the design efficiency of the geometric incompatibility design method.
Consider a linear motion changing the height $l \in [0, 62]$ while applying the forced displacement load perpendicular to the upper and lower surfaces.

\subsubsection{Experimental Results}

\begin{figure}[tbhp]
  \centering
  \includegraphics[page=24,width=0.8\linewidth]{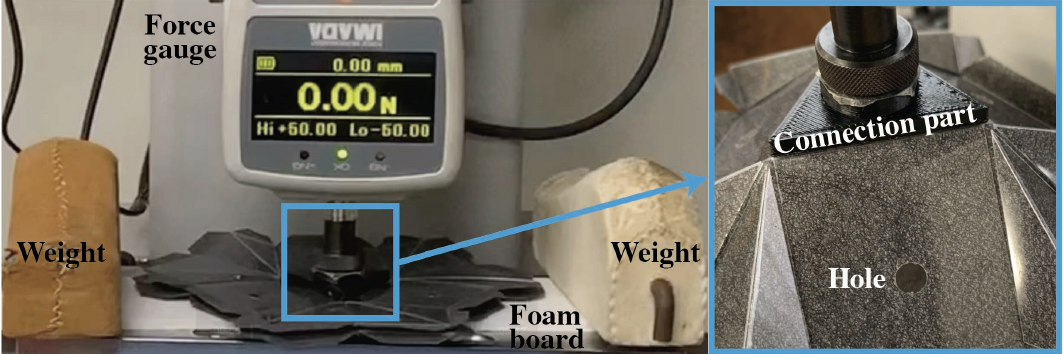}
   \caption{Experiments setup.}
  \label{fig:expsetup}
\end{figure}

We conducted forced displacement cycle loading of deployment and compression using the physical prototypes of bistable modules $A$ and $B$, which are fabricated in Section~\ref{sec:fabrication}.
Figure~\ref{fig:expsetup} shows the setup for experiments.
The bottom triangle of modules was attached to the base made of foam board, and the top triangle of the module is attached to the force gauge (ZTA-50N Imada Corp.).
The connection part between the force gauge and the upper triangular surface of the module is fabricated using the fused filament fabrication (FFF) type 3D printer. 
To eliminate the effect of air pressure, we punched a hole in the rectangular panel of each module (Figure~\ref{fig:buckling and play}).
The speed of the cycle of the force gauge is \SI{50}{\milli \meter \per min}.

\begin{figure}[tbhp]
  \centering
  \includegraphics[page=14,width=\linewidth]{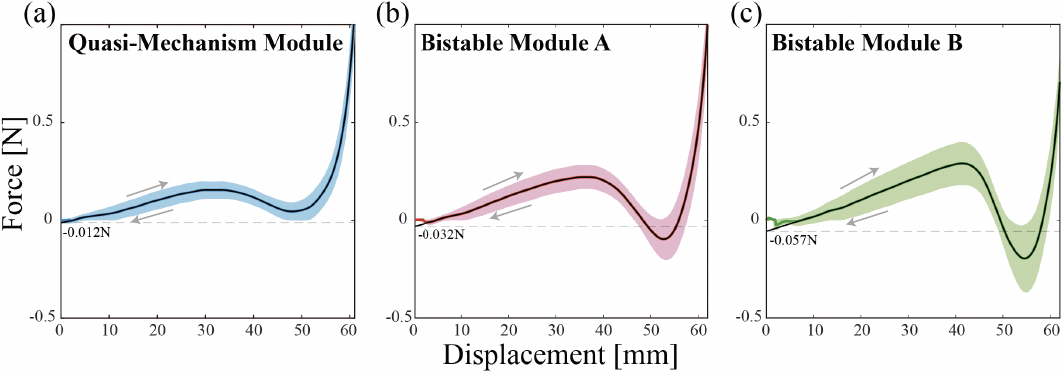}
   \caption{Force--displacement curves of a cycle of extension and compression experiment. Each plot shows the result of each unit Q-bellows block: (a) the quasi-mechanism module, (b) the bistable module $A$, and (c) the bistable module $B$. The black curves represent the averaged load value from which the initial buckling part is replaced by linear extrapolation. The black dashed horizontal lines represent the calibrated zero force.}
  \label{fig:unitexp}
\end{figure}

The outlines of the colored areas in Figure~\ref{fig:unitexp} show the force--displacement curve for a cycle of the extension and compression experiment.
Since the physical experiment contains the effect of the plasticity and elasticity of hinges, we cannot directly compare it with the result from the equivalent strain or the bar-and-hinge model.
We separate the plasticity and elasticity effects due to the hinges between panels in the following way.
As a pre-process, we applied noise removal using Gaussian window filtering of standard deviation $\sigma$ equal to $0.6\%$ of the total displacement of the cycle.

The plasticity or friction of the hinges is observed as the hysteresis curve (colored areas in Figure~\ref{fig:unitexp}) of the single unit curve, which would not exist in a conservative system. 
To separate this effect, we average the load for the forward and backward paths, as shown in the black curves in Figure~\ref{fig:unitexp}.
However, since the plot tends to be noisy due to a series of small buckling near the flat state, we replaced the curve near the zero point 
(\SI{0}{mm}--\SI{2.6}{mm} for the quasi-mechanism module, \SI{0}{mm}--\SI{4.66}{mm} for the bistable module $A$, and \SI{0}{mm}--\SI{7.61}{mm} for the bistable module $B$) 
with a linear extrapolation from the half displacement position of the maximum extreme to the last displacement of the buckling range 
(\SI{2.6}{mm}--\SI{15.83}{mm} for the quasi-mechanism module, \SI{4.66}{mm}--\SI{18.18}{mm} for the bistable module $A$, and \SI{7.61}{mm}--\SI{20.67}{mm} for the bistable module $B$).
The plastic effect leads to the drift of zero points, as shown by the dashed black lines in Figure~\ref{fig:unitexp}.
We calibrated the zero point by letting the average load (black curves in Figure~\ref{fig:unitexp}) become zero. 

\begin{figure}[tbhp]
  \centering
  \includegraphics[page=22,width=\linewidth]{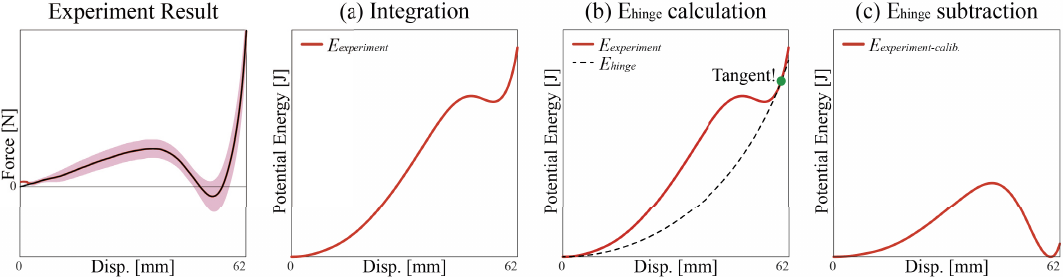}
   \caption{Hinge energy ($E_{hinge}$) calibration process. (a) $E_{experiment}$ is obtained by integrating the force--displacement curve, (b) $E_{hinge}$ is estimated by tangential fitting, and (c) $E_{hinge}$ is subtracted.}
  \label{fig:calibration}
\end{figure}

To remove the elastic stiffness of the creases in Q-bellows, we consider an approximate nonlinear spring represented by angular springs placed at the hinges of the Sarrus linkage (Figure~\ref{fig:quadframe}), simplifying and focusing on the nonlinear relationship between height ($l$) and potential energy.
Therefore, the potential energy ($E_{hinge}$) of the frame is computed as:

\begin{gather}
\label{eq:sarruscalibration}
E_\textrm{hinge} = k \arcsin^2 \frac{l}{2r_0}.    
\end{gather}

% {\color{red}
Figure~\ref{fig:calibration} shows the $E_{hinge}$ calibration process.
From the experimentally obtained average values of force -- displacement plots, the energy plot ($E_{experiment}$) is obtained by integration (Figure~\ref{fig:calibration} (a)).
The stiffness parameter $k$ of $E_{hinge}$ is then determined by tangential fitting to $E_{experiment}$ using the bisection method (Figure~\ref{fig:calibration} (b)).
Finally, $E_{hinge}$ is subtracted from $E_{experiment}$ so that the minimum calibrated potential energy becomes zero (Figure~\ref{fig:calibration} (c)).

The black curves in Figure~\ref{fig:qbellowscomparison} (a) show the calibrated energy plots, and the black curves in Figure~\ref{fig:qbellowscomparison} (b) represent the corresponding force plots obtained by differentiating the energy with respect to displacement ($l$). 
% }

\subsubsection {Bar-and-Hinge Analysis}
For comparison, we performed the bar-and-hinge analysis using Merlin (see Section~\ref{subsec:merlin}) to obtain force--displacement plots by applying vertical displacement at the three points on the top triangle while keeping the bottom triangle fixed.

The material parameters of the bar-and-hinge model were as follows: 
Young's modulus ($E$) is set arbitrarily, as we compare only the relative stiffness. 
We choose the Poisson ratio ($\nu$) and the thickness($t$) of the sheet to be $0.3$ and $0.25$, respectively.
Additionally, we set the hinge stiffness of fold lines to effectively zero by setting the length scale factor $f=100000~[mm]$.
Blue curves in Figure~\ref{fig:qbellowscomparison} show the results from the bar-and-hinge model analysis.

\subsubsection{Estimated Behavior}
Using the equivalent strain energy computed from the geometric incompatibility and its derivative with respect to the displacement, we can obtain the expected plot of energy--displacement and force--displacement plots. 
Red curves in Figure~\ref{fig:qbellowscomparison} show the estimated curves from the geometric model.
Note that the vertical axis for energy or force is relative in our model, so it is transformed to fit the results of the bistable module $A$.

\subsection{Comparison}

\begin{figure}[tbhp]
  \centering
  \includegraphics[page=15,width=\linewidth]{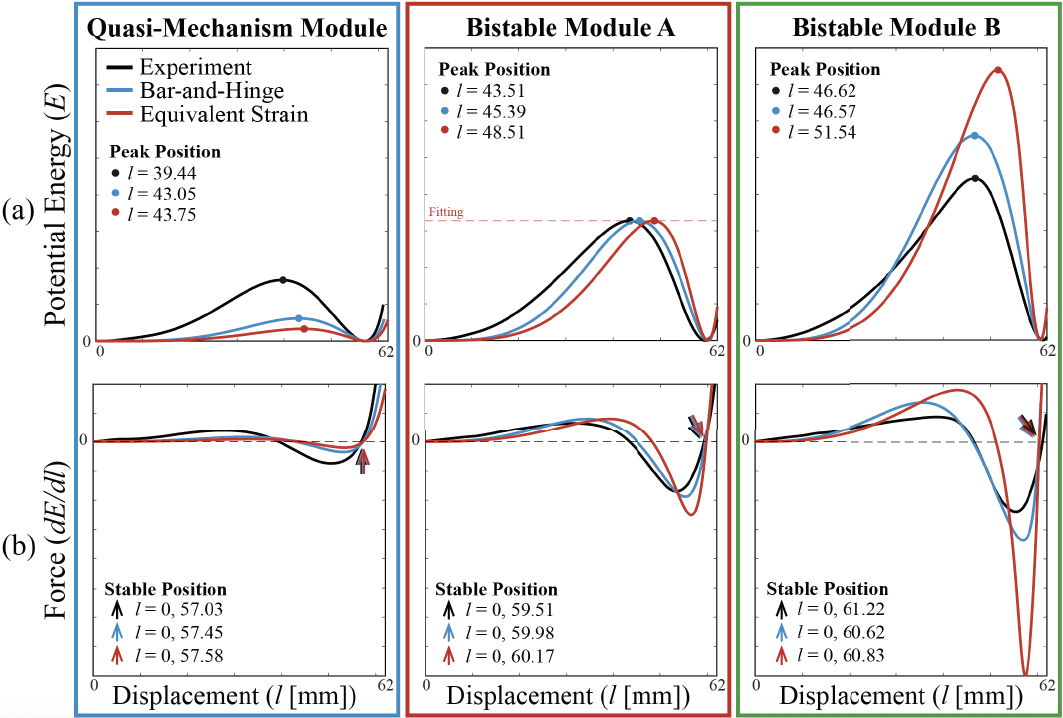}
   \caption{Comparison between the equivalent strain model, bar-and-hinge analysis results, and calibrated experimental results, each indicated by red, blue, and black curves. (a) Potential energy--displacement plot of each module. (b) Force--displacement plot of each module from differentiating from the energy. The point marks the peak position of potential energy.}
  \label{fig:qbellowscomparison}
\end{figure}

Refer to Figure~\ref{fig:qbellowscomparison} for a comparison between Q-bellows and the experiment.
Note that since we did not specify the material property in the model, the values in the potential energy and force are relative.
Therefore, we scaled the highest potential energy in module $A$ to be equal in the three analysis models, to avoid extreme discrepancy of other blocks.

The dots in Figure~\ref{fig:qbellowscomparison} (a) show the displacement of the peak of each energy result.
As designed, we obtained the energy barriers of the quasi-mechanism module being the lowest and the energy barrier of module $B$ being the largest in both the bar-and-hinge model and the experiment.
For all three models, we obtained a force--displacement plot in Figure~\ref{fig:qbellowscomparison} (b), causing the snapping behavior between flat and deployed states.
This validates the design method for differentiating the barriers on a qualitative level.
However, we also observed that the difference in magnitude of the energy barriers between the three modules is smaller in the bar-and-hinge model and even smaller in the experiment compared to the equivalent strain model.
We consider that this is due to the nonlinear effect of the difference between the concentrated strain at the vertices and the actual strain dispersed to panels and crease gaps.

The displacement that takes the peak value and stable points of the potential energy is also similar.
However, the displacement of peak values is slightly shifted to the smaller in the bar-and-hinge model and experiment (Figure~\ref{fig:qbellowscomparison} (top plots)). 
The quasi-mechanism module shifts by $1.13\%$ ($43.75 \rightarrow 43.05$) in the bar-and-hinge result and by $6.95\%$ ($43.75 \rightarrow 39.44$) in the experiment, representing the total stroke. 
Module $A$ shifts by $5.03\%$ ($48.51\rightarrow45.39$) in the bar-and-hinge result and $8.06\%$ ($48.51\rightarrow43.51$) in the experiment, and Module $B$ shifts by $8.02\%$ ($51.54\rightarrow46.57$) in the bar-and-hinge result and $7.94\%$ ($51.54\rightarrow46.62$) in the experiment.

\section{Application}
\label{sec:application}

\subsection{2-bit Mechanical Memory}

\begin{figure}[tbhp]
  \centering
  \includegraphics[page=16,width=0.75\linewidth]{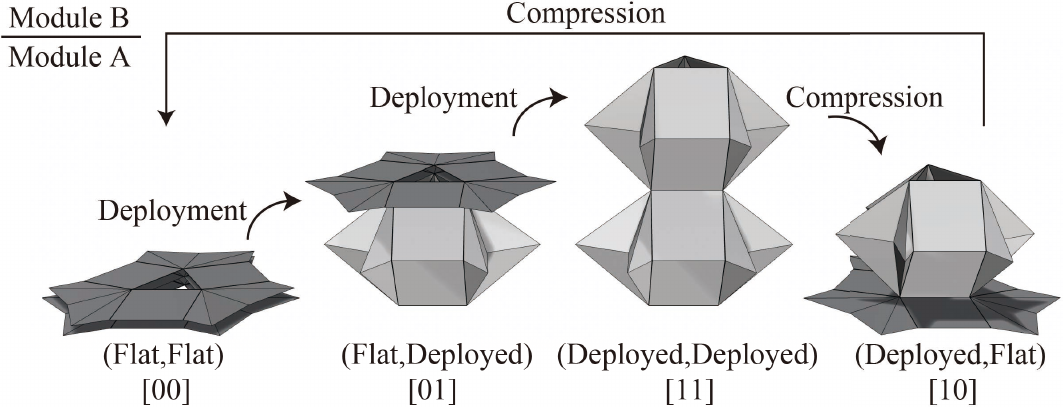}
   \caption{$2^2$ states and 2-bit mechanical memories made of modules~$A$ and~$B$. Module $B$ is attached on top of $A$.}
  \label{fig:mechanical_memory}
\end{figure}

\begin{figure}[tbhp]
  \centering
  \includegraphics[page=17,width=\linewidth]{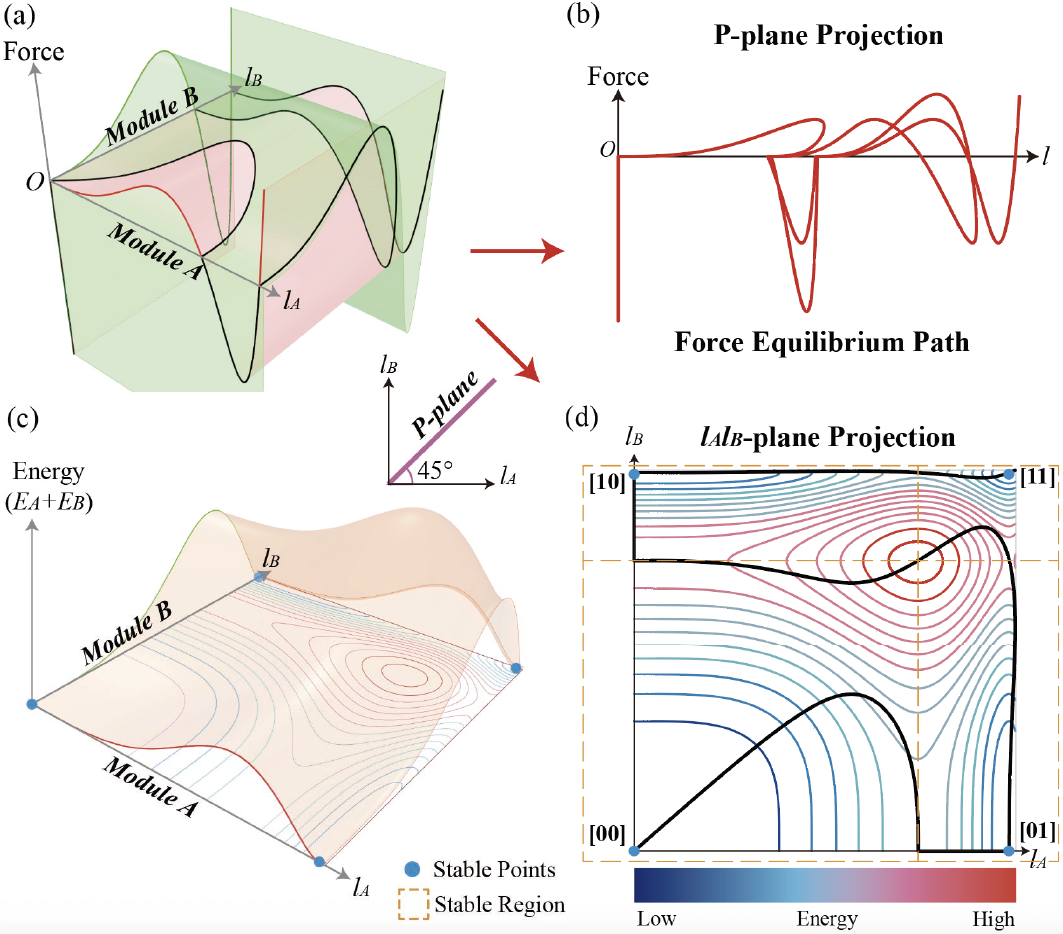}
   \caption{(a) Force--displacement plots of module $A$ (red) and $B$ (green) and extrusion surfaces of each curve. The black curve represents the intersection curve of two extrusion surfaces. (b) The $P$-plane projected intersection curve indicates the force equilibrium path, (c) The total potential energy plot with module $A$ and $B$, (d) The contour map of total energy and the $l_Al_B$-plane projected intersection curve. The dashed orange lines indicate the region dominated by each stable point. The force equilibrium path passes through the four stable states.}
  \label{fig:hysteresis theory}
\end{figure}

\begin{figure}[tbhp]
  \centering
  \includegraphics[page=18,width=\linewidth]{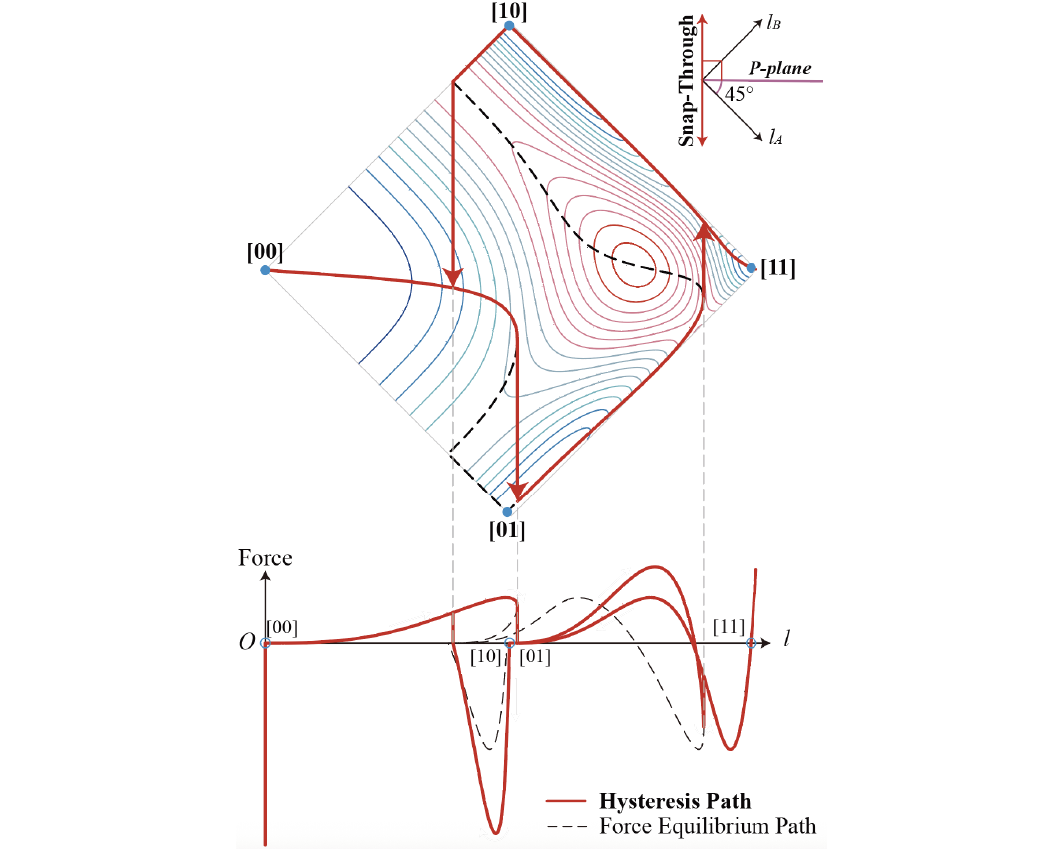}
   \caption{The red curve shows the theoretical hysteresis path (forced displacement path) in the serial system. Snap-back (snapping behavior in forced displacement condition) occurs where the force equilibrium path is tangent to a plane perpendicular to the $P$-plane.}
  \label{fig:hysteresis path}
\end{figure}

The serial composition of Q-bellows blocks with different stiffness can take $2^n$ positions that can be controlled from a sequence of motions on one side; such a behavior can be interpreted as an $n$-bit mechanical memory~\cite{yasuda2017origami}.
Serially connecting modules $A$ and $B$ (Figure~\ref{fig:mechanical_memory}) is expected to be a $2$-bit memory that exhibits a cycle of transitions between $2^2$ states.

\subsubsection{Theory}
First, we make a theoretical prediction of the two-bit memory hysteresis from the force $f$ and the potential energy $E_A$ and $E_B$ of the bistable modules $A$ and $B$.
We consider a three-dimensional space spanned by displacements $l_A$, $l_B$, and the force $f$ as shown in Figure~\ref{fig:hysteresis theory}.
$l_A$ and $l_B$ indicate the heights of bistable modules $A$ and $B$, respectively.
$f$ indicates the equilibrium force of the corresponding displacement, which is identical to the force of the serial system.
The force--displacement plots of modules $A$ and $B$ based on the bar-and-hinge model in Figure~\ref{fig:qbellowscomparison} (b) are plotted in the $l_A$--$f$ and $l_B$--$f$ planes as the red and green curves, respectively, in Figure~\ref{fig:hysteresis theory} (a).
Here, to model the collision of panels at the flat state, we added a vertical curve ($-\infty$ force) to the force--displacement plot at $l=0$, so that a single module cannot take a negative height.

Now, the force equilibrium for module $A$ is described as the surface obtained by the extrusion of the $l_A$-$f$ plot to the $l_B$ axis.
Similarly, the force equilibrium for module $B$ is described as the extrusion surface of the $l_B$-$f$ plot along the $l_A$-axis (Figure~\ref{fig:hysteresis theory} (a)).
Since the equilibrium path of the serial system satisfies that the load on modules $A$ and $B$ is the same, it can be obtained as the intersection of two extrusion surfaces, as expressed by the black curve in Figure~\ref{fig:hysteresis theory} (a). 
We can plot the relationship between the merged length ($l_A + l_B$) as the projection to a plane containing the diagonal direction, called the \emph{P-plane}.
By scaling this projected curve by $\sqrt{2}$ in the horizontal direction, we can obtain the force equilibrium path of the total displacement for the serial system (Figure~\ref{fig:hysteresis theory} (b)).

By observing the total potential energy $E_A(l_A,l_B) + E_B(l_A,l_B)$, we can check the stability of the equilibrium path (Figure~\ref{fig:hysteresis theory} (c)).
The potential energy takes the local minimum at four stable points [00], [11], [10], and [01].
The $l_Al_B$-plane is partitioned into four regions (orange dashed lines in Figure~\ref{fig:hysteresis theory} (d)) corresponding to the four stable points such that the energy gradient descent path from a state in a region ends up in its corresponding stable state.
Figure~\ref{fig:hysteresis theory} (d) shows the projected intersection curve to the $l_Al_B$-plane with the contour map of the total energy plot.
We can check that the force equilibrium path passes through each stable point and region.

The hysteresis path that appears in the forced displacement loading condition can be drawn by choosing the local minimum point for a varying total displacement $l=l_A+l_B$.
Under this forced displacement loading condition, the system cannot follow the equilibrium path backward.
Instead, when an unstable state is encountered, the serial chain system undergoes a snap-through transition internally while keeping the same total displacement.
This behavior is referred to as \emph{snap-back}, and it appears as a sudden perpendicular drop in the force plot, resulting in discontinuous force changes.
Also, the negative force at the flat state makes the hysteresis plot of the serial chain more discontinuous.
Therefore, snap-backs are visualized by straight lines perpendicular to the P-plane on the energy contour plot, as shown in Figure~\ref{fig:hysteresis path}.

Based on the above discussion, we were able to identify the behavior of the forced displacement cycle, which is characterized by the three times snap-back between equilibrium points and hysteresis passing through stable regions [00], [11], [10], [01].
The theoretical hysteresis curves computed from bar-and-hinge and equivalent strain models are shown in Figure~\ref{fig:Experiment memory} (f -- blue curve) and (g -- red curve), respectively.
Note that those actually hit the stable points [00], [11], [10], but not [01]; as a result, the paths pass through $f=0$ at [00], [11], [10], but not at [01].
The estimated behavior of the serial system from equivalent strain energy shows a larger distance between the stable points [01] and [10] and a larger negative force than the bar-and-hinge model.

\subsubsection{Experiment}

\begin{figure}[tbhp]
  \centering
  \includegraphics[page=19,width=\linewidth]{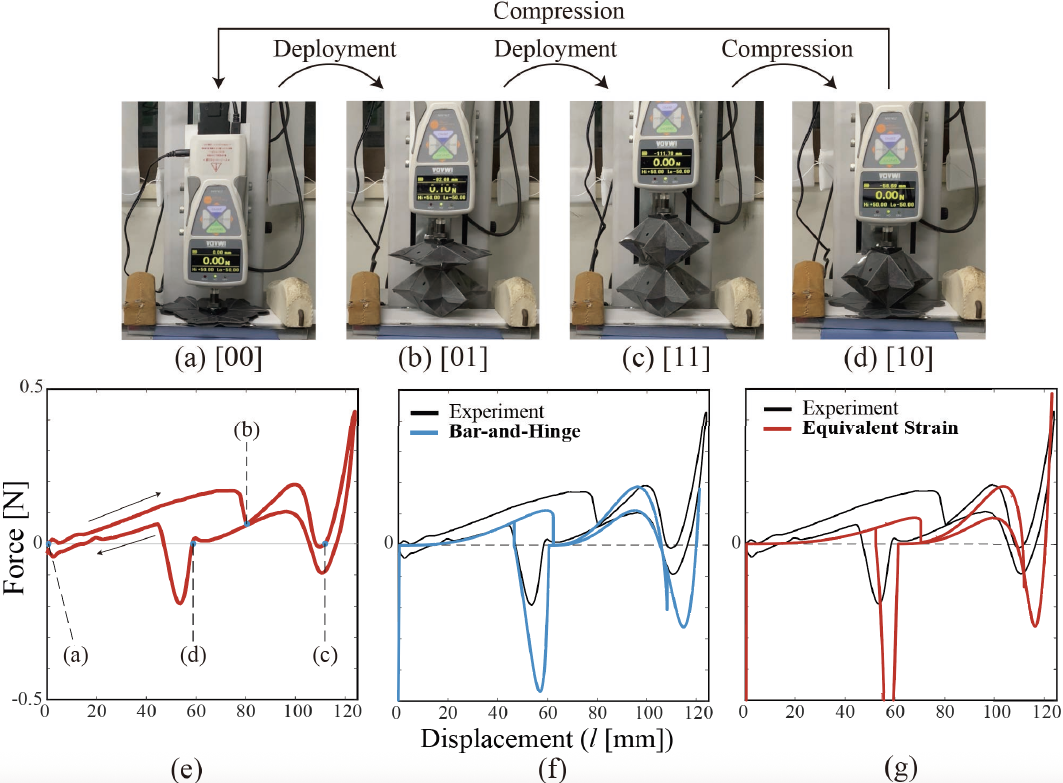}
   \caption{(a)-(d) $2^2$ states of a mechanical memory made of module $A$ and module $B$ were experimentally verified. (e) The results of the force--displacement plot of a reciprocating experiment, which fully deploys from the flat state and returns to the flat state. The graph begins at the origin and proceeds in the direction indicated by the arrows. The blue points indicate the four stable states (a)--(d). The experiment result is compared with (f) Bar-and-hinge analysis results and (g) the estimated behavior from equivalent strain energy.}
  \label{fig:Experiment memory}
\end{figure}

\begin{figure}[thbp]
  \centering
  \includegraphics[page=20,width=0.8\linewidth]{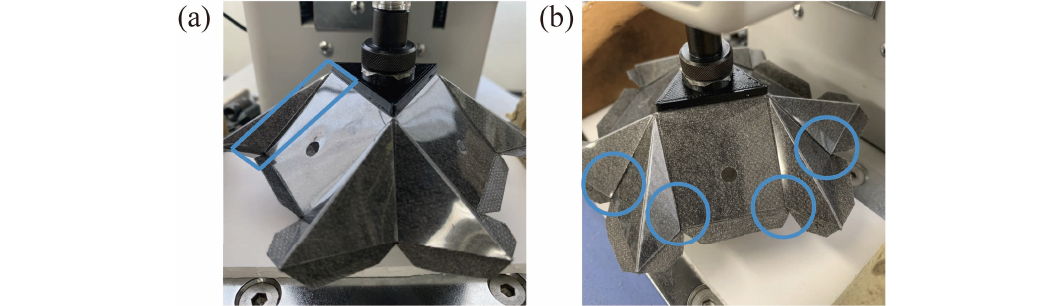}
   \caption{(a) The rectangle panel is slightly bent. (b) The play at the hinge of the nonwoven fabric absorbs deformation.}
  \label{fig:buckling and play}
\end{figure}

We connected modules $A$ and $B$. The bottom triangle of module $A$ was attached to the base made of foam board, and the top triangle of module $B$ is attached to the force gauge (Figure~\ref{fig:Experiment memory} (a)).
We started at the flat state (Figure~\ref{fig:Experiment memory} (a)) and stretched until both modules were deployed (Figure~\ref{fig:Experiment memory} (c)).
Then, the load on the displacement of one cycle returned to the flat state was measured. 
In this experiment, the speed of the force gauge is \SI{50}{\milli \meter \per min}.

We obtained four states, and the hysteresis transitions between the states as designed, as shown in Figure~\ref{fig:Experiment memory} (a--d).
In the deployment process from the flat state ((a) [00]), module $A$ deploys first ((b) [01]), and module $B$ follows ((c) [11]).
When this is in compression, module $A$ first compresses ((d) [10]), then module $B$ compresses and comes back to the flat state ((a) [00]).
During the entire course of the experiment, the local stability problem due to symmetry breaking described in Section~\ref{sec:fabrication} did not occur.
This can be attributed to the fact that the forces applied through the designed connections are evenly distributed.

Figure~\ref{fig:Experiment memory} (e) shows the force--displacement plot of an extension and compression cycle path.
The result shows the $2$-bit memory hysteresis and snapping behavior similar to the expected hysteresis path in Figure~\ref{fig:hysteresis path}. 
We also record that the three stable states of Figure~\ref{fig:Experiment memory} (a), (c), and (d) are at the zero force position, and the stable state of Figure~\ref{fig:Experiment memory} (b) was not at zero force, as we expected.

Figure~\ref{fig:Experiment memory} (f) and (g) compare the hysteresis curves, computed through the bar-and-hinge model and the equivalent strain energy, by fitting the second maximum extreme value to the experimental result. 
Although the presence of plastic and elastic resistances in the hinges and other elements makes a direct comparison difficult, we observed that the experimental model behaved as a two-bit memory system predicted by our models.

The mismatch between the deployment and compression paths in Figure~\ref{fig:Experiment memory} (e) was observed even in the same memory states.
This unintended hysteresis can be attributed to the absorption of energy due to various unexpected physical conditions, bending of the rectangular panel (Figure~\ref{fig:buckling and play} (a)), the play of crease gaps (Figure~\ref{fig:buckling and play} (b)), plasticity, or viscoelasticity in membrane hinges.

\subsection{3-bit Mechanical Memory}

\begin{figure}[thbp]
  \centering
  \includegraphics[page=21,width=\linewidth]{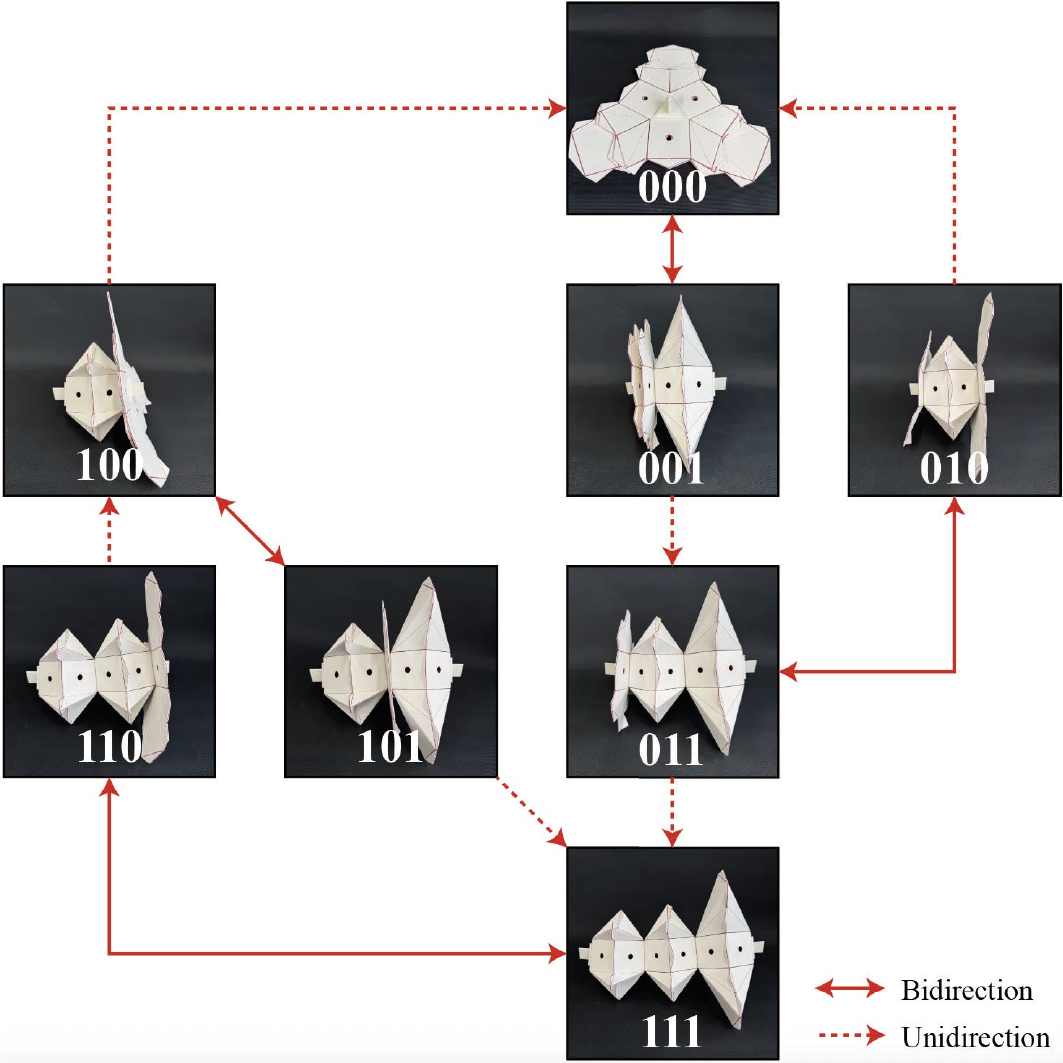}
   \caption{Logic circuit map of 3-bit mechanical memory with physical prototype demonstrations.}
  \label{fig:3bitmemory}
\end{figure}

Figure~\ref{fig:3bitmemory} presents the logic circuit map of a 3-bit mechanical memory, realized through a serial chain assembly of the quasi-mechanism module, bistable modules A and B.
For demonstration, we fabricated larger desktop-scale prototypes of the three designed Q-bellows using the fabrication method 2 described in Section~\ref{subsubsec:deskfabmethod2}, with a unit length ($r_0$) of \SI{50}{mm}.

The prototypes exhibited reliable snapping behavior and deployment stability, and the intended stiffness differences were confirmed through the hand-interaction and 3-bit mechanical memory tests.
Furthermore, the modules consistently maintained their performance under repeated folding–unfolding cycles, demonstrating reproducibility and durability.
When connected in series to form the 3-bit mechanical memory, the system has $2^3$ configurations, and it follows a more complex logic deployment sequence with reversible and irreversible routes, as illustrated in Figure~\ref{fig:3bitmemory}, than the 2-bit system.
This demonstrates that our multistable origami block system can program the deployment sequences and partial deployment(folding), following the prescribed deployment motion.

\section{Conclusion}

\begin{figure}[htbp]
    \centering
    \includegraphics[keepaspectratio,width=\linewidth, page=23]{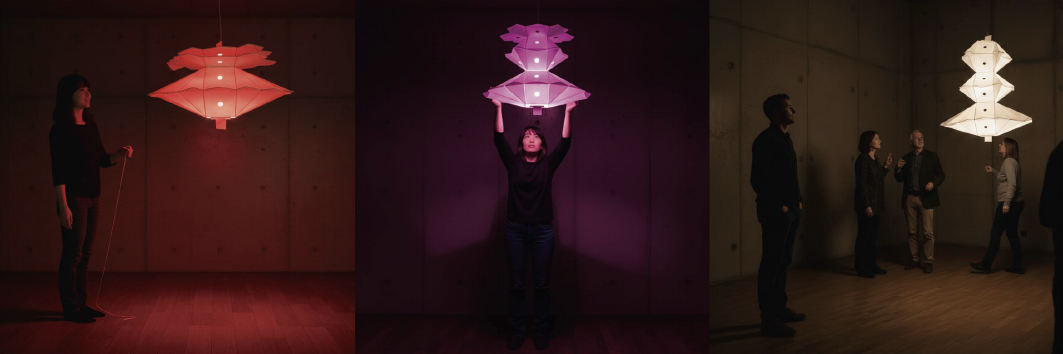}
    \caption{AI-generated illustration of Q-bellows light system. From left, the colors are changed in different bits of $001$, $011$, and $111$ memory configurations. Made by \emph{Gemini 2.5 Pro}}
    \label{fig:AI_Qbellows}
\end{figure}

We have proposed a novel design approach for multistable origami blocks, named Q-bellows, by filling the linkage frame with quadrilateral boundary rigid origami modules.
By controlling the geometric incompatibility between linkage and rigid origami through optimization-based design, the stiffness of Q-bellows was programmed.
Using this incompatibility as the evaluation metric for equivalent strain, we estimated the behavior of the designed Q-bellows and validated it through bar-and-hinge analysis and experiments.
The results showed qualitative agreement; however, the equivalent strain model tended to overestimate the differences in energy barriers.

We assembled Q-bellows prototypes into a serial chain system and performed tests on a $2$-bit mechanical memory, where the observed equilibrium paths and hysteresis transitions were well described by our new prediction method for serial chain behavior. 
Additionally, a 3-bit mechanical memory was demonstrated, following the predicted logic sequence, which highlights the scalability of our approach toward higher-bit systems with programmable deployment sequences.

Potential applications include adaptive building facades, deployable light systems (Figure~\ref{fig:AI_Qbellows})\footnote{We provided the pictures to Gemini 2.5 Pro, and typed the following prompt: Change these structures to various colored light systems in a room, and depict a person pulling them from below.}, and deployable bridges.

\subsection{Discussion}
This chapter is constructed following four primary objectives.
The contributions and limitations corresponding to each objective are as follows.

\subsubsection{Objective 1: Exploring multistable origami design methodologies}

\paragraph{Contributions}
Regarding the geometric incompatibility method, previous work~\cite{melancon2021multistable} demonstrated that multistable origami can be created by combining rigid origami units, but did not provide concrete strategies for controlling stiffness or deployment motion.
Our Q-Bellows design addresses this gap by demonstrating that prescribed deployment motions and tunable stiffness can be achieved through an optimization-based design framework.
This validates that multistability originating from geometric incompatibility can be systematically programmed.

\paragraph{Limitations}
We demonstrated only on linear deployment motions derived from the Sarrus linkage, and the framework has not yet been generalized to other linkages.
Extending the Q-Bellows concept to alternative linkages (Figure~\ref{fig:linkage_example}), arbitrary quadrilateral frames, and more complex deployment motions, such as multi-degree-of-freedom (DOF) linkages~\cite{kamijo2021serial}, has not been explored.
Additionally, the applicability of our method to three-dimensional linkage mechanisms, such as Chebyshev nets~\cite{takatera2003sphere}, scissors mechanisms~\cite{nishimoto2024transformable} with quadrilateral boundary holes, remains an open challenge.
These generalizations will be included in future work.

\subsubsection{Objective 2: Analyzing mechanics across the design parameters}

\paragraph{Contributions}
We compared the analysis results from three different analysis methods: a simplified prediction model (equivalent strain), a low-resolution numerical model (bar-and-hinge), and an actual model.
This comparison makes a significant contribution to understanding how the simplified models deviate from actual behavior, the mechanical features they can capture, and within what range of design stages they remain valid.
Through this, we provide reference examples for selecting an efficient modeling approach required in their design or analysis workflow.

We also developed a prediction method to describe the complex nonlinear behavior of serial chain systems by combining three-dimensional equilibrium paths and total energy contours.
The prediction successfully explained how snapping and hysteresis occur under cyclic loading in serial chain systems and was experimentally validated, contributing to a deeper understanding of the mechanics of the assembled system.

\paragraph{Limitations}

The hinge calibration introduced in Eq.~\ref{eq:sarruscalibration} represents the nonlinear stiffness associated with height and efficiently. 
While the calibrated results well capture the dominant nonlinear behavior of Q-bellows, it should be mentioned that this energy calibration cannot be regarded as governing the total hinge energy. 

% To better preserve the validity and accuracy of this calibration approach, it may be desirable to use stiffer materials for the frame panels.
% This allows the linkage motion of the frame to be maintained without panel deformation, concentrating panel deformation primarily in the modules.

Although the calibrated experimental results showed the stable positions well, there remains room for improvement.
The calibrated results exhibited higher initial stiffness than other models (Figure \ref{fig:qbellowscomparison} (b)).
This suggests that an additional first-order calibration term, such as setting the initial stiffness to zero, might be considered.
However, these differences may also originate from unexpected things such as contact or hinge play; therefore, a simple calibration was retained for clarity.

\subsubsection{Objective 3: Programming global systems by assembling blocks}

\paragraph{Contributions}
By assembling the designed blocks with different stiffness along the prescribed deployment motion, we encoded binary sequences for deployment.
This demonstrated the fundamental feasibility of employing Q-Bellows as programmable elements for controlling the global system behavior.

\paragraph{Limitations}
Further study is required to propose broader architectural or large-scale applications.
Only simple linear serial assemblies were explored, and no assembly methodology was provided for programming more complex logic, deployment motions, or spatial configurations.
Thus, these assembly results are limited to a conceptual demonstration rather than a complete framework for programmable global systems.

\subsubsection{Objective 4: Demonstrating the applicability via desktop and large-scale fabrication}

\paragraph{Contributions}
The desktop-scale prototypes successfully demonstrated the intended mechanical behaviors derived from our design method.
Through close observations during deployment, we identified practical issues, such as local stability, play in the hinge area, and fabrication tolerances that influence the actual performance of the system.
These observations provide important insights for scaling the system up and serve as a basis for understanding the manufacturing considerations required for future large-scale implementations.

\paragraph{Limitations}
Large-scale fabrication was not conducted in this study.
Therefore, whether the intended mechanical performance can be achieved at larger scales remains unverified; further studies on suitable panel materials, hinge structures, and fabrication processes will be required.
Moreover, a more concrete architectural applicability should be discussed based on the large-scale prototypes, and it will be included in future work.

\subsection{Future Work}

Future work will extend our Q-bellows design framework to other linkages (Figure~\ref{fig:linkage_example}), multi-DOF linkages~\cite{kamijo2021serial}, and tessellated frames, enabling more diverse programmable deployment motions. 
The assembly will also be upgraded from simple serial binary sequences toward more complex logics and programmable spatial systems~\cite{nishimoto2024transformable}. 
Finally, large-scale fabrication remains to be validated.
Further investigations into suitable panel materials, hinge strategies, and structural performance are required to provide practical architectural applications.

	\chapter{T-toroid: Multistable Trapezoidal-hedra Origami Block}
\label{chap:ttoroid}

\emph{\textbf{Author's note:}
The content covered in this chapter is partially included in publications~\cite{lee2024multistable}.}

\section{Introduction}

While several studies have explored multistable origami beyond existing patterns and applied them in block assemblies, most have been limited to planar tessellations~\cite{hoberman2019construction, li2023multistable} or simple spatial structures~\cite{ebara2003deployable, melancon2021multistable}.
This gap arises because the geometric design methods for stable configurations of multistable origami, particularly those aimed at assembling complex global geometries, including cylinder-, dome-, and saddle-like three-dimensional curved surfaces, remain underexplored.

To address this gap, we present the design and analysis of multistable origami blocks with a desired polygonal trajectory and profile, termed the \emph{T-toroid}, and showcase a novel modular building system that can construct 3-dimensional curved surfaces by assembling these blocks.
We (1) design \emph{a T-toroid} based on the rigid foldable tubes called T-hedral tubes~\cite {sharifmoghaddam2023generalizing}, 
(2) analyze mechanical behaviors across various design parameters using finite element analysis, and validate the multistability through the desktop-scale prototypes,
(3) investigate achievable curved surfaces based on the geometric features(edge offset) of the T-toroid, and 
(4) demonstrate practical application possibilities through the fabrication of a large-scale prototype.
These objectives are addressed in Sections~\ref{Ttoroid: Geometry Design}, ~\ref{Ttoroid:mechanics},~\ref{Ttoroid: Assembly}, and~\ref{Ttoroid:Application}, respectively.

\section{Geometry Design}
\label{Ttoroid: Geometry Design}

\begin{figure}[htbp]
    \centering
    \includegraphics[keepaspectratio,width=0.9\linewidth, page=1]{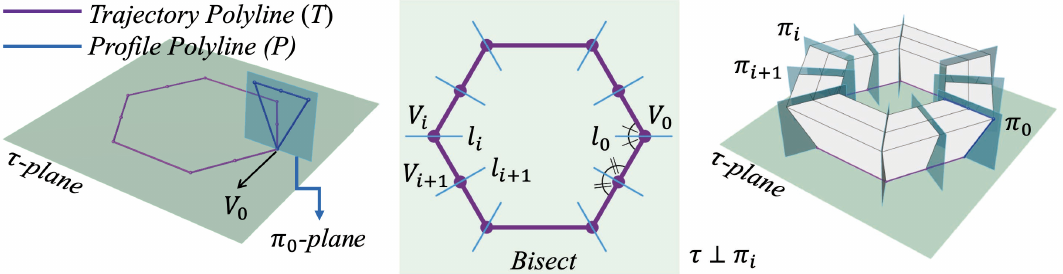}
    \caption{Design process of T-hedral tube (T-tube).}
    \label{fig:Ttube}
\end{figure}

\subsection{T-hedral tube}
\label{Ttoroid: Ttube}
We first construct rigidly foldable tubes using the concept of Trapezoidal polyhedra (T-hedra), which are a special type of planar quad (PQ) meshes with rigid faces, degree-four vertices, and rotational joints on their edges. 
While such PQ meshes are usually geometrically rigid, T-hedral structures exhibit a 1-DOF deformation, preserving the rigidity of the planar faces~\cite{sharifmoghaddam2020using}. 

A T-hedron can be constructed using two planar boundary polylines, referred to as \emph{profile} and \emph{trajectory} polylines (denoted as $P$ and $T$ respectively), lying on two perpendicular planes ($\pi_0$ and $\tau$) that share a vertex $V_0$, along with a series of profile planes ($\pi_i$, $i=0, \dots,k$) perpendicular to $\tau$ (see Figure~\ref{fig:Ttube}). 
Each profile plane $\pi_i$ can be determined by $l_i$, the intersection line of $\pi_i$ with $\tau$, and passes through the corresponding vertex $V_i$ of the trajectory polyline. The construction is straightforward, involving the projection of $P$ along $T$ onto the $\pi_i$-planes and the subsequent connection of corresponding vertices at each projection step.

A cylindrical topology, known as a \emph{T-hedral tube} (T-tube), can be achieved by utilizing a closed polygon $P$ as the profile. While we anticipate the resulting tube to exhibit rigidity, preserving the 1-DOF movement is contingent upon satisfying the loop closure condition outlined in Theorem 3.1 of~\cite{sharifmoghaddam2023generalizing}.

\subsection{Design Variations}

\begin{figure}[ht]
    \centering
    \includegraphics[keepaspectratio,width=0.9\linewidth, page=2]{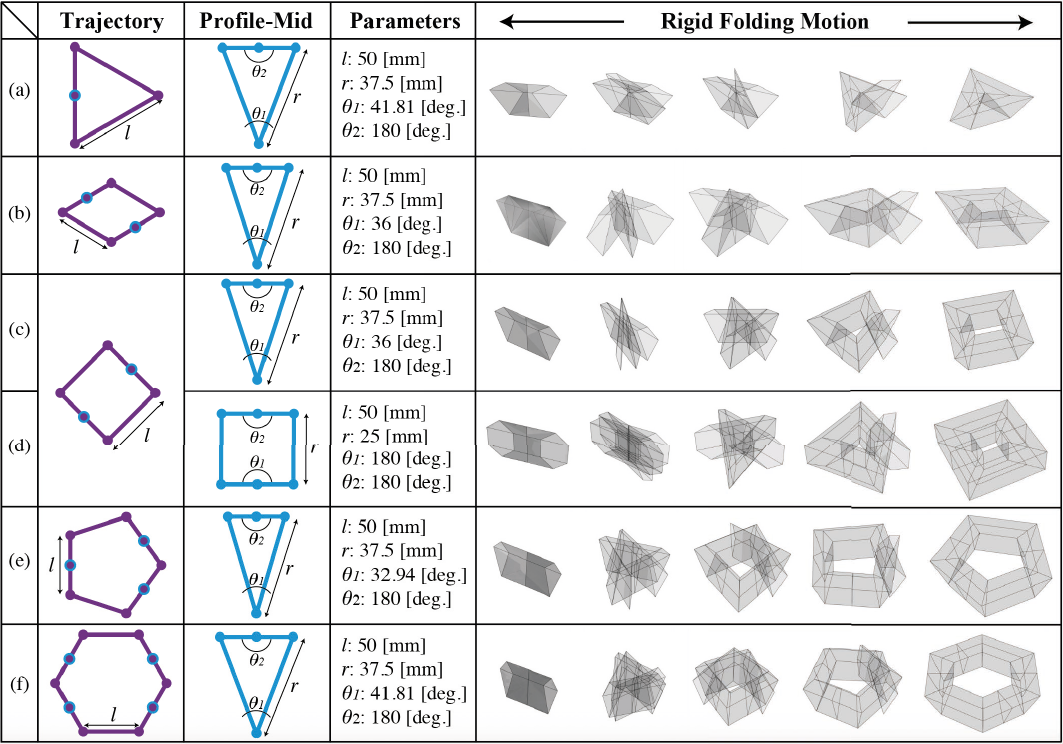}
    \caption{Design parameters and rigid folding motions of various T-tubes. Points on the polylines represent rotational joints. The profile is located at the mid-joint (blue-circled purple points) of the trajectory and projected along it. From the top row: (a) Triangle–Deltoid, (b) Golden rhombus–Deltoid, (c) Square–Deltoid, (d) Square–Hexagon, (e) Pentagon–Deltoid, and (f) Hexagon–Deltoid T-tubes.}
    \label{fig:designvariation}
\end{figure}

We design various \emph{T-tubes} with triangular, quadrilateral, pentagonal, and hexagonal trajectories, with deltoidal and hexagonal profiles. 
Figure~\ref{fig:designvariation} illustrates design parameters ($l, r, \theta_1, \theta_2$) and motions of T-tubes.

A T-tube with a polygonal profile reaches a flexion limit when at least one profile segment becomes parallel to $\tau$ \cite{sharifmoghaddam2023generalizing}. 
Hence, we opt for deltoidal and hexagonal profiles, ensuring that when they reach their flexion limit in the fully deployed state, their deformed cross-sections become triangular and quadrilateral, respectively.
Modifying the length parameters $l$ and $r$, or the angular parameters $\theta_1$ and $\theta_2$, can control both the fully deployed states and the assembled curved surfaces. 
For example, the values in Figure~\ref{fig:designvariation} yield the shapes showcased in Section~\ref{Ttoroid: Assembly}.

To make the flat-foldable trajectory, we split the trajectory polygon at the midpoints of its segments.
Without these additional joints (blue circled purple points on the trajectory polyline in Figure~\ref{fig:designvariation}), the trajectory curve will fold with a total turn angle of $n\pi$, where $n$ is the number of vertices of the polygon.
To make the total turn into $2\pi$, $n-2$ additional joints are added, each of which turns by $-\pi$.
Thus, we have installed the $n-2$ joints at the appropriate positions to enable flat-foldability.

\subsection{Loop-closing}
\label{Ttoroid: loop-closing}

To obtain the multistability, we propose a novel method, called \emph{T-toroidalization}, by closing both ends of the T-tube into tori.
To get closure at the deployed state, (1) the profile polygons at the start and the end need to be congruent.
In addition, to have the flat-folded state, we impose the condition of flat-foldability.
Flat foldability is achieved by making (2) the T-tube (before ends are connected) flat-foldable and (3) the trajectory polygon flat-folded to a line.
Condition (2) is achieved by positioning $\pi_i$-planes to bisect each segment angle around $V_i$, thus endowing the structure with flat-foldability~\cite{sharifmoghaddam2023generalizing}.
Conditions (1) and (3) are achieved by using symmetric trajectory polylines.

While the designed T-tubes with ends not connected are rigid-foldable, their rigid motion does not preserve their topology, as the trajectories open up and often impose self-intersections during the motion (see the rigid folding motion part of Figure~\ref{fig:designvariation}). 

We obtain T-toroids, the toroidal topology blocks with multistability, by hinge-connecting and closing at both ends of the designed T-tubes.
Geometrically, the T-toroid belongs to the rigid origami class at both the deployed and flat-folded states.
Meaning that, geometrically, the T-toroid experiences no strain in these two configurations.
While the motion between these states does not belong to the rigid origami class.
Thus, the panels should be deformed somewhere, and this generates a strain energy barrier and induces snap-through behavior when the structure transitions between these two states.

\section{Mechanics}
\label{Ttoroid:mechanics}

\begin{wrapfigure}{l}{0.65\textwidth}
    \centering
    \includegraphics[keepaspectratio,width=\linewidth, page=3]{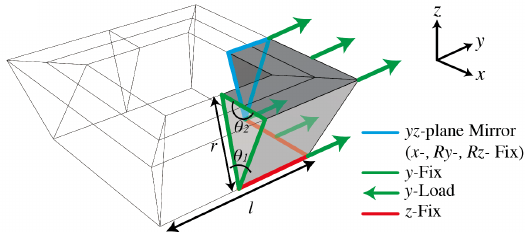}
    \caption{Applied boundary condition for snapping analysis}
    \label{fig:Ttoroidfeasetting}
\end{wrapfigure}

In this section, we investigate the mechanical behavior of T-toroids across various design parameters using finite element analysis.
The parameters considered include panel thickness, hinge stiffness, profile angles ($\theta_1, \theta_2$), and the edge ratio of trajectory to profile ($r/l$).
To reduce computational cost and to focus on the influence of these parameters, the trajectory is fixed to a square, which serves as a representative T-toroid.
Multistability with other polygonal trajectories will be examined later through prototype hand-interaction testing in subsequent Section~\ref{Ttoroid:Application}.

\subsection{Modeling and Boundary Conditions}
\label{Ttoroid:boundary}

The modeling method for FEA follows Section~\ref{subsec:FEA}.
Figure~\ref{fig:Ttoroidfeasetting} shows the boundary conditions applied to the model. 
To simplify the model design and reduce computational costs, we model a biaxially symmetric analysis model ($1/4$ of the full block) with a square trajectory and various parameterized deltoidal profiles.
The analysis models are initialized in the flat state and constrained with mirror symmetry. 
The vertical mid-section along the $yz$-plane is constrained with an $x$-translation fix and restrictions on $R_y$ and $R_z$ rotations. The horizontal mid-section along the $xz$-plane is constrained with a $y$-translation fix.
The bottom edges, corresponding to the trajectory profile, are fixed only in the $z$-translation.
For deploying the T-toroid, we apply a $y$-direction forced-displacement load ($y$) to the outermost and bottom edges of the analysis model, $l/2 + r\sin{\frac{\theta_1}{2}}$ and $l/2$ respectively.

Furthermore, Figure~\ref{fig:Ttoroidfeasetting} also shows the reference model for the comparison across the various geometric parameters. 
We set the parameters for the reference model as $r=$ \SI{400}{mm}, $l=$ \SI{800}{mm}, $\theta_1=36^\circ$, $\theta_2=180^\circ$, thickness($t$) $=$ \SI{10}{mm}, and the hinge scaling factor ($f$) is $10^{10}$.
For the material setting, we used polypropylene material properties with an elastic modulus of \SI{1134}{MPa}, density of \SI{9.05e-10}{ton/mm^3}, and the Poisson's ratio of 0.38~\cite{zhang2025effect}.
The snapping analysis is performed using a quasi-static method in the dynamic implicit solver.

\subsection{Snapping Analysis}
\label{subsec:TtoriSnappingAnalysis}

\begin{figure}[htbp]
    \centering
    \includegraphics[keepaspectratio,width=\linewidth, page=4]{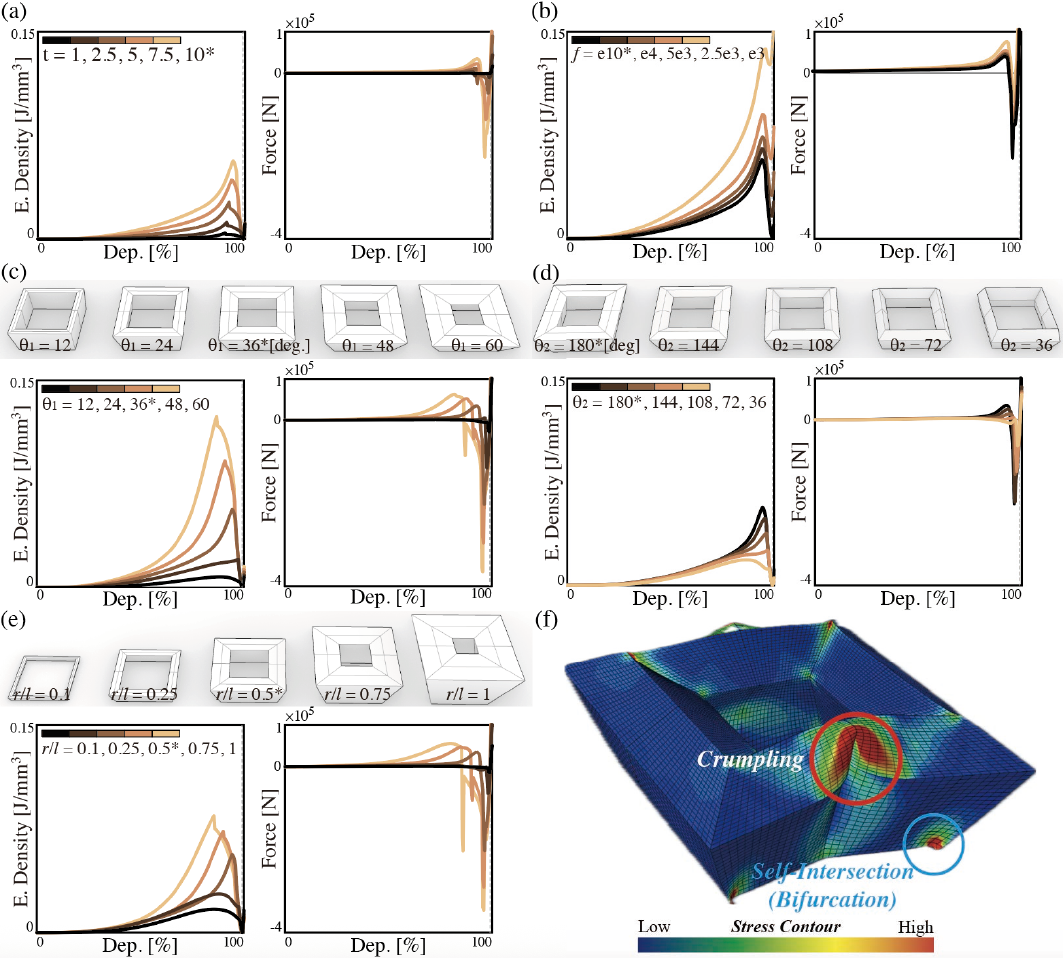}
    \caption{Analysis results of (a) panel thickness ($t$) parameters, (b) hinge stiffness($K_{crease}$) parameters, (c) bottom profile angle ($\theta_1$) parameters, (d) top profile angle ($\theta_2$) parameters, and (e) slenderness ($r/l$) parameters. For the comparison, the results of the reference model are indicated by a $\ast$ mark in the legends. (f) Stress contour plot of $\theta_1=60^\circ$ case. Crumpling and self-intersection behavior were captured during deployment.}
    \label{fig:results1}
\end{figure}

\begin{table}[htbp]
\centering
\caption{Peak locations of \emph{Figure~\ref{fig:results1}}.}
\begin{tabular}{|l||*{10}{c|}}\hline
\label{tab:TtoroidPeak}
\makebox[8em]{Deployment [\%]} &\makebox[4em]{(a)}&\makebox[4em]{(b)}&\makebox[4em]{(c)}&\makebox[4em]{(d)}&\makebox[4em]{(e)}\\\hline
$C_{Black}$       & 91.129 & 95.161 & 89.113 & 95.161 & 86.290 \\
$C_{Dark Brown}$  & 91.936 & 95.163 & 97.985 & 95.968 & 88.710 \\
$C_{Brown}$       & 93.145 & 95.445 & 95.161 & 96.774 & 95.161 \\
$C_{Light Brown}$ & 94.355 & 95.445 & 91.533 & 97.581 & 90.725 \\
$C_{Yellow}$      & 95.161 & 95.960 & 87.500 & 87.097 & 86.290 \\
\hline
\end{tabular}
\vspace{0.5cm}
\centering
\caption{Peak energy densities of \emph{Figure~\ref{fig:results1}}.}
\begin{tabular}{|l||*{10}{c|}}\hline
\label{tab:TtoroidEdensity}
\makebox[8em]{E. Density $\times10^{-2}$} &\makebox[4em]{(a)}&\makebox[4em]{(b)}&\makebox[4em]{(c)}&\makebox[4em]{(d)}&\makebox[4em]{(e)}\\\hline
$C_{Black}$       & 0.425 & 5.583 & 0.758 & 5.583 & 1.664 \\
$C_{Dark Brown}$  & 1.189 & 6.520 & 2.023 & 4.770 & 2.763 \\
$C_{Brown}$       & 2.639 & 7.380 & 5.583 & 3.713 & 5.583 \\
$C_{Light Brown}$ & 4.216 & 8.999 & 9.054 & 2.495 & 7.230 \\
$C_{Yellow}$      & 5.583 & 13.754 & 12.237 & 1.823 & 8.345 \\
\hline
\end{tabular}
\vspace{0.5cm}
\centering
\caption{Force ratio between deployment and folding ($|F_{min} / F_{max}|$) of \emph{Figure~\ref{fig:results1}}}
\begin{tabular}{|l||*{10}{c|}}\hline
\label{tab:TtoroidFratio}
\makebox[8em]{$|F_{min} / F_{max}|$} &\makebox[4em]{(a)}&\makebox[4em]{(b)}&\makebox[4em]{(c)}&\makebox[4em]{(d)}&\makebox[4em]{(e)}\\\hline
$C_{Black}$       & 2.684 & 5.871 & 21.233 & 5.871 & 19.246 \\
$C_{Dark Brown}$  & 3.433 & 4.788 & 19.177 & 7.770 & 23.116 \\
$C_{Brown}$       & 4.308 & 3.276 & 5.871 & 9.202 & 5.871 \\
$C_{Light Brown}$ & 4.878 & 2.341 & 5.669 & 21.728 & 6.114 \\
$C_{Yellow}$      & 5.871 & 1.023 & 5.831 & 21.880 & 6.229 \\
\hline
\end{tabular}
\end{table}

Figure~\ref{fig:results1} (a--e) and Table~\ref{tab:TtoroidPeak}--\ref{tab:TtoroidFratio} show the FEA analysis results of snapping behavior depending on design parameters.
The horizontal axis ($Dep.$) is defined as the $y$-axis forced-displacement ($y$) of the bottom-loaded edge divided by half of the trajectory edge ($l/2$), $2y/l \times 100$ [\%].
The left plots show the energy density, calculated by dividing the total strain energy by the volume of the geometry, and the right plots show the required force during deployment.
By comparing the energy density, we can avoid the influence of panel area differences along the design parameters, providing a uniform comparison across different T-toroid geometries.
The asterisk ($\ast$) marked in figures and legends of Figure~\ref{fig:results1} represents the result of the reference model.
Further result details can be found in Table~\ref{tab:TtoroidPeak}--~\ref{tab:TtoroidFratio}.
Table~\ref{tab:TtoroidPeak} presents the deployment locations of energy density peaks.
Table~\ref{tab:TtoroidEdensity} shows the peak value of energy density.
Table~\ref{tab:TtoroidFratio} shows the absolute ratio between maximum and minimum force values ($|F_{min}/F_{max}|$), representing the required load difference for deploying and folding the T-toroid.

Figure~\ref{fig:results1} (a) shows the influence of panel thickness ($t$) under the same geometry.
As the thickness increases, the peak energy also increases (Table~\ref{tab:TtoroidEdensity} (a)), and the location of the energy peak shifts to a later stage (Table~\ref{tab:TtoroidPeak} (a)).
Also, the clear snapping behavior is observed, requiring larger forces for deployment and even greater forces for folding than for deployment (Table~\ref{tab:TtoroidFratio} (a)).
The noteworthy point is the inconsistent behavior observed at thinner thicknesses.
As the panel becomes thinner, it tends to deform like a membrane, concentrating severe \emph{crumpling} near the mid-joint region (see Figure~\ref{fig:results1} (f)).
This crumpling appears earlier as the thickness decreases, disrupting the clear snapping behavior. 
Even after the energy peak, oscillations continue, making convergence difficult and leading to unstable results.

Figure~\ref{fig:results1} (b) shows the influence of hinge stiffness (Eq.~\ref{eq:Khinge} (b)) under the same geometry.
The moment at the hinge is set to zero when the toroid is in its flat state. 
Even as hinge stiffness increases (i.e., smaller $f$), the location of the energy peak remains nearly constant (slightly shifting back, refer to Table~\ref{tab:TtoroidPeak} (b)). 
While the peak energy value increases with hinge stiffness (Table~\ref{tab:TtoroidEdensity} (b)), the energy of the second stable state also rises, leading to a decrease in stability in the deployed state. 
Consequently, when hinge stiffness becomes too high, it becomes difficult to maintain the deployed state or lose the multistability.
In addition, higher hinge stiffness requires greater force for deployment; however, the difference between the forces required for deployment and folding is reduced (Table~\ref{tab:TtoroidFratio}).

Figure~\ref{fig:results1} (c) shows the influence of the profile angle ($\theta_1$).
As $\theta_1$ increases, the peak energy value increases (Table~\ref{tab:TtoroidEdensity} (c)). 
Additionally, the peak location shifts forward (Table~\ref{tab:TtoroidPeak} (c)), which can be attributed to more severe deformation as the toroid widens, resulting in crumpling that causes the peak to appear earlier. 
Thus, in the 48$^\circ$ and 60$^\circ$ cases, crumpling occurs (see Figure~\ref{fig:results1} (f)), resulting in unstable behavior near the peaks.
Conversely, as $\theta_1$ decreases, a sudden drop in the energy path can be observed. 
This is likely attributed to \emph{self-intersection} (see Figure~\ref{fig:results1} (f)) at the lower edge tips, which introduces bifurcation in the energy pathway.
Also, as $\theta_1$ increases, the difference in force required for deployment and folding tends to decrease (Table~\ref{tab:TtoroidFratio} (c)). 
However, due to the uncertainties introduced by crumpling and self-intersection, the results of peak locations and force ratios are inconsistent and difficult to predict.
Furthermore, as $\theta_1$ approaches zero, the structure converges toward a simple linkage mechanism, and thus bistability is not expected.

Figure~\ref{fig:results1} (d) shows the influence of varying the profile angle $\theta_2$ while fixing $\theta_1$ as 36$^\circ$.
As $\theta_2$ decreases, the energy peak diminishes (Table~\ref{tab:TtoroidEdensity} (d)), and the peak position shifts backward (Table~\ref{tab:TtoroidPeak} (d)), at which bifurcation due to self-intersection is observed. 
Conversely, as $\theta_2$ increases, the structure requires greater force for deployment and exhibits a clearer snapping behavior.
However, the force difference between deploying and folding tends to be smaller (Table~\ref{tab:TtoroidFratio} (d)).

Figure~\ref{fig:results1} (e) shows the influence of the edge-length ratio ($r/l$) between the trajectory edge ($l$) and the profile edge ($r$).
Overall, as the $r/l$ increases, the energy peak also increases (Table~\ref{tab:TtoroidEdensity} (e)). 
A larger $r/l$ widens the toroid, leading to crumpling; 
thus, the peak appears earlier (Table~\ref{tab:TtoroidPeak} (e)).
Conversely, when the $r/l$ decreases, the structure is prone to being affected by self-intersection.
Because of these crumpling and self-intersection, clear snapping behavior becomes difficult to observe, and the trend of the force ratio remains uncertain (Table~\ref{tab:TtoroidFratio} (e)).

\begin{figure}[htbp]
    \centering
    \includegraphics[keepaspectratio,width=\linewidth, page=5]{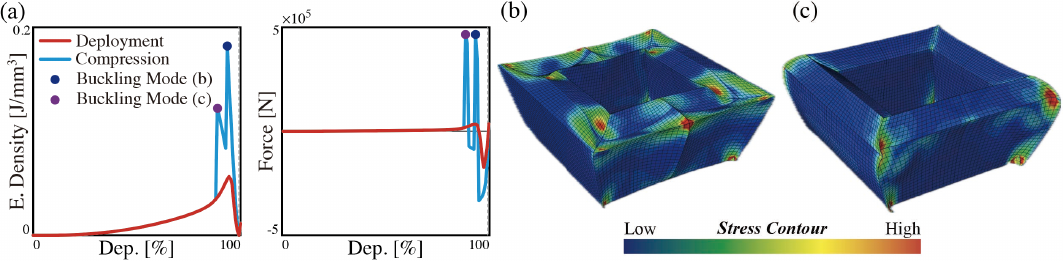}
    \caption{(a) Cyclic loading analysis result of reference model. (b) and (c) show the stress contours of local buckling modes during the compression loading.}
    \label{fig:results2}
\end{figure}

Figure~\ref{fig:results2} shows the T-toroid behavior of the cyclic loading (forced displacement) of the reference model under the same boundary conditions.
It can be observed that the folding process requires greater force than predicted in Table~\ref{tab:TtoroidFratio}. 
This occurs because, in the deployed state, the mid-joint parts straighten and act as bar members subjected to axial loads, thereby resisting compression.
Figure~\ref{fig:results2} (b) and (c) show the deformed configurations of the reference model at the energy peaks under compressive loading.
In (b), local buckling of the panels under compression is observed.
 Subsequently, in (c), the mid-joint parts are aligned, functioning as bars carrying axial loads, while the panel regions near the loading points exhibit noticeable stress.

\subsection{Discussions}
Multistability was consistently observed across all cases, and as intended in our design, the strain energy at the deployed state geometrically approaches zero.
The required forces for deployment and folding are asymmetric, suggesting that under the given boundary conditions, the T-toroid is easy to deploy but relatively more difficult to fold, representing the \emph{quasi-locking} mechanism.

For the practical design stage, ensuring clear snapping behavior while avoiding unstable behavior such as crumpling requires selecting a sufficiently thick panel while, at the same time, avoiding overly wide toroid geometries.
Also, reducing hinge stiffness is preferable to enhance stability in the deployed state.
Although self-intersections do not occur in reality, the regions where they appear in simulations may correspond to areas prone to stress concentration, contact, or potential damage.
Finally, considering load directions and boundary conditions when designing the total system can further improve structural performance.
These aspects should be carefully recognized and incorporated into the design process.

\section{Assembly}
\label{Ttoroid: Assembly}

\subsection{Assembly Conditions}

\begin{figure}[htbp]
    \centering
    \includegraphics[keepaspectratio,width=\linewidth, page=18]{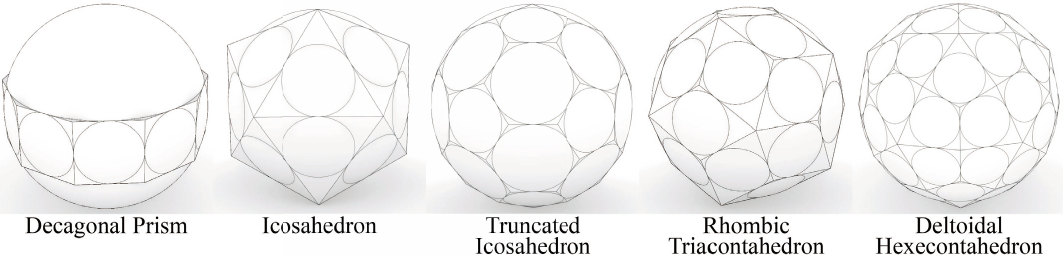}
    \caption{Examples of Koebe polyhedra. All edges of Koebe polyhedra are tangent to the unit sphere, which creates the incircle in each face.}
    \label{fig:koebepolyhedra}
\end{figure}

\begin{figure}[htbp]
    \centering
    \includegraphics[keepaspectratio,width=\linewidth, page=7]{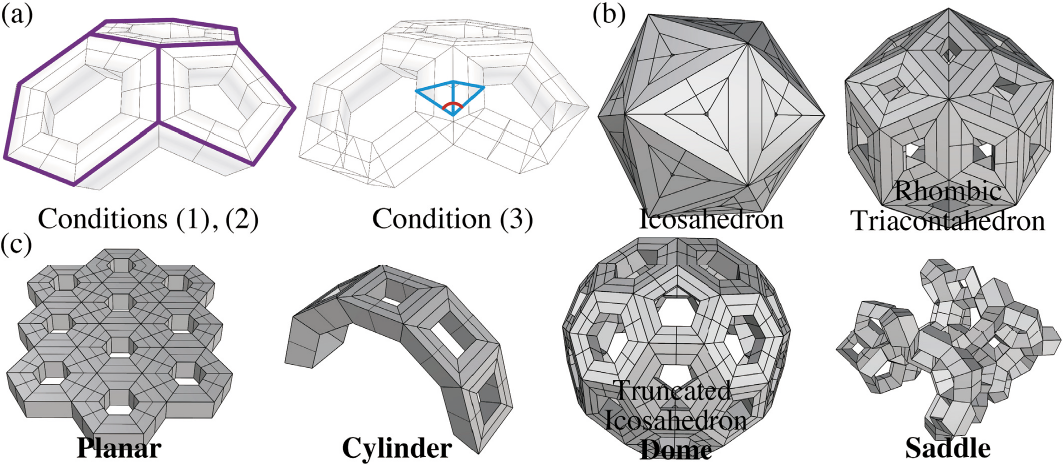}
    \caption{(a) Assembly Conditions. The examples of (b) polyhedra assembly, and (c) standard types of curved-surface assembly with the polygonal trajectory T-toroids.}
    \label{fig:polygonalassembly}
\end{figure}

Using the designed T-toroids, we constructed various curved surfaces and examined the geometric conditions required for continuous face-to-face connections.
Each T-toroid maintains a constant cross-section along its trajectory, resulting in a frustum shape formed by extruding a polygon with a uniform taper.
When multiple T-toroids are assembled by connecting their tapered faces, the distance between the outermost and bottom trajectories remains constant across the entire structure, i.e., they maintain a uniform offset.
This characteristic corresponds to an \emph{edge-offset} mesh, which is parallel to meshes whose edges are tangent to a sphere, known as \emph{Koebe polyhedra}~\cite{pottmann2007geometry,pottmann2010edge, bobenko2017discrete}.
Figure~\ref{fig:koebepolyhedra} shows the examples of Koebe polyhedra, and they serve as the geometric foundation of the T-toroid assembly, especially in cylinder and dome cases.
Based on these geometric properties, we define assembly conditions as follows (Figure~\ref{fig:polygonalassembly} (a)):
(1) The surface mesh aligns with the outermost trajectory of the T-toroids.
(2) Each neighboring T-toroid shares a common face.
(3) The dihedral angle between meshes equals half of the sum of the profile angles ($\theta_1$) of the neighboring T-toroids.

\subsection{Polygonal Trajectory T-toroids Assembly}

Figure~\ref{fig:polygonalassembly} (b) and (c) show the assembled polygon-meshed curved surfaces, including planar, cylindrical, dome, and saddle types.
For the planar surface, we used a hexagonal profile (see Figure~\ref{fig:designvariation} (d)) to form prisms without taper, allowing the application of various 2D polygon tessellation methods.

For the cylindrical surface, we assembled the square-deltoid T-toroid (Figure~\ref{fig:designvariation} (c)) in a row. 
However, tessellation in the transverse direction is not straightforward, as the taper also induces curvature in that direction. 
This limitation can be overcome by attaching another T-toroid on the top side, which will be explained in the next section.
Other cylindrical shapes can also be obtained by using different polygonal trajectories.

Dome surfaces can be achieved through various polyhedral tessellation methods.
In Figure~\ref{fig:polygonalassembly} (b) and (c), examples include an icosahedron assembled from triangle–deltoid T-toroids (Figure~\ref{fig:designvariation} (a)), a rhombic triacontahedron from rhombus–deltoid T-toroids (Figure~\ref{fig:designvariation} (b)), and a truncated icosahedron from hexagon–deltoid and pentagon–deltoid T-toroids (Figure~\ref{fig:designvariation} (e),(f)).
The edge-offset meshes are characterized by vertices where all incident edges are tangent to the same right circular cone~\cite{pottmann2007geometry, pottmann2010edge}.
Because of this property, the edge-offset meshes are parallel to Koebe meshes, defined as polyhedral meshes whose edges are tangent to a sphere and whose faces admit incircles~\cite{bobenko2017discrete, mesnil2015isogonal}.
Through this relationship, a wide variety of polyhedra belonging to the \emph{Koebe polyhedra} class, such as Archimedean solids, Platonic solids, prisms, and isohedra, can be assembled using the edge-offset property of the T-toroid (see Figure~\ref{fig:koebepolyhedra}).
This geometric correspondence provides a foundation for constructing diverse dome-type T-toroid assemblies.

The saddle surface (Figure~\ref{fig:results1} -- saddle) is constructed from pentagon and square meshes. % l=50mm, r=37.5mm
Since all dihedral angles between meshes are identical, the profile angles are set as $\theta_1=\theta_2=43.403^\circ$.
The assembly is achieved by connecting the upper or lower-side faces of adjacent T-toroids along the mesh edges (i.e., the outermost trajectories).
Connections between lower-side faces generate positive Gaussian curvature, while those between upper-side faces produce negative Gaussian curvature, together forming a saddle surface.
However, this case requires relaxing the assembly condition (2), as the connected faces no longer match in shape.
An equivalent geometry can also be obtained by flipping one T-toroid and attaching it to the top face of another with $\theta_2 = 180^\circ$.
Although this method increases fabrication complexity, it constrains the folding motion between adjacent blocks and can enhance structural stability in the assembled state.

\subsection{Quad Trajectory T-toroids Assembly}

\begin{figure}[htbp]
    \centering
    \includegraphics[keepaspectratio,width=\linewidth, page=8]{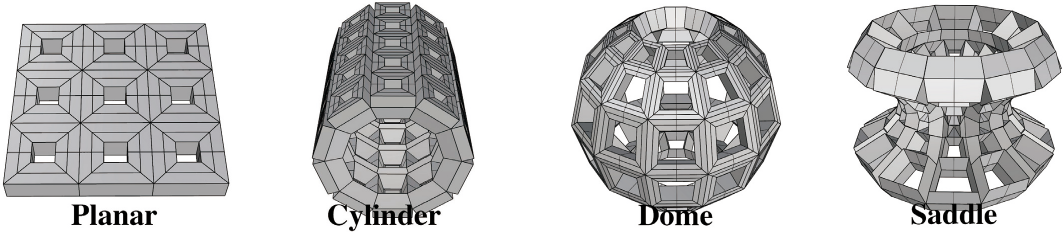}
    \caption{Standard types of curved surface assembly with quad trajectory T-toroids.}
    \label{fig:quadassembly}
% \end{figure}
\vspace{1cm}
% \begin{figure}[htbp]
    \centering
    \includegraphics[keepaspectratio,width=\linewidth, page=17]{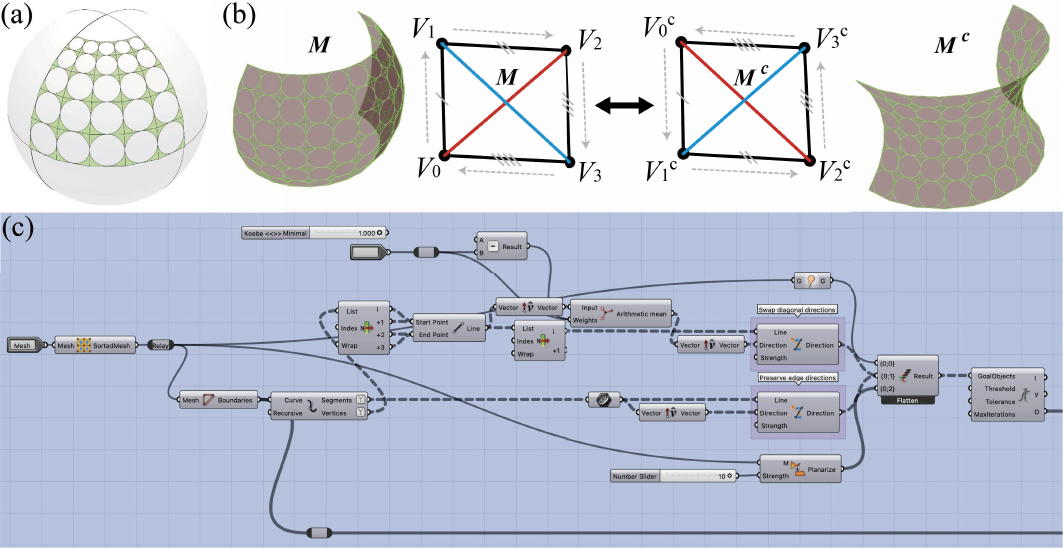}
    \caption{(a) Koebe PQ mesh generation, (b) The Christoffel dual transformation between Koebe PQ mesh ($\textbf{M}$) and minimal surface ($\textbf{M}^{\textbf{c}}$), and (c) this implementation by \emph{Rhino/Grasshopper}.}
    \label{fig:christoffeldual}
\end{figure}

Figure~\ref{fig:quadassembly} shows an assembly of T-toroids constructed based on the planar quad (PQ) meshes.
For the planar surface assembly, this can be realized simply by assembling the square-hexagon T-toroid (Figure~\ref{fig:designvariation} (d)) into a planar arrangement.

For the cylindrical surface assembly, similar to Figure~\ref{fig:polygonalassembly} (c), curvature in the cylindrical direction is generated using square-deltoid T-toroids (Figure~\ref{fig:designvariation} (c)). 
This can extend to the orthogonal direction by attaching square-hexagon T-toroids (Figure~\ref{fig:designvariation} (d)) on top.

For the dome surface, the assembly is constructed using Koebe polyhedra of PQ meshes (Koebe mesh), as shown in Figure~\ref{fig:christoffeldual} (a). 
The construction method for Koebe mesh is the same as the circle packing problem between two meridians of the sphere, which can be calculated using the non-linear recurrence relations by referring to Section 4.1 in~\cite{mesnil2015isogonal}.
We use the constructed PQ meshes as trajectories.
The additional mid-joint is placed at the midpoint of an arbitrary edge of the trajectory, and another mid-joint is added at a position that satisfies flat-foldability from simple length calculation.
The profile angle $\theta_1$ can be calculated from dihedral angles between adjacent meshes(see Figure~\ref{fig:polygonalassembly} (a)), which is then applied to design each T-toroid.
By assembling designed T-toroids, dome-type surfaces are obtained (Figure~\ref{fig:quadassembly}, Dome).

Koebe mesh and minimal surfaces are known to be in a reciprocal duality, which can be transformed by the \emph{Christoffel duality}~\cite {pottmann2010edge, bobenko2017discrete, tovsiae2019design}.
We implemented Christoffel duality using the \emph{Rhino / Grasshopper - Kangaroo2} to transform Koebe polyhedra into minimal (saddle) surfaces (Figure~\ref {fig:christoffeldual} (b)).
By preserving the parallelism of mesh edges $(\overline{V_i V_{(i+1)\bmod 4}} \parallel \overline{V_i^c V_{(i+1)\bmod 4}^c}, (i\in\{0,1,2,3\}))$, while reversing the diagonal components of each mesh $(\overline{V_0 V_2} \parallel \overline{V_1^c V_3^c},\quad \overline{V_1 V_3} \parallel \overline{V_0^c V_2^c})$, the resulting geometry yields saddle-shape (minimal surface) PQ meshes~\cite{bobenko2017discrete, tovsiae2019design}.
Figure~\ref {fig:christoffeldual} (c) shows the actual visual programming codes in \emph{Rhino/Grasshopper}.
The generated saddle surface is used as the trajectories.
Because the Christoffel duality preserves the dihedral angles, we can use the same profile angles of the basic Koebe polyhedron.
For the saddle shape assembly, $\theta_2(=\theta_1)$ needs to be utilized.
The assembled saddle surface can be seen in Figure~\ref{fig:quadassembly}.

\begin{figure}[htbp]
    \centering
    \includegraphics[keepaspectratio,width=\linewidth, page=14]{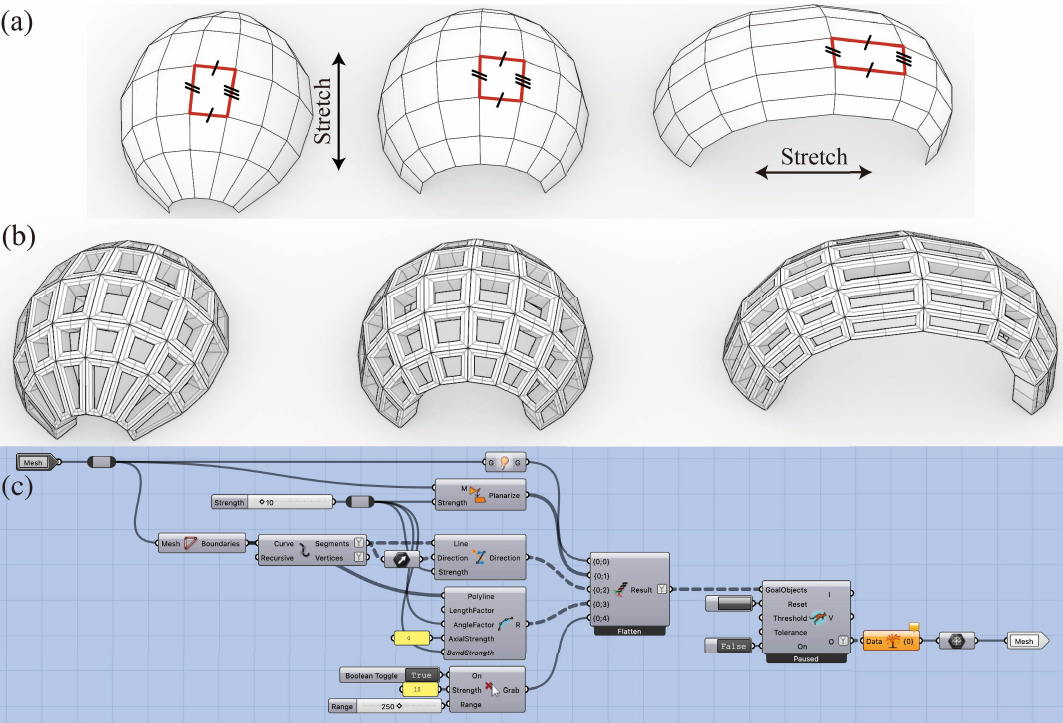}
    \caption{(a) Mesh parallelism between corresponding meshes, (b) The assemblies of deformed meshes. (c) Implementation of mesh parallelism by \emph{Rhino/Grasshopper}.}
    \label{fig:meshparallelism}
\end{figure}

\subsection{Complex Curved Surface}

\begin{figure}[htbp]
    \centering
    \includegraphics[keepaspectratio,width=\linewidth, page=16]{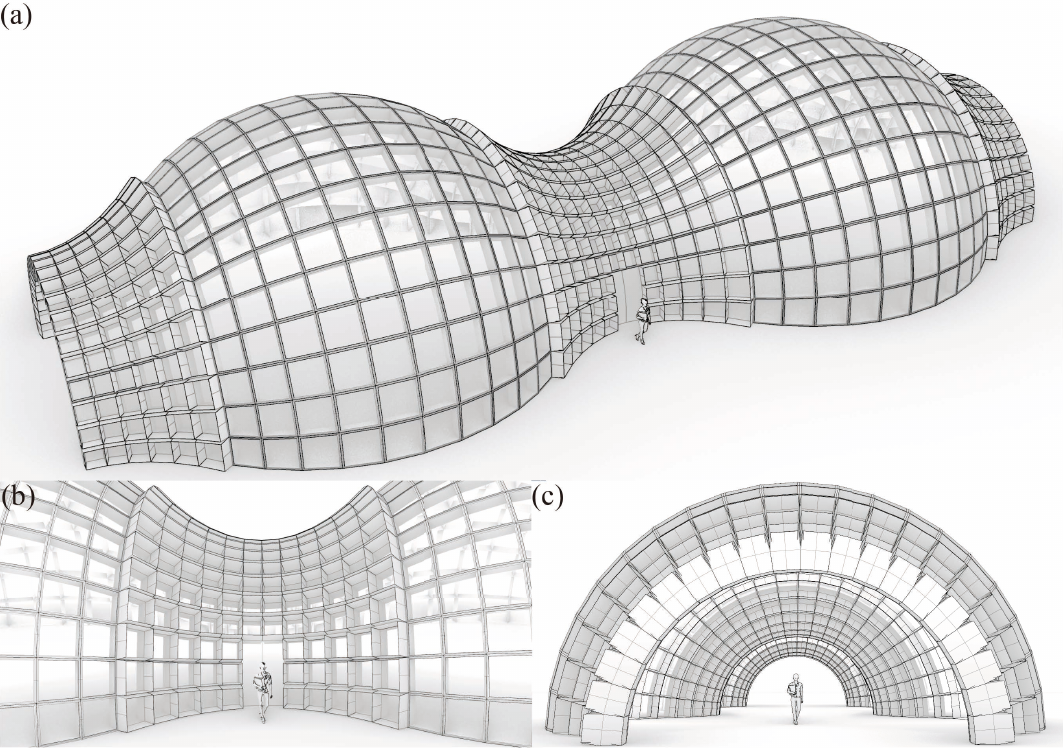}
    \caption{Concept image of the large-scale pavilion by T-toroid assembly from (a) perspective, (b) side, and (c) inner front view.}
    \label{fig:extendsurf}
\end{figure}

Mesh parallelism~\cite{pottmann2007geometry, mesnil2015isogonal}, also called the Combescure transformation~\cite{pirahmad2025area}, allows transforming curved surfaces by adjusting only the edge lengths while preserving edge directions and dihedral angles.
Figure~\ref{fig:meshparallelism} (a) shows the mesh parallelism transformations.
The middle dome represents the base Koebe polyhedron, and other dome configurations are derived from it through mesh parallelism transformations.
This property enables deformed meshes to be assembled using the same profile angles as the base Koebe polyhedron, as shown in Figure~\ref{fig:meshparallelism} (b) for T-toroids assemblies.
We implemented this transformation in \emph{Rhino/Grasshopper -- Kanagroo2} (Figure~\ref{fig:meshparallelism} (c)) by stretching or shrinking the Koebe meshes, while maintaining edge parallelism.

Figure~\ref{fig:extendsurf} (a) illustrates a large-scale pavilion concept assembled from T-toroid blocks.
The resulting curved surface exhibits both positive and negative Gaussian curvature and is biaxially and rotationally symmetric.
To construct this geometry, we continuously combine a Koebe polyhedral mesh with its Christoffel dual.
For the positive Gaussian regions, we set only $\theta_1$ with $\theta_2=180^\circ$, whereas for the negative Gaussian regions, we set $\theta_1=\theta_2$, causing the blocks to appear protruded outward.
Figures~\ref{fig:extendsurf} (b) and (c) show the assembly details from the side and interior views.

Although the curved surfaces in this study are limited to symmetric cases, we anticipate that more complex freeform surfaces could be achieved by incorporating optimization-based approaches, as explored in other edge-offset mesh-related works~\cite{pottmann2007geometry, pottmann2010edge, bobenko2017discrete, tovsiae2019design}.

\section{Application}
\label{Ttoroid:Application}

\subsection{Desktop-scale Prototypes}

\begin{figure}[htbp]
    \centering
    \includegraphics[keepaspectratio,width=\linewidth, page=6]{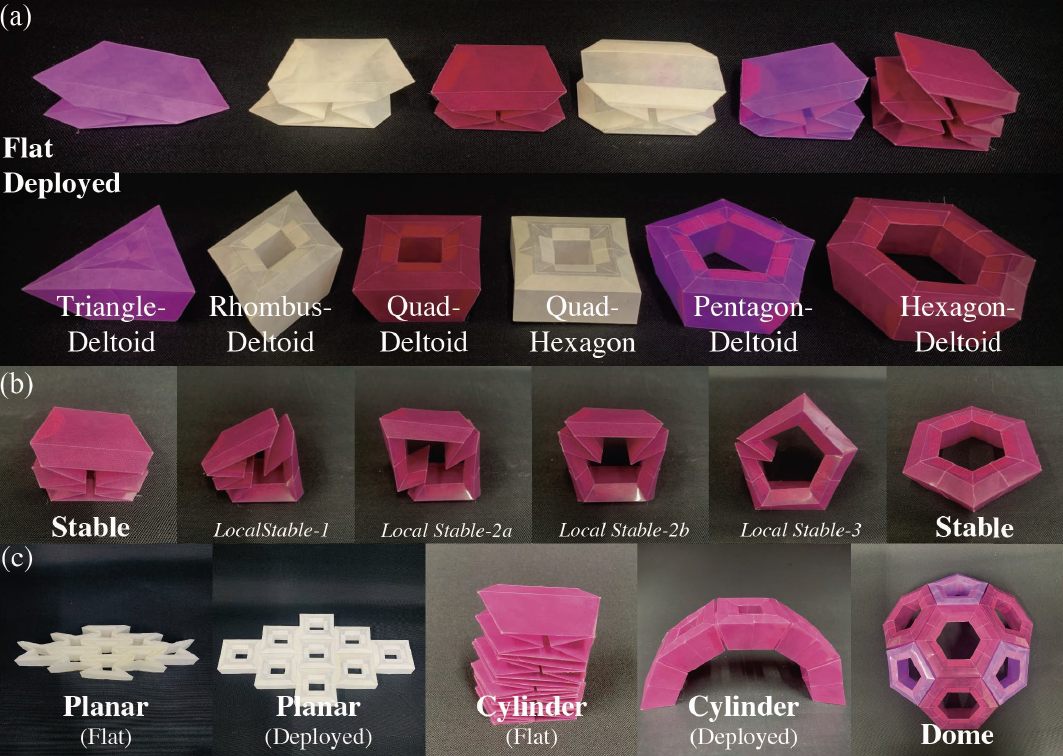}
    \caption{(a) Pictures of flat and deployed states of desktop-scale T-toroid prototypes. (c) Stable and local stable states sequence of the Hexagon trajectory T-toroid. (c) Assembled configurations. The planar and cylindrical surfaces can be globally flat-foldable.}
    \label{fig:deskfab}
\end{figure}

We fabricate the designed T-toroids using the fabrication method 1 in Section~\ref{subsubsec:deskfabmethod1} and validate their multistability through hand-interaction manipulation. 
Figure~\ref{fig:deskfab} (b) shows flat and fully deployed states of physical prototypes of the T-toroids in Figure~\ref{fig:designvariation}. 
We deployed the flat blocks by pulling each midpoint joint part outward and folded the deployed blocks by pushing each midpoint joint part inward (see Figure~\ref{fig:deskfab} (c)).
Through hands-on interaction tests, we experienced multistability in all prototypes.
In the actual behavior, between the two designed stable states of completely flat and deployed, we found multiple local stable states, where each midpoint joint is individually deployed, as shown in Figure~\ref{fig:polygonalassembly} (c).
In addition, prototypes exhibit asymmetric stiffness properties as mentioned in Section~\ref{subsec:TtoriSnappingAnalysis}.
With the same boundary condition as Figure~\ref{fig:Ttoroidfeasetting}, prototypes are relatively flexible in the flat-to-deployment actuation. 
In contrast, they are stiffer in the opposite direction (from deployment to flat). 
This indicates that some blocks are locked at the fully deployed state and can act as an ideal static building block.

Figure~\ref{fig:deskfab} (d) shows the physical prototypes of assembled curved surfaces.
We assembled a planar surface with a quad-hexagonal block in a stretcher-bonding pattern with a constant direction and a cylindrical surface with quad-deltoid blocks, which is the same geometry as the planar surface example of Figure~\ref{fig:quadassembly} (a), and the cylindrical surface example of Figure~\ref{fig:deskfab} (c), respectively.
These assembled structures are globally flat-foldable, meaning they can be folded flat in their assembled state.
The global structure deploys well in one direction by pulling the structure in both horizontal directions. 
However, folding from the deployed state to the flat state requires compressing each block separately.
Also, they are stable enough to sustain their deployed shapes.
About the dome surface prototype, although the overall structure lacks global flat-foldability, each block constrains the deformation of its neighboring blocks, enabling the system to behave as a robust static block.
Therefore, considering the results of the above analysis, we can expect better load-bearing performance of the T-toroid in this type of assembly case.
% It is also stiff in the prototype(Figure~\ref{fig:deskfab} (c) Dome) and was able to support vertical load $39.9$ times its weight (\SI{71.9}{g}).

\subsection{Large-scale Prototype}

\begin{figure}[htbp]
    \centering
    \includegraphics[keepaspectratio,width=\linewidth, page=12]{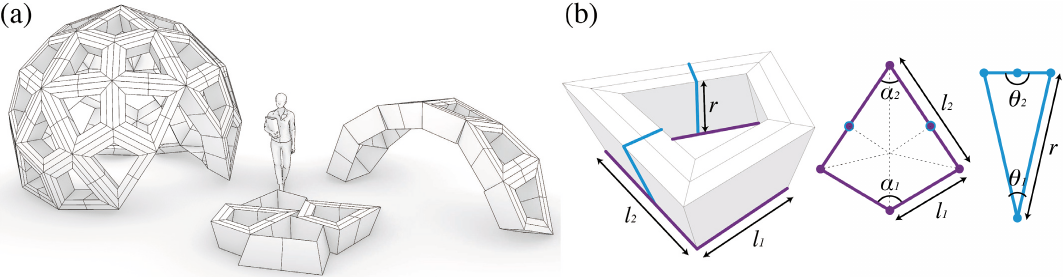}
    \caption{(a) Conceptual illustration of the T-toroid modular system. Using only a single type of deltoidal hexecontahedron-based T-toroid, planar, arch, and dome shell structures can be assembled. (b) Geometric parameters of the trajectory and profile of the deltoidal hexecontahedron-based T-toroid.}
    \label{fig:LargeTconcept}
\end{figure}

To show the feasibility of the T-toroid, we fabricated a full-scale prototype. 
Our concept is the \emph{ALL-IN-ONE} modular system that demonstrates the versatility of a single T-toroid in various assembly configurations. 
Figure~\ref{fig:LargeTconcept} (a) illustrates this concept. 
A single block can be used as furniture, such as a bench or stool; stacking two blocks allows transformation into a desk or standing table. 
By arranging multiple blocks in a planar configuration, the system can function as a bed, while assembly into arch or dome geometries enables use as temporary spatial structures. 
This multifunctionality highlights the potential of the modular system to flexibly adapt according to user needs.

Considering loading conditions in actual usage (e.g., as a bench or bed), we selected the T-toroid based on the deltoidal hexecontahedron as the prototype (Figure~\ref{fig:LargeTconcept} (b)). 
Among the possible alternatives for constructing a spherical polyhedron from a single polygon type, this geometry requires the largest number of faces, which in turn provides the smallest dihedral angle. 
In other words, it requires a deltoid profile with minimized $\theta_1$. 
This property can offer structural advantages under vertical loading in the deployed state compared to other alternatives with relatively wider profiles. 
Moreover, since only a single block needs to be designed and fabricated, the approach can also simplify the production process and improve productivity.

Figure~\ref{fig:LargeTconcept} (b) presents the illustration of the prototype along with the trajectory and profile polyline parameters. 
For the trajectory, the values were set as $l_1 =$ \SI{560}{mm}, $l_2 =$ \SI{862.03}{mm}, $\alpha_1 = 118.27^\circ$, and $\alpha_2 = 67.783^\circ$. 
The mid-joint (blue circled purple points) was defined as the foot of the perpendicular dropped from the intersection of the angle bisectors of each corner to the $l_2$ edges, ensuring that the profile can achieve inward flat-folding.
For the profile, we set $r =$ \SI{410.42}{mm}, $\theta_1 = 25.88^\circ$, and $\theta_2 = 180^\circ$, so that the deployed T-toroid reaches a height of \SI{400}{mm}, considering the bench usage.

\subsubsection{Preliminary Structural Analysis}

\begin{figure}[htbp]
    \centering
    \includegraphics[keepaspectratio,width=\linewidth, page=13]{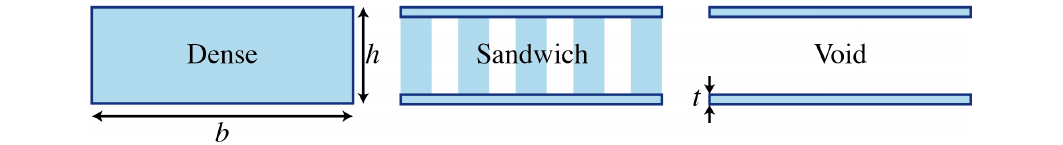}
    \caption{Mid-layer assumption of the sandwich panel.}
    \label{fig:sandwichcore}
\end{figure}

Before fabrication, we conducted a preliminary structural analysis (FEA) to verify whether the designed T-toroid exhibits multistability and to evaluate its feasibility as a bench. 
It should be noted that this analysis was intended as a simplified assessment, focusing on identifying snapping behavior and examining the structural response under loading conditions, to highlight key considerations for fabrication and practical use. 
Additionally, material properties were simplified and defined as a range. 
Other factors, such as hinge stiffness, plasticity, and contact effects, were not considered in this stage (same as Section~\ref{Ttoroid:mechanics}).

For the material setting, we used a \SI{4.9}{mm}-thick \emph{Plapearl (PDPPZ-080)}.
In this sandwich sheet, the core and face layers are made of the same material, and because the elastic modulus heavily depends on the density and alignment of the core, its exact value is difficult to determine. 
To estimate the range of material properties of this sandwich panel, we considered two extreme assumptions.
In the first assumption (Figure~\ref{fig:sandwichcore}, left), the panel is treated as a homogeneous material ($E_{dense}$), in which all polypropylene (PP) is assumed to be homogeneously distributed throughout the entire cross-section.
In the second assumption (Figure~\ref{fig:sandwichcore}, right), the core is treated as a void ($E_{void}$), such that all material is concentrated in the top and bottom face layers.
Ideally, the face-sheet thickness $t$ is assumed to be much smaller than the overall panel thickness $h$ $(t << h)$.
The actual material properties of the sandwich panel are expected to lie between the dense and void assumptions.

For the calculation, we use the average flexural modulus value $EI_{dense}$ (provided by \emph{Kwakami Sangyo}).
By setting the flexural stiffness ($EI_{dense}$) as the fixed given value for all assumption cases, we can get the $E_{void}$.

\begin{gather}
\label{eq:sandwich1}
    EI_{dense} = E\frac{bh^3}{12} = E_{void}I_{void} = E_{void}\frac{bth^2}{2}\\
    E_{void} = E\frac{h}{6t}
\end{gather}

Using this $E_{void}$, we can assume the in-plane stiffness of both panel assumption cases.

\begin{gather}
    EA_{dense} = Ebh\\
    E_{void}A_{void} = E_{void} \times 2bt = E\frac{h}{6t}\times 2bt = \frac{1}{3}Ebh = \frac{1}{3}EA_{dense}
\end{gather}

Therefore, under the void panel assumption, the panel can be considered to have the same flexural stiffness ($EI_{dense}$), but one-third of the in-plane stiffness ($EA_{dense}/3$) compared to the dense assumption.
To incorporate this void panel assumption into the FE analysis without modifying the modeling framework, we further transform it into an equivalent homogeneous panel characterized by an equivalent elastic modulus $E_{eq-void}$ and equivalent thickness $h_{eq-void}$.

\begin{gather}
    E_{void}I_{void} = EI_{dense} = E\frac{bh^3}{12}=E_{eq-void}\frac{b(h_{eq-void})^3}{12}\\
    E_{void}A_{void} = \frac{1}{3}EA_{dense}=\frac{1}{3}Ebh=E_{eq-void}bh_{eq-void}\\
    E_{eq-void}=\frac{E}{3\sqrt{3}}\\
    h_{eq-void} = \sqrt{3}h
    \label{eq:sandwich3}
\end{gather}

Based on equations~\ref{eq:sandwich1}–\ref{eq:sandwich3}, the material properties under the dense panel assumption are as follows: the elastic modulus ($E$) is \SI{273.25}{N/mm^2}, the density ($\rho$) is \SI{1.6d-10}{t/mm^3}, the Poisson ratio is $0.38$, and the thickness ($h$) is \SI{4.9}{mm}.
For the equivalent void panel assumption, the elastic modulus ($E_{eq-void}$) is \SI{52.587}{N/mm^2}, the density ($\rho/\sqrt{3}$) is \SI{9.238d-11}{t/mm^3}, the Poisson ratio is $0.38$, and the thickness ($h_{eq-void}$) is \SI{8.487}{mm}.

\begin{figure}[htbp]
    \centering
    \includegraphics[keepaspectratio,width=\linewidth, page=9]{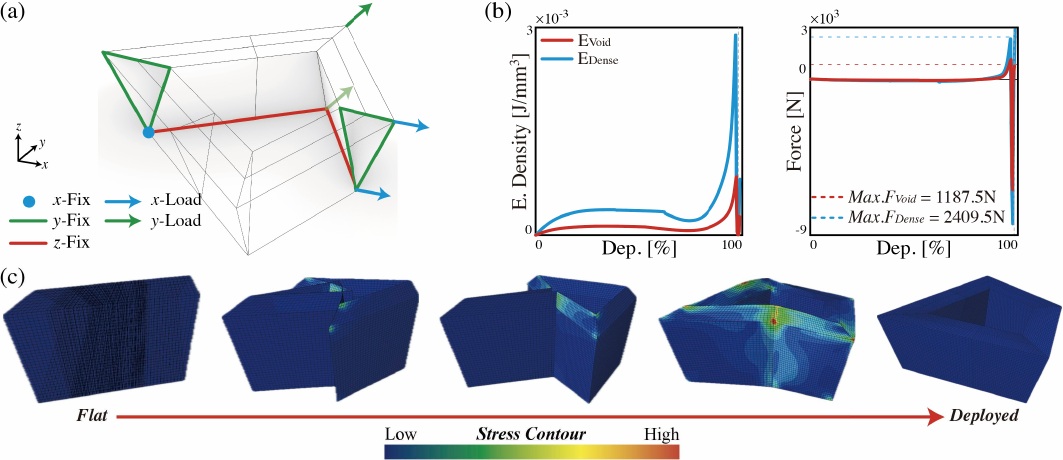}
    \caption{(a) Boundary conditions of snapping analysis of large-scale prototype. (b) shows the snapping analysis results, and (c) shows the deformation and stress contours during deployment.}
    \label{fig:preanalysis1}
\end{figure}

We first conducted a snapping analysis. 
Figure~\ref {fig:preanalysis1} (a) illustrates the boundary conditions applied for this analysis. 
Considering symmetry, only half of the block was modeled and arranged symmetrically with respect to the $xz$-plane. 
The bottom trajectory was constrained against displacement along the $z$-axis, while the cross-section intersecting the $xz$-plane was constrained against displacement along the $y$-axis, thereby realizing hinge behavior with zero rotational stiffness. 
Additionally, the left end of the bottom trajectory was fixed in the $x$-direction to ensure stability.
The analysis started from the flat state. 
To deploy the structure, forced displacements in the $x$ and $y$-directions were applied to the edge points located along the bottom and outermost trajectories.

Figure~\ref {fig:preanalysis1} (b) presents the snapping analysis results. 
As shown in the energy density plot on the left, the designed T-toroid exhibited bistability. 
During the initial stage of deployment, the structure deployed smoothly with relatively low stiffness, followed by a sharp increase in stiffness immediately before snapping. 
At this point, the maximum deployment force was estimated to be in the range of approximately $1200$--\SI{2400}{N}, which means 2--4 men are required for deployment under the given boundary conditions.
Although the influence of boundary constraints is considered significant, the required deployment force is still too large.
Therefore, it is necessary to reduce this deployment force in the actual design and fabrication stage.

Figure~\ref {fig:preanalysis1} (c) shows the stress contour on the panels during the deployment process. 
In the early stages, stress variations were relatively small. 
However, near the snapping point, a pronounced stress concentration was observed around the mid-joint region of the top panels. 
Considering practical use, a single block should ideally be easy enough for one adult to deploy without panel damage. 
Based on snapping analysis results, it is evident that design improvements are required to mitigate strain energy and stress concentration.

\begin{figure}[htbp]
    \centering
    \includegraphics[keepaspectratio,width=\linewidth, page=10]{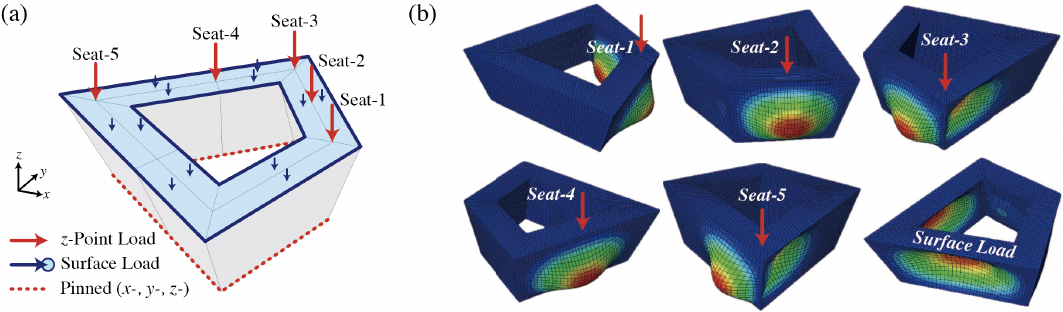}
    \caption{(a) Boundary conditions of linear buckling analysis of large-scale prototype. (b) shows the first linear buckling modes with different seat locations.}
    \label{fig:preanalysis2}
\end{figure}

\begin{table}[htbp]
\centering
\caption{Buckling loads on different material properties.}
\begin{tabular}{|l||*{10}{c|}}\hline
\label{tab:TtoroidLinBuckle}
\makebox[7em]{Load [$N$]} &\makebox[3em]{Seat-1}&\makebox[3em]{Seat-2}&\makebox[3em]{Seat-3}&\makebox[3em]{Seat-4}&\makebox[3em]{Seat-5}&\makebox[5em]{Surface Load}\\\hline
$F_{Void}$          & 1100.7 & 600.57 & 981.65 & 472.28 & 876.87 & 2371.1 \\
$F_{Dense}$         & 1181.7 & 643.85 & 1065.5 & 504.33 & 950.25 & 2528.6 \\\hline\hline
\makebox[7em]{Avg. Weight [$kg$]}  & 116.449 & 63.491 & 104.446 & 49.827 & 93.220 & 249.985 \\
\hline
\end{tabular}
\end{table}

Figure~\ref{fig:preanalysis2} (a) illustrates the boundary conditions and loading cases considered in the linear buckling analysis, conducted to examine whether the deployed T-toroid can sustain seating loads and to estimate the maximum seating capacity. 
In the simulations, the bottom trajectory was pinned, and concentrated or distributed loads were applied to the top panels. 
The \emph{Seat-1} to \emph{Seat-5} cases represent concentrated loads applied at the corners (\emph{Seat-1, 3, 5}), the mid-joint (\emph{Seat-4}), and the edge center (\emph{Seat-2}) of the mid-trajectory on the top panel. 
The surface load case applied a distributed load over the top panel.
Figure~\ref{fig:preanalysis2} (b) shows the results of the first buckling modes for each loading case, and the corresponding buckling loads are summarized in Table~\ref{tab:TtoroidLinBuckle}.

The results indicate that loads applied on the edge (\emph{Seat-2, 4}) produce lower buckling capacities compared to corner loads (\emph{Seat-1, 3, 5}). 
This is because, under corner loading, the force is distributed through multiple intersecting panels and edges, creating a more effective load-sharing mechanism. 
Consequently, \emph{Seat-4}, loads applied at intermediate edge locations to relatively wide panels, exhibited the lowest buckling load.
From the average weight values in Table~\ref{tab:TtoroidLinBuckle}, it can be inferred that buckling is possible to occur if an adult (approximately \SI{70}{kg}) sits at non-corner locations. 
Therefore, for practical use as a bench, it is recommended that users sit near the corners.
According to the surface load case results, it is suggested that the T-toroid can sufficiently support two adults sitting together or lying down. 
% However, simultaneous seating at \emph{Seat-3} and \emph{Seat-5} should be avoided, as buckling may occur.

% {\color{red}
\subsubsection{Fabrication}
\label{subsec:LargeTtoroidFabrication}

\begin{figure}[htbp]
    \centering
    \includegraphics[keepaspectratio,width=\linewidth, page=19]{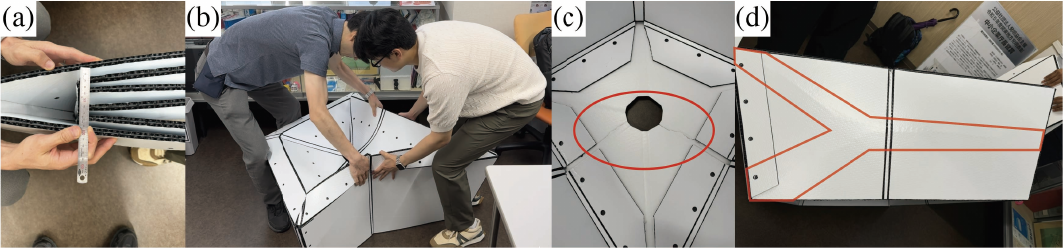}
    \caption{(a) Thickness accommodation design failure, (b) severe deformation of upper panel during folding, (c) fractures in upper panel, (d) local buckling in side panels.}
    \label{fig:Ttoroidlargefabfailure}
\end{figure}

In the first trial prototype, we encountered several failures from design, mechanical, and application standpoints.
From a design standpoint, the primary issue was insufficient accommodation of panel thickness.
As shown in the Figure~\ref{fig:Ttoroidlargefabfailure} (a), although a double-crease pattern was introduced, the flat-foldable geometry was not properly realized in fabrication.
Additional thickness interference arose from the tabs attaching the upper and lower panels, as well as from adhesive layers, which were not fully accounted for in the geometric design.
As a result, the folding pattern could not achieve the intended geometrically flat configuration.

From a mechanical standpoint, as confirmed in the preliminary snapping analysis (Figure~\ref{fig:preanalysis1} (b)), the first trial prototype exhibited excessively high strain energy, making it difficult to fold even by two people (Figure~\ref{fig:Ttoroidlargefabfailure} (b)).
In addition, the opening in the top panel was designed too narrowly, further concentrating deformation in this region.
As anticipated in snapping analysis (Figure~\ref{fig:preanalysis1} (c)), deformation was concentrated at the mid-joint region of the upper panel, which eventually led to fracture (Figure~\ref{fig:Ttoroidlargefabfailure} (c)).

From an application standpoint, as anticipated by the linear buckling analysis (Figure~\ref{fig:preanalysis2} (b), Table~\ref{tab:TtoroidLinBuckle}), seat loading near the mid-joint region (seat-4) caused local buckling failure on the side panels (Figure~\ref{fig:Ttoroidlargefabfailure} (d)).

\begin{figure}[htbp]
    \centering
    \includegraphics[keepaspectratio,width=\linewidth, page=11]{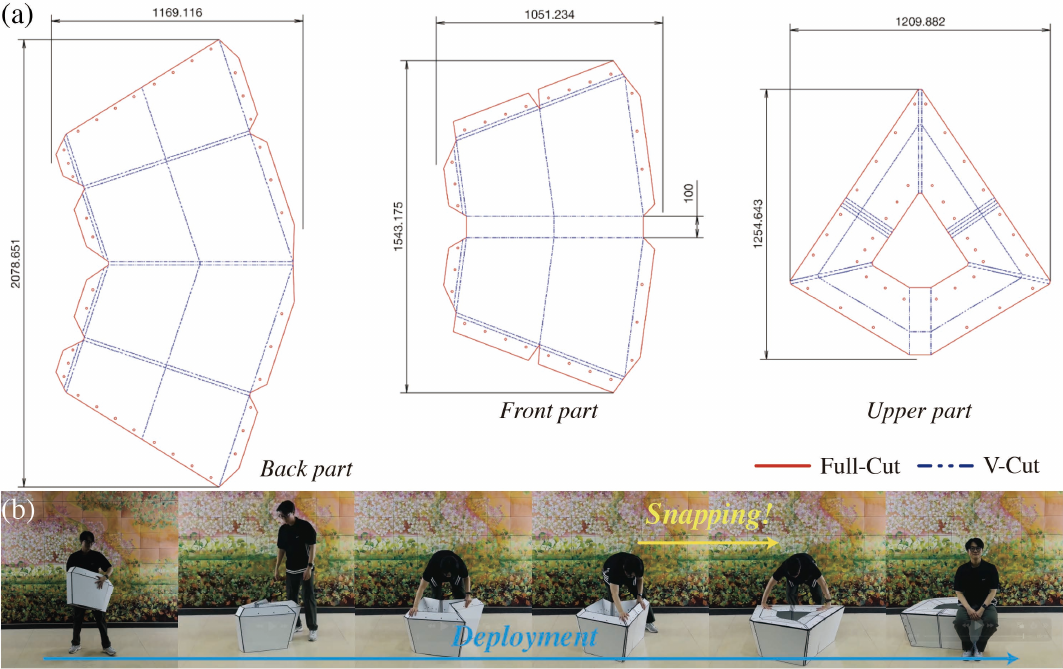}
    \caption{(a) Cutting patterns of the back, front, and upper panels. (b) Deployment motion of Large-scale T-toroid prototype.}
    \label{fig:Ttoroidlargefab}
\end{figure}

Based on these first trial prototype failures, the final cutting pattern (Figure~\ref{fig:Ttoroidlargefab} (a)) was refined, measuring thickness interferences (Figure~\ref{fig:Ttoroidlargefabfailure} (a)) that arose during the actual assembly process (e.g., interference at the tabs connecting the upper panels, thickness of tape).
We used \SI{4.9}{mm}-thick \emph{Plapearl (PDPPZ-080)}. 
The outer boundaries were processed with full cuts (red curves), while the creases were processed with V-cuts (blue dash-dotted curves). 
Due to the sheet size limitation, the pattern was divided into three parts for cutting. 
Circular red curves were inserted into the pattern for alignment during assembly. 
For the thickness accommodation, the double-crease technique was applied to the valley folds~\cite{hull2017double}.
Additionally, four V-cut creases were introduced in the mid-joint region of the upper panel part to create \emph{compliant zones}, thereby accommodating thickness, reducing and dispersing the strain energy and stress concentrations through the play and plasticity.
Each part was then assembled using high-strength double-sided tape, resulting in the completion of the final prototype.

Figure~\ref{fig:Ttoroidlargefab} (b) shows the deployment motion of the completed prototype. 
Deployment was performed sequentially by unfolding each mid-joint, and snap-through occurred when the second mid-joint was deployed.
Deployment was easy enough for a single person, and the T-toroid stands stable in the fully deployed state.
Due to the stiffness in creases and compliant zones, the block exhibited a slight tendency to open in the folded state; however, overall, it folded flat, making storage convenient. 
In the deployed state, the block can support up to two people seated on each corner. 
Also, the block supported stably when a panel was placed on top, and a person lay down on it.
The weight of a single block was approximately \SI{3.286}{kg}, light enough to be carried by a person. 
Further detailed prototyping records are included in Appendix~\ref{chap:AppendixB}.
% }

\section{Conclusion}

\begin{figure}[htbp]
    \centering
    \includegraphics[keepaspectratio,width=\linewidth, page=15]{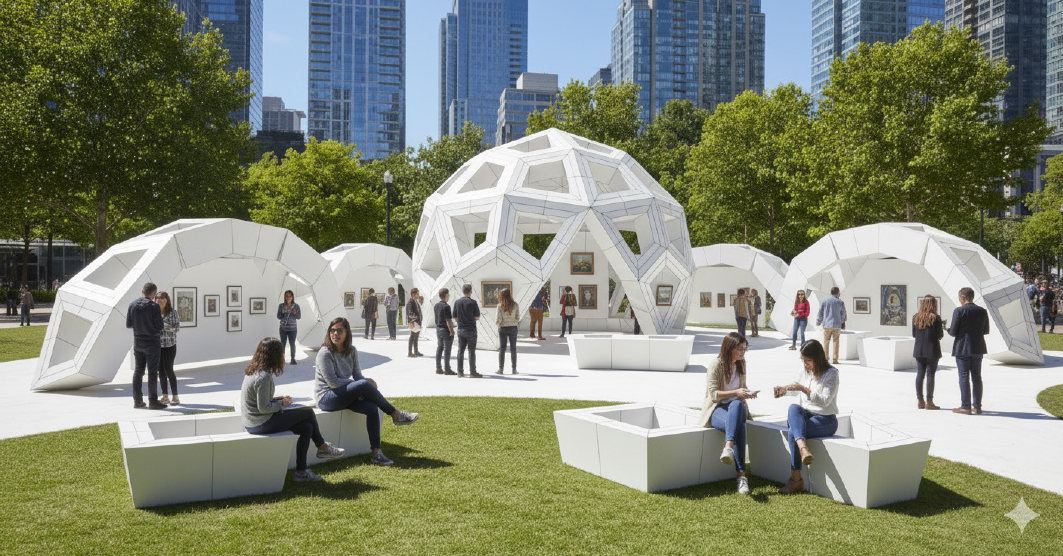}
    \caption{AI-generated illustration of T-toroid pavilion. Made by \emph{Gemini 2.5 Pro}.}
    \label{fig:AI_Ttoroid}
\end{figure}

We presented a design methodology for multistable origami blocks (\emph{T-toroid}) by applying the loop-closing method to rigid origami tubes (T-tubes) with various polygonal trajectories and profiles.
The designed T-toroids were analyzed through finite element analysis to confirm their multistability and the influence of design parameters, and their multistability was validated through the hand-interaction test of desktop-scale prototypes.

We designed T-toroids with various polygonal trajectories and profiles and assembled them into standard curved surfaces, such as planar, cylindrical, spherical, and saddle shapes. 
By leveraging the edge-offset property, we demonstrated that the T-toroid assembly system can be extended to more complex curved surfaces based on planar quad-meshed geometries.
We also verified assembled geometries using desktop-scale prototypes, and we found that the overall structure can achieve global flat-foldability depending on the assembly strategy.

Furthermore, a large-scale T-toroid prototype was analyzed and fabricated to validate multistability and applicability for practical usage. 
By applying these blocks both individually and in assemblies, we proposed the feasibility of an all-in-one modular system adaptable across multiple scales, serving as benches, beds, and pavilions as shown in Figure~\ref{fig:AI_Ttoroid}\footnote{We provided the Figure~\ref{fig:LargeTconcept} to Gemini 2.5 Pro, and typed the following prompt: Make this illustration into pavilions for exhibition in the city park.}.

\subsection{Discussion}
This chapter is constructed following four primary objectives.
The contributions and limitations corresponding to each objective are as follows.

\subsubsection{Objective 1: Exploring multistable origami design methodologies}

\paragraph{Contributions}
We extended the loop-closing approach beyond its previous use in Miura-based star-shaped bistable origami~\cite{kamrava2019origami} by applying it to rigid-origami tubes defined by a wider variety of trajectory and profile polygons.
This extension enabled the design of stable configurations on toroidal topologies while preserving flat-foldability, demonstrating that T-toroids can serve as versatile building blocks for constructing complex global shapes.

\paragraph{Limitations}
The current T-toroids are restricted to rotationally symmetric trajectories and mirror-symmetric profiles.
For broader applicability, extending the framework to non-symmetric trajectories and profiles will be necessary.
Establishing design guidelines for such generalized T-toroids would allow a wider range of global geometries to be realized and improve the flexibility of the assembly system.

\subsubsection{Objective 2: Analyzing mechanics across the design parameters}

\paragraph{Contributions}
Through snapping analyses, we identified how the geometric design parameters affect the mechanical response, providing practical guidance for controlling the energy barrier and stability characteristics.
We also performed preliminary analyses to predict the behavior of a large-scale prototype, resulting in local buckling near seating regions and stress concentrations during deployment.
These results were reflected in the second design and contributed to the successful fabrication of the large-scale prototype, demonstrating the utility of the analysis framework.

\paragraph{Limitations}
The mechanical behavior trends could not be fully understood because the snapping analysis produced severe local deformations, such as crumpling and self-intersection of panels, which introduced unexpected peaks and irregularities in the equilibrium paths.
Since these issues arise partly from the idealized modeling assumptions, it remains unclear whether such concentrated deformations will occur, or how they may be relieved, in physical prototypes.
Clarifying these severe behaviors will require further experiments, and it will be included in future work.

The load-bearing results (Figure~\ref{fig:results2}) remain insufficient for a comprehensive understanding of structural performance, as only a limited case was tested and the presence of unstable modes obscured the overall behavior.
Nevertheless, based on geometric considerations, we expect the deployed configuration to exhibit high load-bearing capacity.
A more comprehensive evaluation of load-bearing-related structural performance across parameters will be addressed in future work.

As shown in Figure~\ref{fig:deskfab} (c), if the geometry becomes more complex or asymmetric, multiple local stable states emerge, and branching in the deployment sequence can occur, making the actual behavior difficult to predict.
However, if properly exploited, these local stabilities may allow control over the strain energy characteristics through local deployment orders.
The systematic analysis of such local stability remains an open problem.

\subsubsection{Objective 3: Programming global systems by assembling blocks}

\paragraph{Contributions}
We clarified that the assemblability of T-toroids originates from the edge-offset relationship between the top and bottom panels, and demonstrated through this framework that diverse curved surfaces can be constructed via mesh-guided assembly using T-toroids.
We also showcased that some assembled systems can achieve global flat-foldability depending on the chosen assembly method, suggesting positive prospects for future deployable architectural applications.

\paragraph{Limitations}
The assembly demonstrations in this study were limited to simple symmetric cases.
Whether more complex curved surfaces, even freeform shapes, can be achieved remains unresolved, and additional research on mesh-guided (edge-offset) assembly optimization will be required.
However, we have partially addressed this challenge in one of our related publications~\cite{snappingtoroids-2025}, where edge-offset meshes were optimized to target surfaces and demonstrated surface construction by assembling multistable origami blocks along the mesh edges.

Our understanding of the mechanical behavior of the assembled configurations also remains limited.
Key questions such as stability under self-weight, performance under practical loading, and ultimate load-bearing capacity were beyond the scope of this study.
These issues, together with the extension to optimized complex surface assemblies, will be addressed in future work.

\subsubsection{Objective 4: Demonstrating the applicability via desktop and large-scale fabrication}

\paragraph{Contributions}
The multistability of the T-toroid was successfully validated through both desktop-scale and large-scale prototypes.
Also, the assemblability in practice was demonstrated through the desktop prototypes.
During large-scale fabrication, we identified practical issues, such as thickness accommodation and stress concentration, and then we introduced experimental structural solutions, including double V-cuts and compliant zones, to mitigate these issues without significantly altering the intended geometric or mechanical behavior.
These solutions provide useful references for future large-scale fabrication of multistable blocks.

\paragraph{Limitations}
A more systematic design methodology is required for complex geometries such as the T-toroid.
Specifically, numerical guidelines for designing thickness accommodation and fitted compliant zones for stress relief are still lacking, as this study relied primarily on experiential adjustments.
Furthermore, only a single large-scale T-toroid was fabricated; assembling multiple blocks is likely to introduce additional challenges, including block connection, cumulative geometric tolerance, and load-bearing under self-weight.
These issues will require further investigation in future work.

\subsection{Future Work}
Future work will focus on generalizing the T-toroid assembly framework and realizing large-scale construction. 
Although our recent work on snapping toroids~\cite{snappingtoroids-2025}, including the T-toroid geometry, partially addresses remaining challenges, such as generalized block design and optimization-based mesh-guided assembly, the assembly with only the T-toroid remains unsolved. 
Further experimental studies are also required to compare the actual behaviors of severe local behaviors observed in the simulations, such as crumpling and self-intersection, and to understand the influence of design parameters on load-bearing performance. 
In addition, large-scale structural realization will require numerical design strategies for thickness accommodation, compliant-zone design, and block connections. 
Finally, we plan to construct a full-scale T-toroid spatial structure, including verification of its load-bearing capacity under assembled systems, which can be integrated with air-pressure-based actuation for autonomous deployment.

	\chapter{CC-block: Multistable Curved-Crease Origami Block}
\label{chap:ccblock}

\emph{\textbf{Author's note:}
The content covered in this chapter is partially included in publications~\cite{lee2024lightweight, lee2025multistable}.}

\section{Introduction}

Large-scale applications of curved-crease origami are particularly prominent in cylindrical topologies, with proposed applications including column design~\cite{lalvani2020form, koschitz2016designing}, deployable space reflectors~\cite{soykasap2004new}, beams~\cite{lee2016folded}, bridges~\cite{rihaczek2022timbr, maleczek2024fold}, self-locking coiled tubes~\cite{lee2024self}, and curved wall modules~\cite{lee2024lightweight}.
Since these curved-crease origami inherently store bending stress in their panels, they require additional supporting equipment to maintain a deployed shape.
This trade-off between shape stability and deployability of curved-crease origami remains a key challenge for large-scale applications.

To overcome this challenge, we present the design and analysis of multistable curved-crease origami blocks, termed \emph{CC-Blocks}, and introduce a modular building system capable of generating reconfigurable, quasi-continuous wall surfaces with a self-supporting feature.
Specific study objectives are to: 
(1) establish the CC-Block design method via a loop-closing method applied to cylindrical curved-crease origami with a desired polygonal trajectory and elastica curvature profile, 
(2) analyze mechanical behaviors across various geometric parameters using finite element analysis, and validate the multistability through the desktop-scale prototypes, 
(3) assemble the blocks through interlocking curved surfaces and integrate block and assembly design processes using a shape grammar–based system
and (4) demonstrate practical application possibilities through the fabrication of large-scale prototypes.
These objectives are addressed in Sections~\ref{sec:geodesign},~\ref{sec:mechanics},~\ref{sec:ccassembly} and ~\ref{sec:ccfabrication}, respectively.

\section{Geometry Design}
\label{sec:geodesign}

\begin{figure}[htbp]
    \centering
    \includegraphics[keepaspectratio,width=\linewidth, page=1]{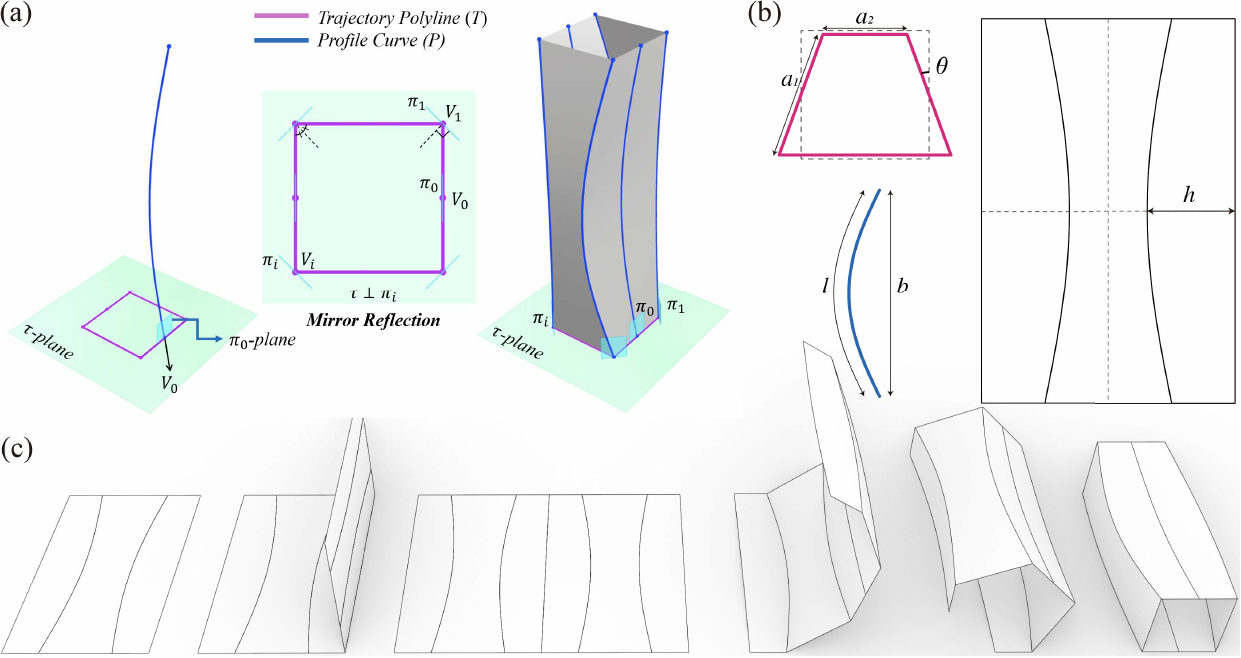}
    \caption{(a) Design process, (b) design parameters, and (c) deployment motion of curved-crease origami.}
    \label{fig:design}
\end{figure}

In this section, we present the design of multistable curved-crease origami blocks (CC-blocks).
We generate the cylindrical curved-crease origami by mirror-reflecting the profile curve along the polygonal trajectory.
To introduce multistability, we used a \emph{loop-closing} method that connects both ends of the cylindrical form.
The resulting configuration shows stability in both flat-folded and deployed states.

\subsection{Curved-crease Origami}
\label{subsec:CCoridesign}
Following the design method in~\cite{lee2024lightweight, mitani2012column, lee2018elastica}, we construct curved-crease origami with cylindrical developable surfaces using a \emph{trajectory polyline} and a \emph{profile curve}, as illustrated in Figure~\ref{fig:design} (a).
Here, we define each vertex of the trajectory polyline as $V_i$ $(i\in[0,n])$.
Reflection planes from $\pi_0$ to $\pi_i$ are generated at polyline vertices with normal vectors of planes coincident with the bisector of the corner angle (Figure~\ref{fig:design} (a)).

The profile curve is modeled as an elastica curve, which describes the naturally stable post-buckled shape of a slender elastic rod. 
Such elastica profiles have been shown to accurately represent the geometry of elastically bent curved-crease origami with an enforced, fixed-length boundary condition~\cite{lee2018elastica}. 
In our case, the profile is defined as a pinned–pinned elastica segment, characterized by an arc length $l$ and a deformed height $b$, corresponding to the first mode configuration (Figure~\ref{fig:design} (b)).
In this paper, we define the profile curvature ($c$) as the ratio $c = b/l$.
To construct the curved-crease origami block, this profile is placed on the initial reflection plane $V_0$, and then mirrored successively across the adjacent planes along the trajectory, as illustrated in Figure~\ref{fig:design} (a)-right.
The motion of the designed curved-crease origami is illustrated in Figure~\ref{fig:design} (c).

\subsection{Loop-closing}
\label{subsec:loopclosing}

\begin{figure}[ht]
    \centering
    \includegraphics[keepaspectratio,width=\linewidth, page=2]{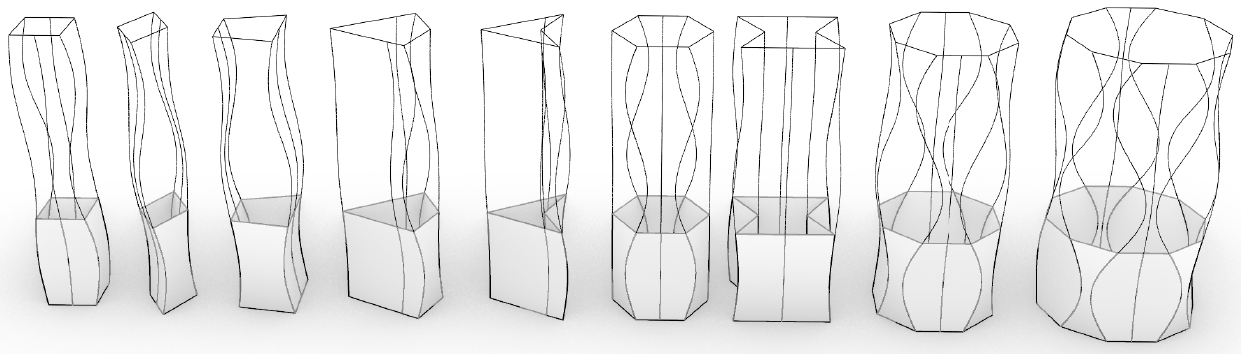}
    \caption{Design variations of CC-blocks with various trajectory polylines and $k$-mode elastica profile curves.}
    \label{fig:variation}
\end{figure}

To achieve multistability in the designed cylindrical curved-crease origami, we adopt the \emph{loop-closing} method in Section~\ref{method:loop}.
While the method has primarily been used for rigid origami, we extend this method to curved-crease origami.
To get the closure cylindrical shape at the deployed and flat states:
(1) The start vertex ($V_0$) and the end vertex ($V_n$) of the trajectory polyline should be matched to complete the loop.
(2) The trajectory vertices should be an even number to allow connectivity and continuity of the curved surface.
(3) At least two vertices should lie on straight edges rather than being corners, to create linear creases. One of the edge-on vertices is set as $V_0 (=V_n)$.
Finally, (4) this trajectory must be flat-foldable by folding these edge-on vertices in the outward direction.
By following the design process (Figure~\ref{fig:design} (a)) while satisfying the above trajectory conditions, multistability can be achieved.

Figure~\ref{fig:loadbearing} (c) shows the deployment motion of the CC-block. 
While the motion deviates from the original curved-folding motion (Figure~\ref{fig:design} (c)), it still reaches both flat and deployed states. 
Compared to the original curved-folding, an increased panel distortion generates a distinct energy barrier between the two configurations.

Based on this design method, we can design flat-foldable and multistable CC-blocks with various trajectories and $k$-mode elastica profiles; such design variation is illustrated in Figure~\ref{fig:variation}.
Although various design configurations can be considered, in this paper, we limit our scope to first-mode elastica profiles to simplify the mechanical analysis and to trapezoidal trajectories to keep consistency with the assembly method employed in our previous work~\cite{lee2024lightweight}.

The column with the first-mode elastica profile has alternating convex and concave sides.
In our design, we chose to put the linear creases on the concave side, similarly to~\cite{maleczek2024fold}.
This is because, through physical models, we observed no snapping when the creases were added to the convex side~\cite{soykasap2004new}.

\section{Mechanics}
\label{sec:mechanics}

\begin{wrapfigure}{l}{0.5\textwidth}
    \centering
    \includegraphics[keepaspectratio,width=\linewidth, page=3]{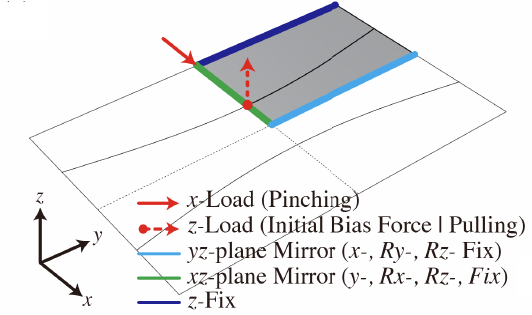}
    \caption{Applied boundary conditions for snapping analysis.}
    \label{fig:feasetting}
\end{wrapfigure}

Here, we analyze the mechanical behaviors of CC-blocks across various design parameters and loading conditions through the finite element method, using \emph{Abaqus 2024 (Dassault Systèmes)}~\cite{abaqus}.
We consider a flat-folded tube as shown in Figure~\ref{fig:feasetting}, where it lies on the $xy$-plane.
When deploying, we have two options to apply compression to the $x$ direction (pinching) or tension to the $z$ direction (pulling).
In Section~\ref{subsec:snappinganalysis}, we analyze the snapping behaviors based on the pinching deployment and elaborate on the effect of different parameters on the snapping behavior.
Additionally, we discuss the differences in snapping behavior when objects are pinched or pulled.
In Section~\ref{subsec:loadbearinganalysis}, we analyze the load-bearing behavior of the deployed block under compression in $y$ directions, the typical governing load condition when used as a column or wall element.

\subsection{Modeling and Boundary Conditions}
\label{subsec:feamodeling}

The modeling method for FEA follows Section~\ref{subsec:FEA}.
Figure~\ref{fig:feasetting} shows the boundary conditions applied to the model.
We used a mirror-symmetric condition to simplify the model design and reduce computational costs. 
The analysis of Figure~\ref{fig:results} (b, d, e, f, i), which considers triaxially-symmetric CC blocks with a rectangular trajectory, $1/8$ of the full block was modeled. For the analysis of Figure~\ref{fig:results} (c, g, h), which considers a biaxially-symmetric CC block with a trapezoidal trajectory, $1/4$ of the full block was modeled.
The analysis models are initialized in the flat state and constrained with mirror symmetry. 
The vertical mid-section along the $yz$-plane is constrained with an $x$-translation fix and restrictions on $R_y$ and $R_z$ rotations. The horizontal mid-section along the $xz$-plane is constrained with a $y$-translation fix and restrictions on $R_x$ and $R_z$ rotations.
The outermost edges along the $y$-axis, corresponding to the midpoint joints of the trajectory, are fixed only in the $z$-direction, as this allows ideal hinge behavior without introducing rotational stiffness.

For the snapping analysis, forced displacements of $h$ value were applied along the $x$-direction at the center point of the concave panel.
For breaking the initial flat configuration, we applied a \SI{1}{N} initial bias load in the $z$-direction for $0-0.1s$, after which it was released.
For the material setting, we used polypropylene material properties with an elastic modulus of \SI{1134}{MPa}, density of \SI{9.05e-10}{ton/mm^3}, and the Poisson's ratio of 0.38~\cite{zhang2025effect}.
The snapping analysis is performed using a quasi-static method in the dynamic implicit solver.

\subsection{Snapping Analysis}
\label{subsec:snappinganalysis}

\begin{figure}[ht]
    \centering
    \includegraphics[keepaspectratio,width=\linewidth, page=4]{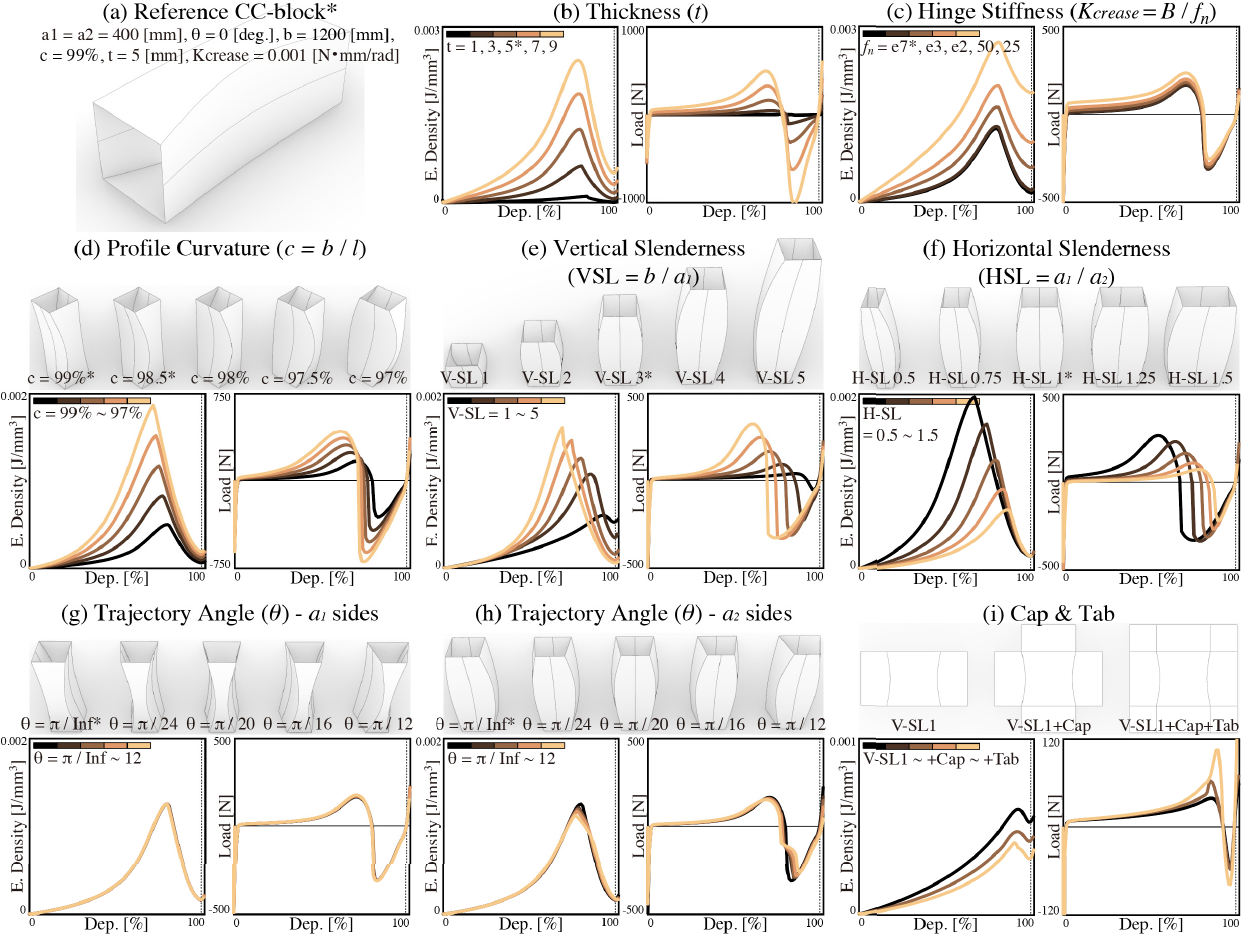}
    \caption{(a) Reference CC-block and design parameter values. Analysis results of (b) thickness parameters, (c) hinge stiffness parameters, (d) vertical slenderness parameters, (e) elastica profile curvature parameters, (f) horizontal slenderness parameters, (g) trajectory angle parameters, which include the additional creases located in the $a_1$-side panels, or (h) $a_2$-side panels. (i) Behavior differences with and without the cap and tab. }
    \label{fig:results}
\end{figure}

\begin{table}[ht]
\centering
\caption{Peak locations of \emph{Figure~\ref{fig:results}}.}
\begin{tabular}{|l||*{10}{c|}}\hline
\label{tab:peaks}
{\begin{footnotesize}\emph{Peak Dep.[\%]}\end{footnotesize}}
&\makebox[3em]{(b)}&\makebox[3em]{(c)}&\makebox[3em]{(d)}&\makebox[3em]{(e)}&\makebox[3em]{(f)}&\makebox[3em]{(g)}&\makebox[3em]{(h)}&\makebox[3em]{(i)}\\\hline
$C_{Black}$ & 84.204  & 80.046 & 80.046 & 92.665 & 67.540 & 80.046 & 80.046 & 92.665 \\
$C_{Dark Brown}$ & 81.085 & 79.947 & 77.162 & 85.079 & 74.737 & 79.813 & 78.608 & -- \\
$C_{Brown}$ & 80.046 & 79.947 & 75.035 & 80.046 & 80.046 & 79.293 & 78.658 & 92.665\\
$C_{Light Brown}$ & 79.006 & 80.972 & 73.291 & 74.514 & 83.607 & 79.859 & 77.599  & -- \\
$C_{Yellow}$& 79.006 & 80.972 & 71.735 & 69.062  & 86.559 & 80.024 & 76.657 & 90.606  \\
\hline
\end{tabular}
% \end{table}
\vspace{0.5cm}
% \begin{table}[ht]
\centering
\caption{Energy ratio ($E_{peak}/E_{stable}$) of \emph{Figure~\ref{fig:results}}.}
\begin{tabular}{|l||*{10}{c|}}\hline
\label{tab:energyratio}
{\begin{footnotesize}\textbf{$E_{peak}/E_{stable}$}\end{footnotesize}}
&\makebox[3em]{(b)}&\makebox[3em]{(c)}&\makebox[3em]{(d)}&\makebox[3em]{(e)}&\makebox[3em]{(f)}&\makebox[3em]{(g)}&\makebox[3em]{(h)}&\makebox[3em]{(i)}\\\hline
$C_{Black}$ & 14.934 & 7.821 & 7.821 & 1.142 & 12.344 & 7.821 & 7.821 & 1.142 \\
$C_{Dark Brown}$ & 10.414 & 6.289 & 8.586 & 3.073 & 10.309 & 7.748 & 7.515 & -- \\
$C_{Brown}$ & 7.821  & 2.738 & 9.063 & 7.821 & 7.821  & 7.745 & 7.427 & 1.133 \\
$C_{Light Brown}$ & 6.051  & 1.950 & 9.421 & 15.904 & 5.736  & 7.735 & 7.345 & -- \\
$C_{Yellow}$ & 4.853  & 1.463 & 9.620 & 27.139 & 4.321  & 7.649 & 6.879 & 1.284 \\
\hline
\end{tabular}
% \end{table}
\vspace{0.5cm}
% \begin{table}
\centering
\caption{Force ratio between deployment and folding ($|F_{min}/F_{max}|$) of \emph{Figure~\ref{fig:results}}.}
\begin{tabular}{|l||*{10}{c|}}\hline
\label{tab:forceratio}
{\begin{footnotesize}\textbf{$|F_{min}/F_{max}|$}\end{footnotesize}}
&\makebox[3em]{(b)}&\makebox[3em]{(c)}&\makebox[3em]{(d)}&\makebox[3em]{(e)}&\makebox[3em]{(f)}&\makebox[3em]{(g)}&\makebox[3em]{(h)}&\makebox[3em]{(i)}\\\hline
$C_{Black}$ & 2.723 & 1.886 & 1.886 & 1.240 & 1.245 & 1.886 & 1.886 & 1.240 \\
$C_{Dark Brown}$ & 2.252 & 1.858 & 1.777 & 2.442 & 1.441 & 1.863 & 1.843 & -- \\
$C_{Brown}$ & 1.886 & 1.635 & 1.719 & 1.886 & 1.886 & 1.853 & 1.830 & 0.862 \\
$C_{Light Brown}$ & 1.912 & 1.428 & 1.675 & 1.338 & 2.440 & 1.832 & 1.802 & -- \\
$C_{Yellow}$ & 2.022 & 1.093 & 1.633 & 1.001 & 2.981 & 1.793 & 1.753 & 0.738  \\
\hline
\end{tabular}
\end{table}

Figure~\ref{fig:results} (b--i) show the FEA results of the pinching loading behavior depending on design parameters.
The horizontal axis (deployment) of plots is defined as the forced-displacement ($d$) divided by the insertion depth ($h$) of the concave panel in deployed states, $d/h$.
The left plots show the energy density, dividing total strain energy by the volume of the geometry, and the right plots show the required force during deployment.
By comparing the energy density, we can exclude the effect of panel size difference along the design parameters, providing a more uniform comparison across different CC-block geometries.
Note that energy density plots in Figure~\ref{fig:results} have a small initial peak, which occurs due to the initial bias $z$-load, and because of this, the negative force is captured in the initial step of the force plots.
We set the reference model for the comparison, which is shown with design parameter values in Figure~\ref{fig:results} (a).
The asterisk $\ast$ marked in figures or legends represents this reference model.
Further result details can be found in Table~\ref{tab:peaks}--~\ref{tab:forceratio}.
Table~\ref{tab:peaks} presents the deployment locations of energy density peaks.
Table~\ref{tab:energyratio} shows the energy ratio between the peak energy and the second stable state energy value ($E_{peak}/E_{stable}$), indicating how stable a model is at the deployed state.
Table~\ref{tab:forceratio} shows the absolute ratio between maximum and minimum force values ($|F_{min}/F_{max}|$), representing the required load difference for deploying and folding the CC-block.

Figure~\ref{fig:results} (b) shows the influence of panel thickness ($t$). 
As thickness increases, the peak appears slightly earlier (Table~\ref{tab:peaks} (b)) and becomes more rounded, which can weaken the snapping behavior.
Although both the energy barrier and stable state energy rise, the stability in the deployed state is reduced (Table~\ref{tab:energyratio} (b)).
This is due to the thicker panel requiring more energy at the same curvature.
Therefore, the thicker panels demand greater force during deployment and folding.
While folding generally demands greater force than deployment, no consistent trend is observed with varying thickness (Table~\ref{tab:forceratio} (b)).

Figure~\ref{fig:results} (c) shows the influence of hinge stiffness.
The hinge stiffness was defined in equation~\ref{eq:Khinge}.
Here we set the length scale factor ($f = L_{crease} \times f_n$), and used the normalized scale factor ($f_n$) as the parameter. 
As stiffness increases, although both the energy barrier and stable state energy rise, the stability in the deployed state is reduced (Table~\ref{tab:energyratio} (c)).
Energy peaks are almost unaffected by hinge stiffness (Table~\ref{tab:peaks} (c)).
As the hinge stiffness increases, the difference in the required force between deployment and folding decreases (Table~\ref{tab:forceratio} (c)).
Therefore, excessive stiffness may eliminate multistability, leading to a monotonic increase in energy.

Figure~\ref{fig:results} (d) shows the influence of profile curvature ($c = b / l$) with a fixed height $b=1200$.
A higher $c$ raises the energy barrier and shifts the load peak earlier (Table~\ref{tab:peaks} (d)), while the stable state energy remains similar, indicating improved stability (Table~\ref{tab:energyratio} (d)).
While the larger curvature increases the required forces, the difference in the required force between deployment and folding decreases (Table~\ref{tab:forceratio} (d)).

Figure~\ref{fig:results} (e) shows the influence of the height-to-edge length ratio ($b/a_1$) while maintaining a constant $c$ of $99\%$.
When the vertical slenderness (VSL) increases, there is a higher energy barrier with a sharper and earlier peak during deployment (Table~\ref{tab:peaks} (e)).
The energy density at the stable state decreases, enhancing stability in the deployed state (Table~\ref{tab:energyratio} (e)).
Absolute force values become greater, while the force difference between deployment and folding becomes smaller (Table~\ref{tab:forceratio} (e)).

Figure~\ref{fig:results} (f) shows the influence of the trajectory edge length ratio ($a_1/a_2$) while maintaining a $c$ of $99\%$.
A smaller horizontal slenderness (HSL) leads to a higher energy barrier with earlier(Table~\ref{tab:peaks} (f)) and sharper peaks, indicating clear snapping behavior.
Since the energy density at the stable state is the same, the stability in the deployed state is improved (Table~\ref{tab:energyratio} (f)).
Additionally, while the required force increases for both deployment and folding, the force difference between the two becomes smaller (Table~\ref{tab:forceratio} (f)).

Figures~\ref{fig:results} (g, h) show the influence of trajectory angle ($\theta$).
In (h), where the linear crease is on the $a_1$ side, $\theta$ has almost no effect.
In (g), with the crease on the $a_2$ side, a geometric difference between front and back panels is observed. 
So, plots show the averaged energy density and load.
As $\theta$ increases, the energy peak very slightly decreases (Table~\ref{tab:peaks} (g)).
In general, a larger $\theta$ slightly reduces stability (Table~\ref{tab:energyratio} (g)) at the deployed state and the force ratio between deployment and folding (Table~\ref{tab:forceratio} (g)).
The slight kink in the force curve reflects differences in snap-through timing and energy barriers between the front and back panels, indicating multistability in this geometry.

Figure~\ref{fig:results} (i) compares the behavior of the least stable case (VSL1 in Figure~\ref{fig:results} (e)) with and without cap and tab additions.
When only the cap is added, the peak location remains unchanged (Table~\ref{tab:peaks} (i)), and stability in the deployed state is slightly reduced (Table~\ref{tab:energyratio} (i)).
Adding both the cap and the tab shifts the peak to an earlier stage (Table~\ref{tab:peaks} (i)) and further improves stability in the deployed state (Table~\ref{tab:energyratio} (i)).
In both cases, the required deployment force increases, while the folding force remains nearly the same.
As a result, folding becomes easier than deployment in both cases (Table~\ref{tab:forceratio} (i)).

In Figure~\ref{fig:results} (i), because the areas of the cap and tab are added, the energy density looks to be decreased. 
However, the actual panel deformation mainly occurred in the body part, not the cap and tab parts.
Thus, the cap and tab act more like boundary constraints than as contributors to strain energy storage.

Figure~\ref{fig:loadbearing} (a) compares the snapping behavior of the reference CC-block under $x$-direction (pinching) and $z$-direction (pulling) loading, showing energy density (left) and force (right) plots.
The $x$-load boundary condition is shown in Figure~\ref{fig:feasetting}, while the $z$-load applies the forced-displacement only along the $z$-axis.
The same geometry results in identical energy at both the peak and deployed states.
However, $z$-loading produces a delayed and sharper energy peak, resulting in more distinct snapping behavior.
It also requires higher deployment and folding forces, with a force ratio ($|F_{\min}/F_{\max}|$) of $3.769$, nearly double that of the $x$-load ($1.886$).
These results suggest that the CC-block in the $z$-direction has better resistance to flattening, emphasizing the need to consider loading direction in CC-block assembly and applications.

\subsection{Load-bearing Analysis}
\label{subsec:loadbearinganalysis}

\begin{figure}[htbp]
    \centering
    \includegraphics[keepaspectratio,width=\linewidth, page=7]{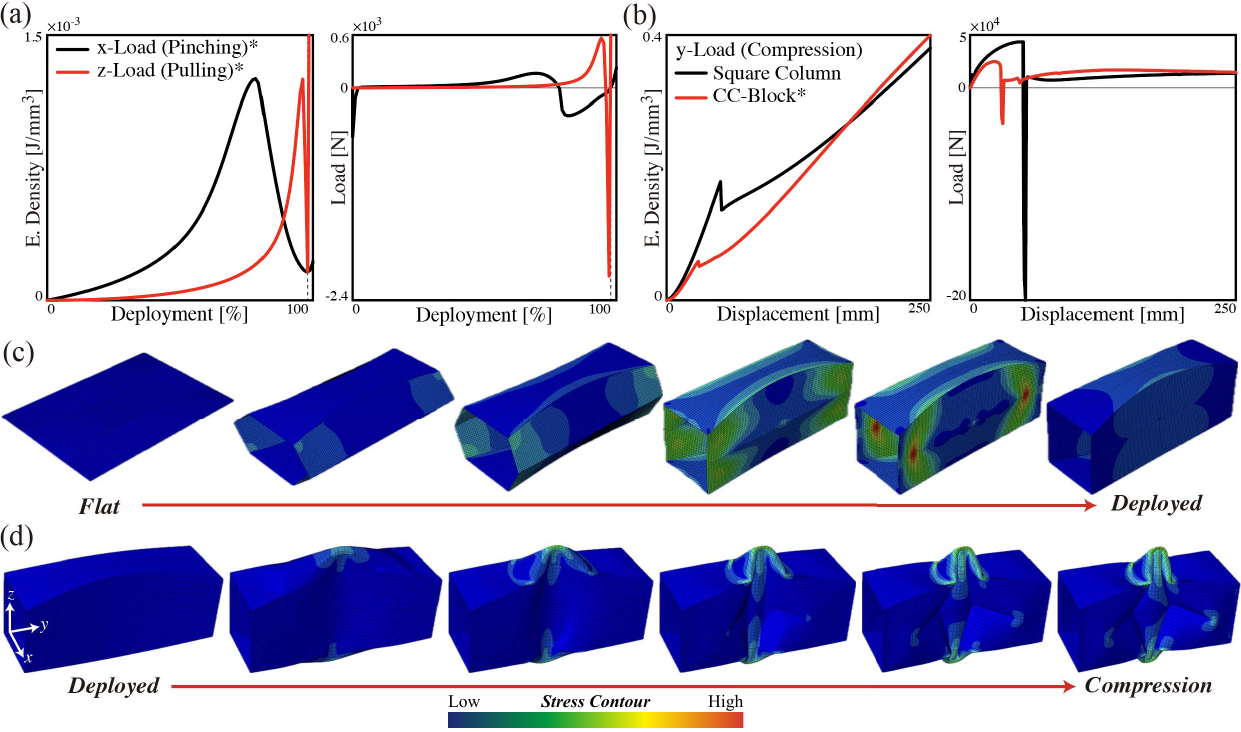}
    \caption{(a) Snapping behavior under different loading directions. (b) Compressive load-bearing behavior under $y$-direction compression and comparison with a square column. The simulation and stress contour of (c) deployment (pinching) and (d) compression of the reference model.}
    \label{fig:loadbearing}
\end{figure}

Figure~\ref{fig:loadbearing} (b) and (c) illustrate the load-bearing behavior of the deployed reference model under $y$-direction compression. 
The simulation begins with deployment from the flat state, followed by boundary release to reach a naturally stable configuration. 
Then, the top and bottom edges were constrained, and a compressive forced displacement of \SI{250}{mm} (corresponding to approximately $21\%$ of the total height) was applied to the top edges in the y-direction.
Here, we added a material property of $1\%$ Rayleigh (both mass and stiffness) damping based on the first eigenmode for the stable analysis.
For comparison, a simple square column with the same trajectory and height as the reference model was also analyzed under the same conditions.
The compression analysis is performed using a quasi-static method in the dynamic implicit solver, and the compression motion is shown in Figure~\ref{fig:loadbearing} (c).

Figure~\ref{fig:loadbearing} (b) presents the energy density (left) and force (right) plots.
Compared to the square column, the CC-block shows earlier buckling and lower load capacity.
However, due to its curved panel, the snapping behavior appears to be significantly mitigated. 
Buckling initiates at the mid-height of the CC-block, where curvature (stress) from panel bending is highest (Figure~\ref{fig:loadbearing} (d)). 
While the panels do not intersect up to a displacement of \SI{100}{mm} (Figure~\ref{fig:loadbearing} (d)), back panel intrusion occurs beyond this point, indicating that panel contact could further influence the actual post-buckling behavior.

\section{Assembly}
\label{sec:ccassembly}
The selected profile curvature forms the curved surfaces, allowing for flexible assembly and interlocking of adjacent blocks, which form quasi-continuous wall surfaces by repeating components.
To support this, we develop a generative shape grammar for parametric wall surface description and integrate it with the block design process.

\subsection{Shape Grammar}
\label{subsec:shapegrammar}

\begin{figure}
  \centering
  \includegraphics[width=\linewidth,page=6]{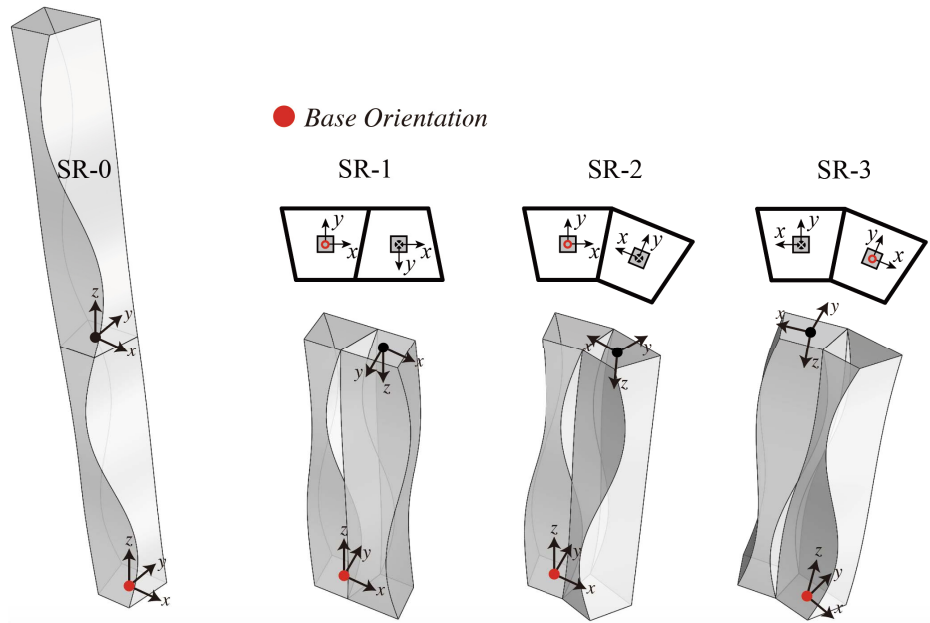}
  \caption{Shape rules for generative assembly of the curved-crease column: From left, shape rule(SR)-0 (adding stories), SR-1 (linear wall segment), SR-2 (curved wall segment, positive $x$ direction), SR-3 (curved wall segment, negative $x$ direction).}
  \label{fig:shapegrammar}
\end{figure}

The trapezoidal trajectory and bidirectional surface curvature of the CC-block enable it to form a quasi-continuous wall surface via surface-to-surface connection with adjacent columns.
Four valid orientations for adjacent column assembly are illustrated in Figure~\ref{fig:shapegrammar}, which can be concisely represented as a generative shape grammar.
Shape grammars comprise a set of shapes, a set of shape transformation rules, and an initial shape~\cite{stiny1982spatial}.
In the context of origami design, shape grammars have been applied to design crease patterns~\cite{yu2021rethinking} and curved surface profiles~\cite{gattas2018generative}. 

A generative shape grammar for wall assembly requires only a single shape, the CC-column with a trapezoidal trajectory and $2n$-mode elastica profile, with base plane and orientation as shown in Figure~\ref{fig:shapegrammar} (a).
Taking this column also as the initial shape, three shape rules are defined to add adjacent columns as illustrated in Figure~\ref{fig:shapegrammar}, denoted as shapes rule 0 to 3 (SR-0 to SR-3), respectively. 
SR-0 stacks a base block on top of the base block, constructing a multi-layer.
SR-1 attaches a block in the ground plane following a linear axis, with the attached block rotated by $180^{\circ}$ in both $xy$ and $xz$ planes. 
SR-2 attaches a block in the ground plane following a curved axis, with the block rotated by $2 \theta$ in the $xy$ plane and $180^{\circ}$ in the $xz$ plane. 
SR-3 performs a similar operation to SR-2, except that it attaches the next block to the opposite leg surface of the preceding block, that is, in the negative $x$ direction of the preceding block's orientation plane. 
A straight wall is thus specified with recursive application of SR-1, a curved wall is specified with alternating application of SR-2 and SR-3, and the direction to which the wall curves is determined by the orientation of the preceding block. 

\subsection{Implementation}
\label{subsec:shapegrammarimplementation}

\begin{figure}
  \centering
  \includegraphics[width=\linewidth,page=10]{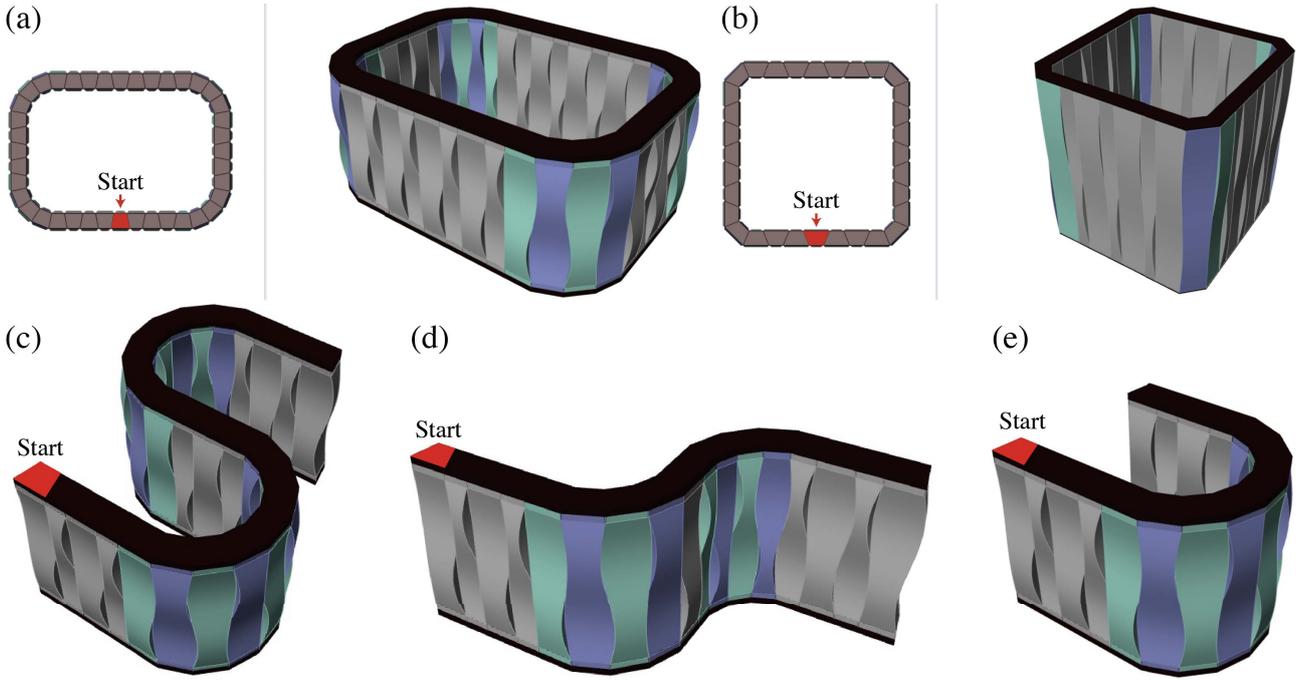}
  \caption{Various wall surfaces generated by the shape grammar logics. (a)  and (b) show the closed wall surfaces, each of which is constructed with different CC-block designs. The left is the top view, and the right is the perspective view. Various wall surface assemblies: (c) S-shape, (d) Tilde-shape, and (e) U-shape.}
  \label{fig:shapegrammarimplementation}
\end{figure}

We integrated the assembly design based on the generative shape grammar with the parametric design of CC-blocks and implemented this integrated design system in \emph{Rhino/Grasshopper}.
Figure~\ref{fig:shapegrammarimplementation} shows the implementation of shape grammar assemblies with different CC-block design parameters.
The gray, green, and purple blocks represent SR-1, SR-2, and SR-3, respectively.
Figure~\ref{fig:shapegrammarimplementation} (a) shows the desktop-scale small house concept design.
Here we name this CC-block as \emph{CC-Column}, which will be used again in the following Section~\ref{sec:ccfabrication}.
The design parameters of the CC-column for trapezoidal trajectory parameters as $a_1 = a_2 =$\SI{46}{mm}, with an angle $\theta =$\SI{11.25}{deg.}; 
profile parameters for a first-mode elastica with $b =$\SI{118.8}{mm} and $l =$\SI{120}{mm}, for $c = 99\%$ (see Figure~\ref{fig:design} (b)).
Here, we utilize the second-mode elastica by doubling $b$ and $l$, while keeping $c$.
After choosing the design parameters, we input the SR-codes $11112323111123231111 \times 2$ and the Figure~\ref{fig:shapegrammarimplementation} (a) is generated sequentially.

Figure~\ref{fig:shapegrammarimplementation} (b) also shows the actual-scale small temporary shelter concept design.
The design parameters of the CC-block for trapezoidal trajectory parameters as $a_1 =$\SI{125}{mm} $a_2 =$\SI{95}{mm}, with an angle $\theta =$\SI{22.5}{deg.}; 
profile parameters for a first-mode elastica with $b =$\SI{1139}{mm} and $l =$\SI{1140}{mm}, for $c = 99.9\%$, and we doubled vertically to make the second-mode elastica.
Then we input the SR-codes $11123111 \times 4$ and the Figure~\ref{fig:shapegrammarimplementation} (b) is generated sequentially.

% {\color{red}
Figure~\ref{fig:shapegrammarimplementation} (c--e) presents various opened wall assemblies generated using the CC-block design of Figure~\ref{fig:shapegrammarimplementation} (a).
The corresponding SR codes are $11123232323111232323231111$ for (c), $111232312323$-$1111$ for (d), and $111232323231111$ for (e).
These results demonstrate that the proposed shape-grammar logic enables the generation of diverse, complex, and continuous curved wall surfaces.
% }

\section{Applications}
\label{sec:ccfabrication}

Here, we fabricate both desktop- and large-scale CC-block prototypes to validate their multistability, modular assembly, and feasibility for application as a modular building system.

\subsection{Prototype: CC-Column}
\label{subsec:cccolumn}

\begin{figure}[ht]
    \centering
    \includegraphics[keepaspectratio,width=\linewidth, page=5]{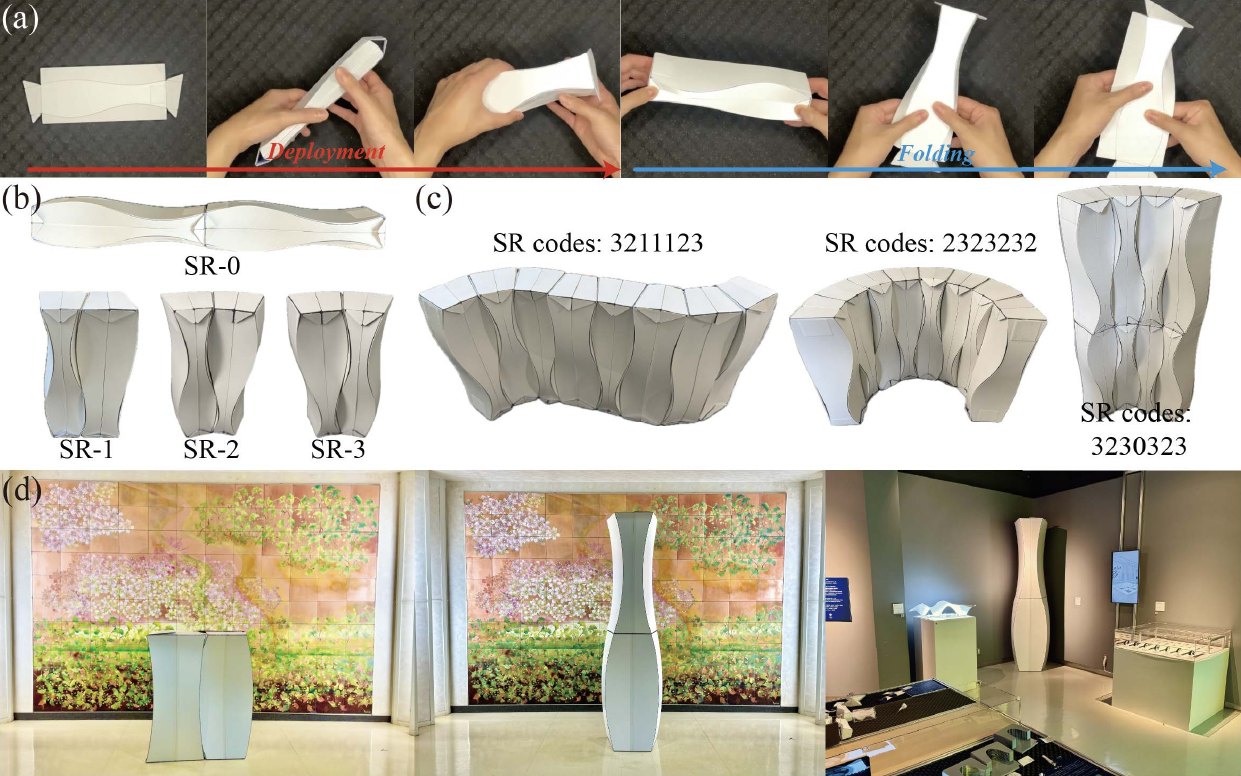}
    \caption{(a) Deployment and folding process of the desktop-scale CC-column. (b) Assembly rules and (c) Assembled configurations (starting from the leftmost (1st floor) block). (d) Large-scale CC-column.}
    \label{fig:cccolumn}
\end{figure}

\subsubsection{Desktop-scale Fabrication}
\label{subsubsec:desktop}

For the desktop-scale fabrication of the \emph{CC-Column}, we used the same design parameters as in Section~\ref{subsec:shapegrammarimplementation} and applied fabrication method 2 described in Section~\ref{subsubsec:deskfabmethod2}.
Figure~\ref{fig:cccolumn} (a) shows the deployment and folding motion of the CC-column.
The deployment was achieved by pressing the concave region (as shown in Figure~\ref{fig:feasetting}), while folding was performed by initiating compression from the convex region.
Snap-through behavior was observed only in the concave regions.
As analyzed in Figure~\ref{fig:results} (h), the prototype exhibits multistability through different snapping between the front and back panels.
Qualitatively, model deployment was consistent with the results of (Figure~\ref{fig:results} (f)), with a narrower concave panel side demanding greater deployment force and being more stable in the deployed state.
Figure~\ref{fig:cccolumn} (b) demonstrates the shape assembly rules from Figure~\ref{fig:shapegrammar}, and (c) shows the implementation of various shape grammar assemblies using only a single type of CC-column.

\subsubsection{Large-scale Fabrication}
\label{subsubsec:large_cccolumn}

Figure~\ref{fig:cccolumn} (d) presents the large-scale CC-column, assembled in both horizontal and vertical configurations, where the geometry parameters related to the length are scaled up by 10 from the desktop-scale.
For the panel material, we employed a \SI{3}{mm}-thick \emph{Plapearl (PCPPZ-050)}.
Due to panel material size limitations, the CC-column was fabricated by dividing it into quarters.
We used the CNC cutter, and all curved creases were made with half-cuts (refer to Section~\ref{subsec:largefab}), while linear creases were made with full-cuts and then reconnected using duct tape (Gorilla Tape) membrane hinge. 
Each block has caps and tabs to fix the deployed state and to enable vertical stacking.
The connection surfaces between blocks were attached with Velcro tape.
As observed in Section~\ref{subsec:loopclosing}, only the block where the linear crease is located on the concave panel showed snapping behavior, while the other did not.
To get snapping as well, the linear crease should be repositioned to the concave side (Figure~\ref{fig:results} (g)).
The weights of the top and bottom blocks in Figure~\ref{fig:cccolumn} (d)-mid are the same as \SI{1.837}{kg}, and they are rigid enough to bear the weight of another block in the deployed state.
Both desktop and large-scale CC-column were exhibited at \emph{Connecting Artifact 04}~\cite{Web_connectingartifacts} in Tokyo, Japan (Figure~\ref{fig:cccolumn} (d)-right).

\subsection{Prototype: CC-Box}
\label{subsec:ccbox}

\subsubsection{Desktop-scale Fabrication}
\label{subsubsec:desktop_ccbox}

\begin{figure}[ht]
    \centering
    \includegraphics[keepaspectratio,width=\linewidth, page=8]{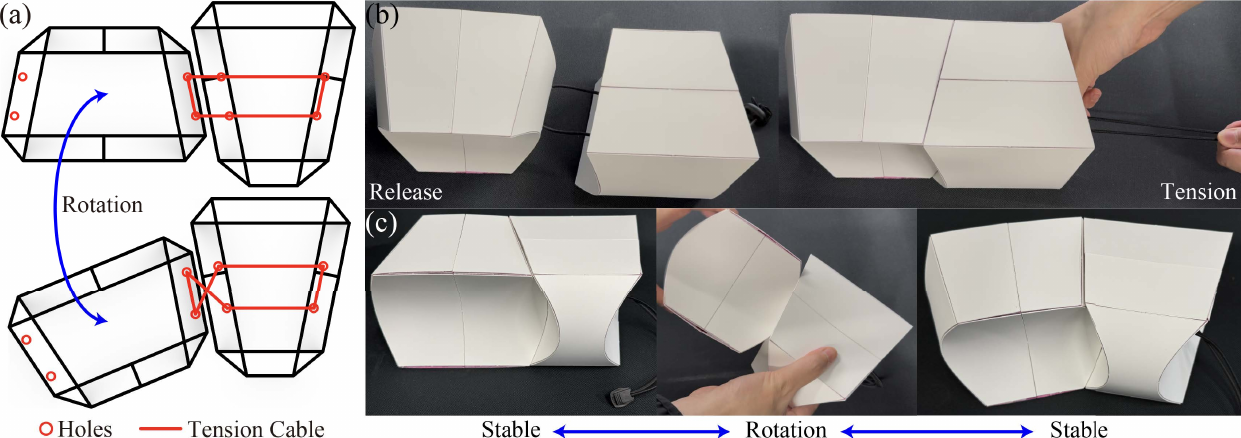}
    \caption{(a) Assembly method with tension cable, (b) CC-box prototypes and tension introducing, (c) motion of CC-box reconfiguring.}
    \label{fig:ccbox}
\end{figure}

The design parameters for trapezoidal trajectory parameters as $a_1=$\SI{100}{mm}, $a_2=$\SI{80.491}{mm}, with an angle $\theta =$\SI{11.25}{deg.}; and profile curve parameters for a first-mode elastica with $b =$\SI{100}{mm} and $l =$\SI{108.108}{mm}, for $c = 92.5\%$. 
End caps and tabs were included.
For the reconfigurable prototype, we utilized two blocks as in Section~\ref{subsubsec:large_cccolumn}, with linear creases placed on the concave panels.
Here, we introduce a tension cable for secure block assembly, which also allows for the relative rotation of adjacent blocks within the assembly. 
Figure~\ref{fig:ccbox} (a) illustrates the concept, and it can be extended by inserting additional blocks in between. 
Figure~\ref{fig:ccbox} (b) shows the process of introducing tension using rubber bands and stoppers.
After introducing the tension, the blocks interlock rigidly. 
Figures~\ref{fig:ccbox} (a, c) demonstrate reconfiguration between blocks, showing how rotation enables a transition from linear (left) to curved (right) wall assemblies.
An interesting observation is that the mismatch during the rotation causes cable extension, creating snapping behavior when the surfaces realign.
This highlights that our new assembly method can achieve \emph{reconfigurability} with multistability through simple rotation, without disassembly.

% {\color{red}

\subsubsection{Large-scale Fabrication}
\label{subsubsec:large_ccbox}

\begin{figure}[ht]
    \centering
    \includegraphics[keepaspectratio,width=\linewidth, page=9]{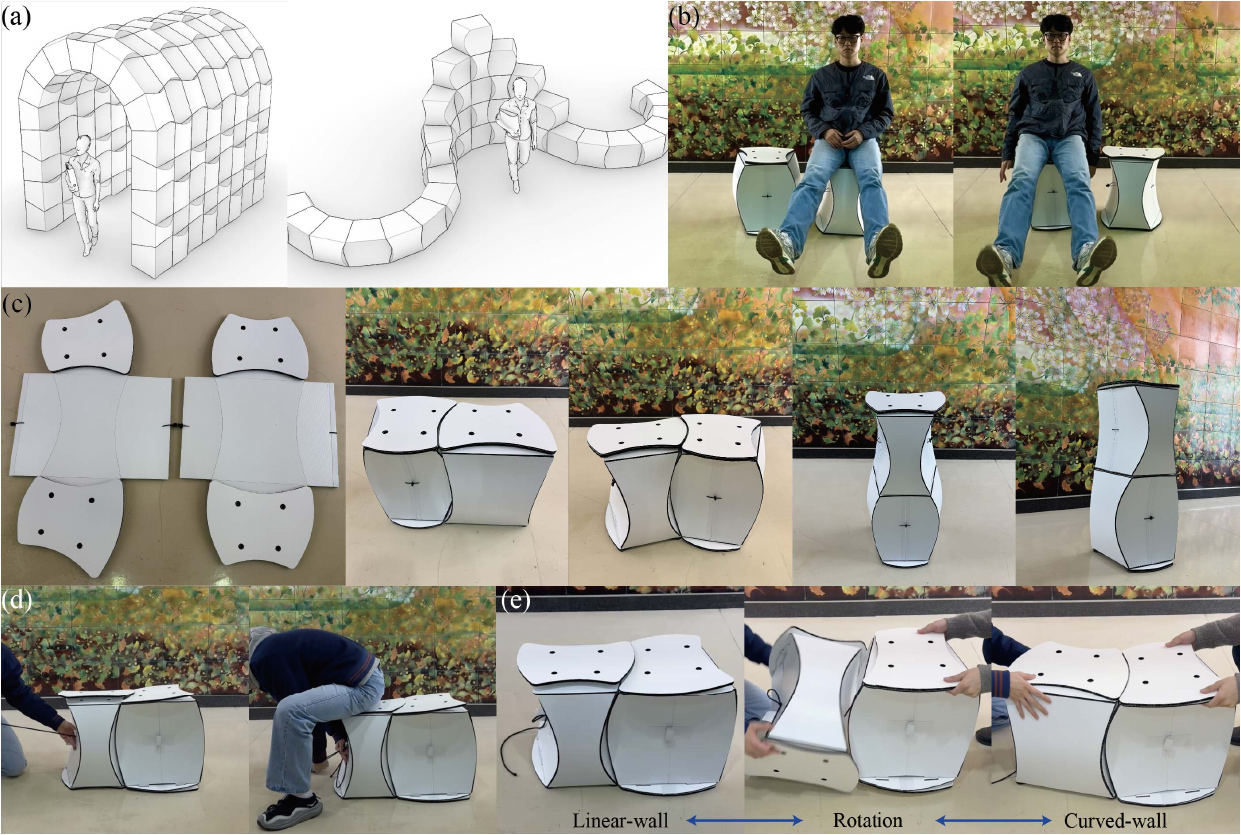}
    \caption{(a) Conceptual illustrations of CC-box assembly applications: arched roof (left) and bench with partitions (right). (b) Seating demonstration using the CC-box prototypes. (c) A pair of CC-box prototypes: from the left, flat configurations, horizontal assemblies forming linear and curved walls, and vertical assemblies shown from different perspectives. (d) Introducing tension between CC-boxes, and (e) reconfiguration of CC-boxes from linear to curved wall assembly by rotation.}
    \label{fig:largeccbox}
\end{figure}

We propose the \emph{ALL-IN-ONE} modular building system concept that demonstrates the versatility of two types of CC-Box in various assembly configurations.
Figure~\ref{fig:largeccbox} (a) shows conceptual illustrations by assembling two types of CC-box.
A single block can function as a stool(Figure~\ref{fig:largeccbox} (b)); and by assembling many, it can be used as a standing table, desk, bench (Figure~\ref{fig:largeccbox} (a)-right), bed, wall, arch roof (Figure~\ref{fig:largeccbox} (a)-left), and other various spatial structures.

Figure~\ref{fig:largeccbox} (c) shows the fabricated CC-boxes. 
For the panel material, we employed a \SI{5}{mm}-thick \emph{Plapearl (PDPPZ-100)}, and other fabrication processes are the same as in Section~\ref{subsubsec:large_cccolumn}.
The design parameters for trapezoidal trajectory as $a_1=$\SI{300}{mm}, $a_2=$\SI{241.473}{mm}, with an angle $\theta =$\SI{11.25}{deg.}; and profile curve parameters for a first-mode elastica with $b =$\SI{400}{mm} and $l =$\SI{412.371}{mm}, for $c = b/l = 97\%$.
This geometry is designed to consider stool use by fixing the height ($b$) at \SI{400}{mm}, and can form a circular assembly with $16$ blocks.

Through several prototyping trials (Appendix~\ref{chap:AppendixB}), we found that the panel material buckled at $c = 95\%$.
Therefore, considering the balances between clear snapping behavior, local buckling, and additional panel deformation under sitting loads, we set $c = 97\%$ and $VSL~(b/a_1) = 1.\dot{3}$ as the profile curvature (see Appendix~\ref{chap:AppendixB}).

In earlier prototypes, both tabs and caps were introduced for stability (Figure~\ref{fig:results} (i)); however, while the caps effectively acted as boundary constraints along the panel edges, the tabs, contrary to analysis predictions, degraded shape stability due to hinge stiffness when being folded.
Accordingly, only the caps were included in the CC-box design (see Appendix~\ref{chap:AppendixB}).

To improve load-bearing capacity and ensure uniform load transfer along the edges of the CC-box, the seat parts were fabricated separately using rigid \SI{10}{mm}-thick \emph{Plapearl (PGPPZ-300)} panels. 
The seat parts are designed as curved hourglass shapes along trapezoidal boundaries with filleted corners, allowing for smooth interlocking.

Figure~\ref{fig:largeccbox} (c) shows the flat and deployed states of CC-boxes. 
The weights of the left and right blocks in Figure~\ref{fig:largeccbox} (a), the most left picture, are \SI{1.510}{kg} and \SI{1.480}{kg}, respectively.
They have sufficient load-bearing capacity for the weight of a \SI{70}{kg} adult sitting and standing on it (Figure~\ref{fig:largeccbox} (b)).
Both the horizontal and vertical assemblies are well-matched as shown in Figure~\ref{fig:largeccbox} (c).

Our CC-boxes were actually used in \emph{Connecting Artifact 05} exhibition~\cite{Web_connectingartifacts} as stools for 8 weeks.
During the exhibition, rubber bands and stoppers were added to prevent popping out and severe deformation, ensuring structural stability (see Appendix~\ref{chap:AppendixB}, version 5).
This practical use demonstrates that our CC-box design is sufficiently robust and applicable under everyday loading conditions.

In addition, by introducing a tension cable (\SI{2.5}{mm}-radius rubber band) between CC-boxes (Figure~\ref{fig:largeccbox} (d)), we successfully demonstrated the reconfiguration from linear to curved (also conversely) wall assembly (Figure~\ref{fig:largeccbox} (e)).
Further detailed CC-box prototyping trials records are included in Appendix~\ref{chap:AppendixB}.

% }

\section{Conclusion}
\label{sec:conclusion}

\begin{figure}[htbp]
    \centering
    \includegraphics[keepaspectratio,width=\linewidth, page=11]{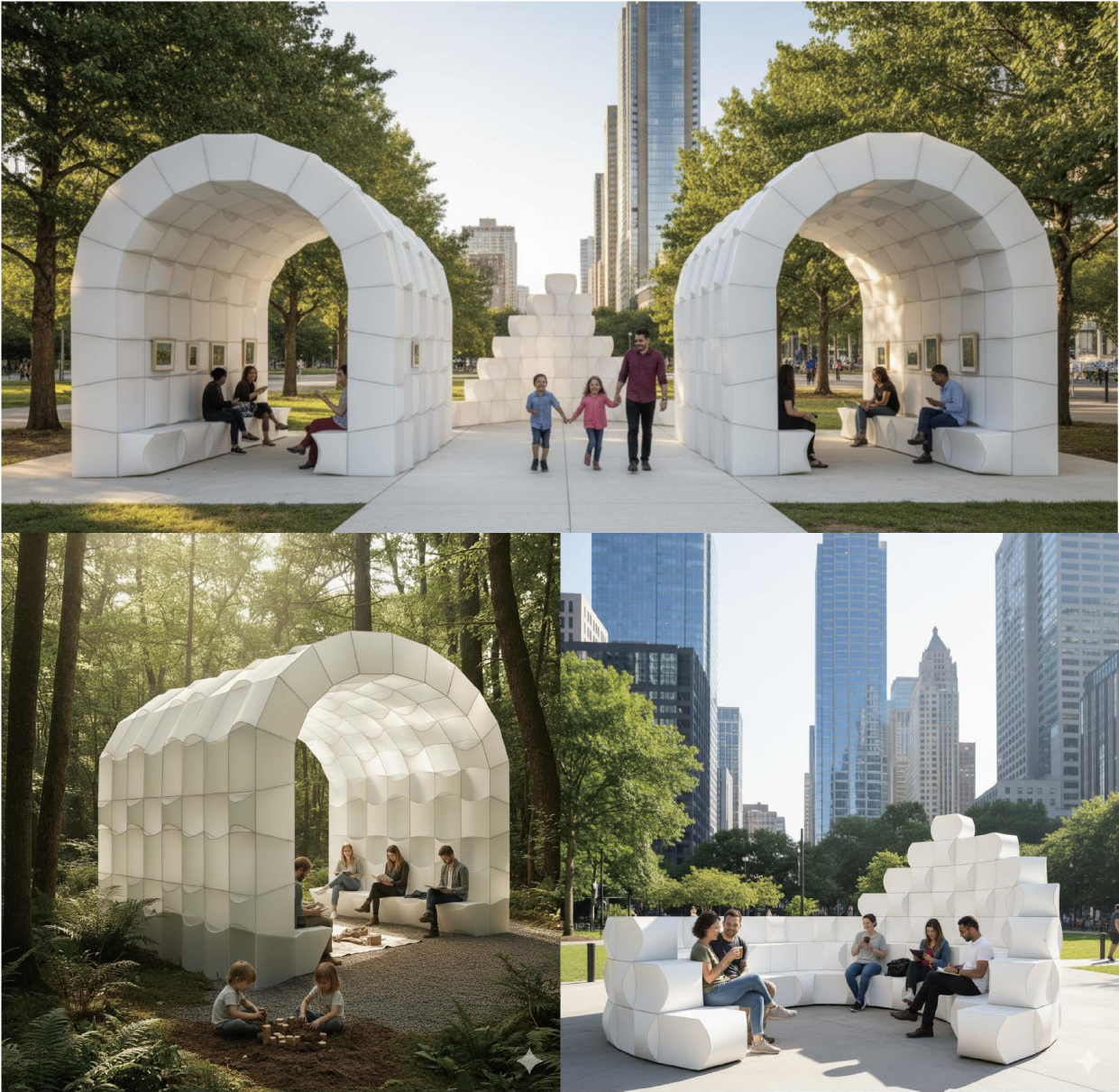}
    \caption{AI-generated illustration of CC-block pavilion. Made by \emph{Gemini 2.5 Pro}.}
    \label{fig:AI_CCblock}
\end{figure}

In this study, we extend the loop-closing method to curved-crease origami and propose cylindrical multistable curved-crease origami blocks (CC-blocks) with various polygonal trajectories and elastica curvature profiles.

We analyzed the designed CC-blocks through FE simulations under different design parameters and validated their multistability with desktop-scale prototypes.
We also examined the influence of loading directions on snapping behavior and load-bearing capacity.

By assembling identical CC-blocks using generative shape grammars, we demonstrated the programmability of multilayer, quasi-continuous curved wall surfaces and an integrated framework that links block design to wall assembly.
Additionally, we improved the assembly method by introducing a tension cable, allowing for post-deployment reconfigurability through snapping.

Furthermore, large-scale prototypes, CC-column and CC-box, were fabricated.
From the large-scale prototypes, we confirmed the snapping behavior, shape stability, and load-bearing capacity, thereby validating the scalability of the CC-block design.
By using these blocks both individually and in assemblies, we proposed and demonstrated the feasibility of an all-in-one modular building system, serving as stools, benches, beds, and spatial structures, as shown in Figure~\ref{fig:AI_CCblock}\footnote{We provided the Figure~\ref{fig:largeccbox} (b) to Gemini 2.5 Pro, and typed the following prompt: Make these illustrations into pavilions in the city park and nature.}.

\subsection{Discussion}
This chapter is constructed following four primary objectives.
The contributions and limitations corresponding to each objective are as follows.

\subsubsection{Objective 1: Exploring multistable origami design methodologies}

\paragraph{Contributions}
We extended the loop-closing approach beyond its previous design ~\cite{kamrava2019origami} by applying it to curved-crease origami defined by a wider variety of trajectory polygons and profile curves.
This enabled the design of stable configurations on cylindrical topologies while preserving flat-foldability, demonstrating that CC-blocks can serve as versatile building blocks for constructing curved wall surfaces.
Additionally, our CC-block features simple origami geometry, consisting only of mountain creases; it is highly accessible, productive, and has a low computational cost.

\paragraph{Limitations}
The current CC-block design is limited to a coaxial tube configuration, where the trajectory is either rotational and mirror-symmetric, and both upper and lower trajectories are identical shapes.
Breaking these symmetries, particularly by varying the scale or shape of the upper and lower trajectories, may enable curved-surface assemblies similar to T-toroids in Chapter~\ref{chap:ttoroid}. 
Such generalization could allow CC-blocks to represent more diverse curved-wall or even spatial surfaces. 
Achieving this, however, requires further understanding of curved-crease origami kinematics and its constraints, and this direction will be addressed in future work.

\subsubsection{Objective 2: Analyzing mechanics across the design parameters}

\paragraph{Contributions}
From the snapping analyses, we obtained reliable equilibrium paths for most parameter sets, which allowed us to identify meaningful tendencies in how the geometric design parameters influence the energy barrier of the CC-block.
These tendencies offer practical guidance for tuning stability characteristics and informing early-stage design decisions.

\paragraph{Limitations}
The load-bearing results (Figure~\ref{fig:loadbearing}) are insufficient for a comprehensive understanding of structural performance; more systematic evaluations across design parameters will be required.

The effect of adding tabs remains unclear. 
While they can contribute geometrically, the actual prototypes exhibited reduced stability when the tabs were folded.
This discrepancy is likely related to the moment by hinge stiffness when folding tabs, but a more detailed investigation is required.

Additionally, the structural performance of assembled configurations is beyond the scope of this study and will be addressed in future work.

\subsubsection{Objective 3: Programming global systems by assembling blocks}

\paragraph{Contributions}
We integrated total workflows, from parametric block design to shape-grammar-based generative assembly, demonstrating the feasibility of adaptable, replaceable, and reconfigurable modular building systems.
This provided a simple yet powerful framework for sequentially assembling multi-layer curved-wall systems.

\paragraph{Limitations}
The current shape-grammar implementation is limited to basic rule sets, making it difficult to achieve target-shape optimization or avoid issues such as self-intersection in generated global shapes.
Also, simultaneous assembly generation and more complex logic structures remain unexplored, and expanding these capabilities will be addressed in future work.

\subsubsection{Objective 4: Demonstrating the applicability via desktop and large-scale fabrication}

\paragraph{Contributions}
Both desktop and large-scale prototypes were successfully fabricated, and each demonstrated clear multistability with stable deployed configurations.
The deployed geometries interlocked as intended, and the introduction of tension cables enabled reliable assembly and reconfigurability, allowing reconfiguration with snapping transitions through simple block rotation.
The large-scale CC-column and CC-box prototypes performed as designed, exhibiting sufficient rigidity to support their own weight.
Even the CC-box was used as a stool,  demonstrating the practical feasibility of our CC-block concept.
Furthermore, the simplicity of the block geometry contributed to a rapid and efficient fabrication process.

\paragraph{Limitations}
Further exploration of panel materials and curved-hinge systems is required to improve the structural performance and manufacturing process in larger prototypes.

\subsection{Future Work}
Future work will focus on generalizing the CC-block design to assemble more complex curved surfaces.
This includes extending the current CC-block to non-symmetric geometries and upgrading our logic-guided assembly methods (e.g., shape grammar) to support optimization for assemblies.

We will also conduct detailed studies on the load-bearing behavior of CC-blocks, both individually and in assembled systems, analyzing parameter-dependent structural performance with experimental verification.

Ultimately, we aim to develop a comprehensive CC-block system with cable-driven assembly and actuation, enabling reconfigurability and practical applicability in architectural contexts.

    \chapter{Conclusion}
\label{chap:conclusion}

In this study, three multistable origami blocks were designed and investigated under four research objectives (geometry design, mechanics, programmable assembly, and application) to address the challenges of applying deployable origami to large-scale architectural systems by leveraging multistable origami as modular building blocks.

We investigated three design methods of multistable origami, each developed for specific purposes, including nonlinear stiffness control and stable configuration control, based on over-constraining approaches applied to rigid and curved-crease origami geometries.
The designed multistable origami blocks were analyzed through simulations and experiments to investigate the relationship between geometric parameters and mechanical behaviors, and their multistabilities were also validated through desktop-scale prototype tests.
By assembling the designed blocks, we demonstrated that we could program the deployment sequence and overall geometries, such as three-dimensional curved surfaces or curved wall structures, of global systems.
Finally, by fabricating the large-scale prototypes, we verified the feasibility of our multistable origami modular systems and showcased their scalability, transportability, deployability, self-supporting, and reconfigurability.

The conclusions of each chapter are as follows:

\begin{itemize}
    
    \item In Chapter~\ref{chap:methodology}, we reviewed existing design approaches for multistable origami and introduced numerical simulations to analyze their mechanics. We also proposed materials and fabrication methods for each desktop and large-scale fabrication. Through this chapter, we established the fundamental workflow for the subsequent chapters.
    
    \item In Chapter~\ref{chap:qbellows}, we present a stiffness-programmable multistable origami (\emph{Q-Bellows}) design method based on geometric incompatibility. By optimizing the incompatibility between linkage and rigid origami modules, we controlled the energy barriers and validated them via bar-and-hinge simulations and experiments. Through the serial chain assemblies, we demonstrated programmability of deployment sequences of global systems.
    
    \item In Chapter~\ref{chap:ttoroid}, we propose a design method for multistable origami blocks (\emph{T-toroids}) using the loop-closing approach applied to rigid origami tubes with the desired toroidal stable configurations. We investigated the mechanical behaviors of T-toroids by FE analyses across various geometric parameters, and diverse curved surfaces were achieved by assembling them based on their edge-offset properties. Finally, a large-scale prototype validated the practical applicability of our modular system.
    
    \item In Chapter~\ref{chap:ccblock}, we extend the loop-closing method to curved-crease origami and present cylindrical multistable origami blocks (\emph{CC-blocks}). The designed CC-blocks were analyzed across various design parameters through FE simulations. By integrating the shape grammar process into CC-block design, we demonstrated various multilayer quasi-continuous curved wall assemblies, and large-scale fabrications verified the practical applicability of our modular system.
    
\end{itemize}

\section{Discussion}

\subsection{Methodologies}
This study was structured around four primary objectives, each investigated through the methodologies introduced in Chapter~\ref{chap:methodology}.
The corresponding achievements, experimental insights, and limitations of methodologies associated with each objective are summarized as follows.

\subsubsection{Geometry Design Methods}

\begin{wrapfigure}{l}{0.35\textwidth}
    \centering
    \includegraphics[keepaspectratio,width=\linewidth, page=4]{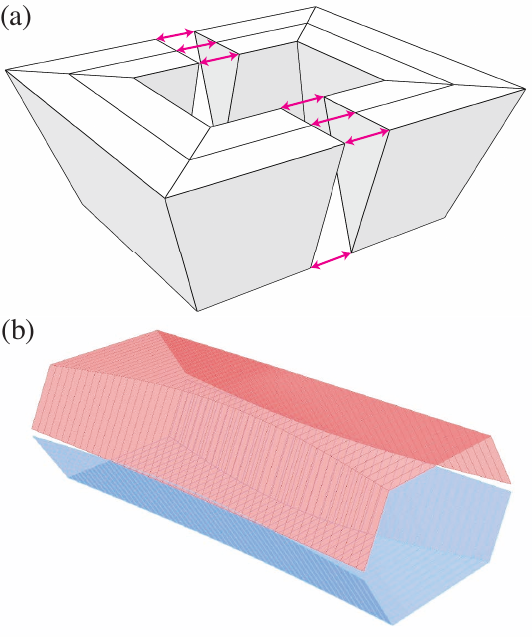}
    \caption{Conceptual illustrations of geometric incompatibility approaches for (a) T-toroid and (b) CC-block.}
    \label{fig:GIxLoop}
\end{wrapfigure}

The geometric incompatibility method can be extended to other linkages with quadrilateral boundaries for realizing complex spatial deployment motions~\cite{bernardes2022design}, including various 1-DOF linkages (Figure~\ref{fig:linkage_example}), multi-DOF linkages~\cite{kamijo2021serial}, origami tubes \cite{tachi2017capping, filipov2015origami}, scissor mechanisms~\cite{nishimoto2024transformable}.

The key idea of the loop-closing method is the ability to design a target stable shape by closing an origami loop. 
This can be extended to a systematic inverse-design method for shaping the stable shape to match a desired geometry~\cite{snappingtoroids-2025}.

These two methods can be integrated in some cases. 
As shown in Figure~\ref{fig:GIxLoop}, by dividing a loop-closed origami into multiple origami units, the geometric incompatibility concept can be applied to estimate the energy barrier during deployment.
We have applied this idea to evaluate T-toroids with square trajectory and parallelogram profile in~\cite{snappingtoroids-2025}.
However, it appears to apply only to symmetric configurations, thereby further validations are required for its generalization.

Overall, these insights offer practical guidelines for selecting an appropriate design strategy depending on the intended application of multistable origami. 
If the goal is to control the mechanical behavior of a single multistable origami, a geometric incompatibility method is recommended. 
If the goal is to prescribe target stable shapes and organize them into assemblies, the loop-closing method provides a more suitable framework.

\subsubsection{Mechanics Analysis Methods}

\paragraph{Numerical Simulations}
From our experience, the bar-and-hinge analysis (\emph{MERLIN2})~\cite{filipov2017bar, Liu2017Mer1EN} presented the following limitation.
About the defining boundary conditions, only translational constraints at nodes are allowed, which prevents the application of rotational constraints or symmetry conditions.
Therefore, the whole structure should be modeled.
This makes it difficult to converge the analysis of multistable origami when snapping behaviors coincide at different locations within a system, and unexpected unstable behavior may arise.
Furthermore, curved-crease origami was difficult to analyze directly using the original MERLIN2 software.
Although a bar-and-hinge method for curved-crease origami was proposed in~\cite{woodruff2020bar}, we adopted a more general structural analysis method that can explicitly capture detailed panel deformation during snapping and can also accommodate more complex conditions such as cyclic loading and load-bearing.

For this reason, we developed a novel crease modeling method based on a pin-joint array for the finite element analysis of origami structures.
We successfully performed modal, linear buckling, and implicit analyses, obtaining reasonable results with relatively low computational cost.
Furthermore, our modeling method can also be applied to other deployable systems, such as kirigami and scissor mechanisms.

\begin{table}[htbp]
    \centering
    \includegraphics[keepaspectratio,width=0.75\linewidth, page=1]{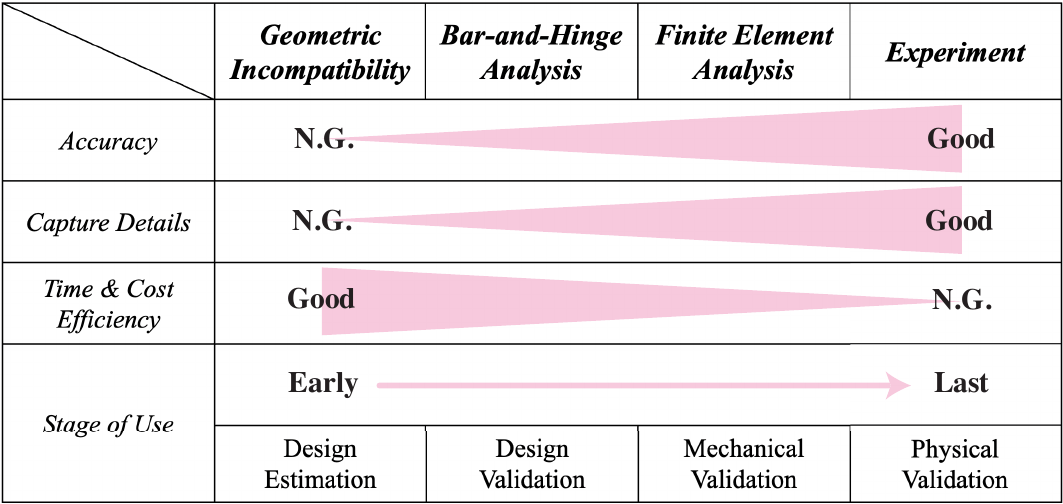}
    \caption{Comparison of mechanical analysis methods.}
    \label{tab:comparison_analysismethod}
\end{table}

\paragraph{Comparison of Analysis Methods}
Across the four analysis methods employed in this study, geometric incompatibility estimation, bar-and-hinge analysis, finite element analysis, and experiments, we observed qualitative agreement in capturing the mechanical trends of multistable origami blocks.

However, these analysis methods differ significantly in their ability to resolve local details. 
The geometric incompatibility estimation tends to overestimate the energy barrier and cannot comprehend where the deformation occurs.
The bar-and-hinge analysis can provide reliable results but is limited in representing detailed local stress concentrations. 
Finite element analysis, while more detailed, relies on modeling assumptions that are usually difficult to fully reflect physical conditions.

Based on the above insight, we can provide guidelines for selecting appropriate analysis methods according to the geometric complexity and the level of detail required (see Table~\ref{tab:comparison_analysismethod}).
When multistable origami was designed based on geometric incompatibility, and the early design stage requires only an estimation of energy barriers or relative stiffness differences between blocks, the geometric incompatibility estimation method can be an efficient approach.
If the geometry is relatively simple and exhibits bistability, and the analysis is intended for early-stage evaluation, a bar-and-hinge model is recommended.
For multistable origami with complex geometries, where detailed deformation and stress distributions must be captured, finite element analysis is more appropriate.
Finally, when deeper insight into physical phenomena such as contact, play, plasticity, or fabrication-induced effects is required, experimental investigation is essential.

\subsubsection{Programmable Assembly Methods}

The logic-guided assembly methods demonstrated in this study represent only the simplest case.
Both the binary logic and the shape-grammar logic have the potential to be significantly expanded.
For example, extending the binary system could lead to mechanical computing applications~\cite{yasuda2021mechanical}.
Enhancing the shape grammar with additional rules, such as parallel generation~\cite{rossi2018voxels}, collision avoidance, or path-targeting, could yield a richer generative design system capable of producing complex systems and even optimizing toward target shapes~\cite{kodnongbua2025design}.

Regarding the mesh-guided method, this study did not fully demonstrate assemblies matched to complex and specific target geometries.
However, PQ-mesh generation for arbitrary freeform surfaces is already a well-addressed problem~\cite{bobenko2006minimal, liu2006geometric, pottmann2007geometry}, making such extensions straightforward.
Also, we have partially addressed this issue in ~\cite{snappingtoroids-2025}.

These two assembly methods, logic-guided and mesh-guided, offer complementary strategies.
The logic-guided method offers a bottom-up approach, suitable for designers aiming to explore sequential behaviors and various emergent global patterns through combinatorial assembly.
In contrast, the mesh-guided approach provides a top-down approach, ideal for applications where a target geometry is prescribed first, and the assembly must conform to a global system.
Thus, the choice depends on whether the final objective prioritizes behavior-driven growth (bottom-up) or geometry-driven synthesis (top-down).

\subsubsection{Physical Prototype Methods}

\paragraph{Desktop-scale}
Both desktop-scale fabrication methods (Section~\ref{subsubsec:deskfabmethod1}) demonstrated being suitable for realizing the intended behaviors.
Through our experiences, we confirmed that Polyester and Polypropylene-based sheet material provide an appropriate balance of elasticity and stiffness between snapping and load-bearing. 
Additionally, the membrane-based hinges offer low stiffness and minimal play, thereby enabling kinematics and mechanical responses close to the ideal origami assumptions.
Based on our experience, we can provide practical guidelines for selecting the appropriate desktop-scale fabrication method, considering the desired scale, geometric complexity, durability, and visual preferences.

Desktop-scale fabrication method 1 utilizes thin panels and single-layer membrane hinges, making it advantageous for relatively small-scale and geometrically intricate designs.
The transparency of the panels also enables refined visual expressions when paired with colored fabrics.
However, the thin panels and a single-layer hinge also make it more vulnerable to fatigue; repeated twisting or snap-through motions can cause gradual delamination at panel–hinge boundaries, and single-layer hinges may stretch and tear over time, increasing unwanted play.
Additionally, this method requires more fabrication time, reflecting another practical trade-off between geometric versatility and production efficiency.

Desktop-scale fabrication method 2, which utilizes thicker panels and double-layer hinges composed of nonwoven fabric and double-sided tape, demonstrated better durability.
These prototypes resist tearing and hinge stretching even under repeated cycles, and generally offer better load-bearing performance than Method 1.
These benefits make it well-suited for mid-scale or load-bearing demanding prototypes.
However, as the geometry becomes smaller or more complex, the relatively higher panel stiffness and thickness begin to suppress ideal origami motions, leading to interference between panels, showing a trade-off between mechanical robustness and kinematic finesse.

\paragraph{Large-scale}
About the large-scale fabrication, the following limitations remain.
The sandwich panel used in our large-scale prototype offered advantages in terms of lightweight and manufacturability; however, it also presented several practical challenges. 
Because the snapping behavior depends on storing and releasing elastic energy through panel bending, the sandwich panel tends to concentrate stress in its thin face layer sheets during bending. 
This makes the system vulnerable to local buckling of the face layer, which is difficult to control and predict the mechanical behaviors.

A more systematic understanding of the material and structural characteristics of sandwich panels will, therefore, be essential for reliable large-scale applications.
The current tape-based hinge and panel connections serve as temporary solutions, lacking the durability required for long-term or structurally demanding applications, and often fail due to fatigue-related peeling or tearing.
These issues significantly degrade the snapping behavior and shape stability.
Therefore, more robust panel-to-panel and block-to-block connection systems must be developed.

Moreover, we did not construct full assemblies using large-scale blocks in this study.
Such assemblies are expected to introduce further challenges, including cumulative geometric tolerances, self-weight-bearing, and reliable face connections, that require detailed structural analysis and experimental validation.
These aspects extend beyond the scope of the present study due to practical constraints in material, cost, and fabrication time, and will be addressed in future work.

\subsection{Comparison between Designed Blocks}

\begin{table}[htbp]
    \centering
    \includegraphics[keepaspectratio,width=\linewidth, page=2]{figure/Chapter6.pdf}
    \caption{The keywords of each multistable origami block.}
    \label{tab:keywordtable}
\end{table}

Table~\ref{tab:keywordtable} summarizes the important keywords of each multistable origami block system.
As shown in the Table~\ref{tab:keywordtable}, the designed blocks differ basically in their design objectives.
Q-bellows is developed to address the challenge from a mechanical standpoint, whereas T-toroid and CC-block focus on geometric design challenges.
Therefore, they have different characteristics across geometry, mechanics, assembly, and application.

Despite these differences, the blocks share several important commonalities.
First, all blocks are based on one-degree-of-freedom origami structures and achieve multistability through intentional over-constraint.
Second, each block can maintain stable deployed configurations, with geometric parameters directly influencing its mechanical behavior.
Third, the assembled global systems of all blocks demonstrate reconfigurability.
Finally, prototype demonstrations confirm both the multistability of the blocks and their scale extension.

Furthermore, by developing these three blocks, we can obtain insights that are not evident from only one or two designs.

From the geometry design aspect, the proposed design methods are not limited to rigid origami but can be extended to curved-crease origami, showcasing the extension of the design mechanism.

From the analysis aspects, finite element analysis clarified where stress and deformation are concentrated, which could not be captured by simplified models, such as equivalent strain and bar-and-hinge models.
Furthermore, through the experiments, we can find hysteresis and unexpected nonlinearities, such as play and plasticity, which are difficult to predict by numerical analysis alone.

From assembly aspects, the nonlinear stiffness variations observed across geometric parameters in T-toroid and CC-block indicate that similar deployment-sequence control is also feasible as Q-bellows.

Finally, from a prototyping perspective, failures during large-scale fabrication revealed design issues that could not be identified through desktop-scale prototypes alone, including thickness accommodation, tolerance accumulation, stress concentration, and load-bearing limitations. 

Although the three blocks were developed with different objectives and geometric conditions, they were all realized within a common framework, demonstrating that it can be utilized not merely as case-specific solutions but as a systematic design. 
Accordingly, the framework presented in this thesis establishes a foundation and practical guidance for designing further multistable origami modular systems beyond the three cases studied here.

% {\color{red}

\subsubsection{Application Domains}

This study demonstrates that multistable origami blocks can form a new class of modular systems that strike a balance between deployability, self-supporting, load-bearing, and scalability.
Overall, they are well-suited to architecture, robotics, and metamaterial-related applications where lightweight, transformability, adaptability, and reconfigurability are prioritized.
Based on Table~\ref{tab:keywordtable}, the application domains of each multistable origami block can be considered.

\emph{Q-bellows} is specialized for stiffness programming.
Assembling blocks with a different stiffness into a binary memory system enables programmable deployment sequences and hysteresis.
These characteristics make Q-bellows more suitable for applications that exploit tuned nonlinear stiffness and hysteresis itself rather than spatial structure systems.
Representative examples include meta-materials, mechanical computing systems, sequentially deployable robotic arms, and energy-dissipation structures, such as dampers, that utilize programmable hysteresis.

\emph{T-toroid} is well-suited for programming various three-dimensional curved surfaces based on its edge-offset geometric property.
Also, depending on the assembly strategy, the global system can achieve either global flat-foldability or rigid configuration.
These characteristics make T-toroid suitable for spatial structures, such as 3-dimensional curved shells, roofs, wall surfaces, kinetic façades, and temporary shelters, while individual blocks can also function as furniture elements, including stools, desks, and beds.

\emph{CC-block}, based on its interlocking geometry between blocks, enables the programming and generation of cylindrical curved surfaces through a generative design method.
Also, depending on the assembly strategy, CC-blocks can achieve global flat-foldability, and their reconfigurability can be further enhanced by introducing tension cables into the assembly.
These characteristics make CC-blocks suitable for spatial structures such as cylindrical curved walls, benches, roofs, kinetic façades, and temporary shelters, while individual blocks can also serve as furniture elements, including stools, beds, and partitions.

\subsection{Design Trade-offs}

These large-scale prototyping trials (see Appendix~\ref{chap:AppendixB}) highlighted that realizing multistable origami at architectural scale requires balancing multiple trade-offs across geometry, mechanics, and fabrication. 
A recurring trade-off was observed between achieving clear multistability and maintaining material elastic range.
An energy barrier that is too high can cause local buckling or tearing of panels and hinges, whereas an energy barrier that is too low leads to loss of multistability due to hinge stiffness and play (e.g., duct-tape hinge, double-hinge).
To achieve this balance, we conducted multiple prototyping trials.
For the T-toroid, compliant zones were introduced to mitigate excessive energy barriers and stress concentration that led to hinge and panel failures.
For the CC-block, the profile curvature and slenderness were repeatedly tuned, based on the analysis results (Figure~\ref{fig:results}), and various hinge systems were tested to secure reliable shape stability. 

However, these modifications introduced further trade-offs between geometry and load-bearing capacity at the application level. 
For example, the compliant zones in the T-toroid reduced load-bearing performance in specific regions. 
At the same time, the CC-block required additional rigid seat parts made from different panel materials to achieve sufficient load-bearing capacity. 
Moreover, adding materials, parts, and complex geometry to improve structural performance increased complexity, time, and cost, creating a trade-off with manufacturability. 

Overall, the large-scale prototyping clarified that the feasibility of multistable origami block systems is governed not by a single design challenge, but by a coupled set of trade-offs among geometry, mechanics, assembly, and fabrication that must be negotiated through an iterative design--analysis--fabrication feedback cycle.

\subsection{Comparison with Conventional Materials}

\begin{table}[htbp]
    \centering
    \includegraphics[keepaspectratio,width=\linewidth, page=3]{figure/Chapter6.pdf}
    \caption{The comparison between conventional architectural materials, general origami, and multistable origami.}
    \label{tab:conventional_comparison}
\end{table}

Table~\ref{tab:conventional_comparison} compares conventional architectural materials, general origami systems (rigid and curved-crease), and the proposed multistable origami blocks from the perspectives of strength, weight, transformability, and reconfigurability.
Conventional materials, such as concrete, steel, and timber, provide high strength but are heavy, fixed geometry (solid), and require disassembly for reconfiguration.
In contrast, origami are lightweight and deployable through kinematics, but they typically remain mechanically weak and lack stable load-bearing configurations.

Multistable origami occupies an intermediate and complementary regime between these two.
While its strength is lower than that of conventional materials, it is quite higher than that of origami, allowing it to function as self-supporting and load-bearing modular blocks.
At the same time, unlike conventional materials, which are geometrically fixed, and unlike origami, which is governed by kinematics, multistable origami combines both transformability and solidity, enabling structures that can be adaptable to the requirements, such as storage and transportation.
Moreover, in terms of reconfigurability, conventional systems require physical disassembly, and origami is limited to predefined kinematics, whereas multistable origami allows reconfiguration through both assembly logic and state switching, offering higher adaptability.

From the perspective of construction energy, these characteristics of multistable origami are particularly advantageous.
Lightweight, compact pack-and-stack, and post-assembly reconfigurability can reduce the energy required for transportation, lifting, and on-site construction compared to conventional materials.

However, although multistable origami blocks can be used as structural members such as walls, roofs, and shell-based spatial structures, their use under large external loads remains more limited than that of conventional materials.
Therefore, rather than replacing existing construction methods, the proposed systems are best understood as a complementary class of architectural materials, particularly suited to applications where lightweight, deployability, and reconfigurability are more critical than load-bearing capacity, such as temporary structures, adaptive façades, enveloping layers, exhibition spaces, and emergency shelters.

% }

\section{Future Work}
Building on the above conclusions and discussions, future work will focus on more practical studies toward realizing our multistable origami modular building systems.

In terms of design methodology, we plan to extend the proposed design methods beyond symmetric configurations to more generalized forms.

% {\color{red}
In terms of analysis methodology, detailed numerical analyses and experimental validations of snapping and load-bearing behavior will be conducted under realistic physical conditions, including contact and material nonlinearity, to clarify the mechanics of multistable origami blocks.
In particular, to calibrate the numerical analyses for enabling a direct comparison between simulations and experiments, at least material tests need to be performed to characterize the elastic modulus and elastic range of the panel materials, as well as the hinge stiffness.
Structural evaluations will also be conducted on full-scale blocks and their assemblies to verify scalability and load-bearing performance.
% }

In terms of assembly methodology, both bottom-up and top-down methods will be explored for assembling multistable blocks into more complex or target global systems using the optimization-based approaches.

In terms of prototyping methodology, we will investigate effective fabrication and assembly strategies for large-scale prototypes, including the use of reusable panel materials, durable membrane hinges, and the integration of compliant and mechanical hinge systems.
In addition, design and fabrication will be developed as an integrated system that accounts for thickness accommodation, hinge regions, and part-to-part connections, enabling automated workflows from geometry design to fabrication.

Looking forward, by incorporating cable-driven or pneumatic actuators, the multistable origami modular system may evolve into actively deployable structures capable of controlled and adaptive transformations.

	\bibliographystyle{ieeetr}
	\bibliography{bibliography}
	\addcontentsline{toc}{chapter}{Bibliography}
    \addcontentsline{toc}{chapter}{Appendix}
    \appendix
\setcounter{chapter}{0}
\setcounter{section}{0}
\renewcommand{\thesection}{\thechapter.\arabic{section}}

\chapter{Crease Modeling Codes for FEA}
\label{chap:AppendixA}

This appendix provides the detailed Python-based code for the crease modeling method introduced in Section~\ref{subsec:FEA}, developed for finite element analysis of origami structures.
Three Python scripts are included, all designed to operate within \emph{Abaqus}.

First, the meshing codes~\ref{AppendixA:Meshing} divide the edges of shell or solid elements proportionally to their length to generate meshes.
Next, within the \emph{Abaqus} interface, the user should manually select the edges intended to act as creases and create an edge set named \emph{Selected Edges-n} (where n is an index number).
The set name can be modified according to user requirements. 
Although this process was performed manually in our study, additional scripts can be developed to automate edge-set generation if desired.
Using the connector creator codes~\ref{AppendixA:Connector}, nodes located at identical positions are connected with pin joints, thereby modeling creases.

If an incorrect crease configuration is generated, the reset codes~\ref{AppendixA:Reset} can be used to delete and rebuild the connector system.
Detailed codes are as follows:

\section{Meshing Codes}
\label{AppendixA:Meshing}

\begin{lstlisting}[
  language=Python,
  caption={Meshing codes for shell or solid models. The meshing was conducted based on edge division, which is based on the division of edge lengths.},
  breaklines=true,
  breakatwhitespace=false,
  columns=fullflexible,
  keepspaces=true,
  breakindent=1em,
]
from abaqus import *
from abaqusConstants import *
from caeModules import *
import regionToolset
import math
import numpy

model = mdb.models['Model-1']
assembly = model.rootAssembly

all_edge_lengths = []
for part_name in model.parts.keys():
    part = model.parts[part_name]
    edges = part.edges
    
    for edge in edges:
        edge_length = edge.getSize(printResults=False)
        all_edge_lengths.append(edge_length)
        
min_edge_length = min(all_edge_lengths)

######## If you want to control edge division scale, set(or add) scaling factor constraints
Edge_scaling_factor = 50000
Mesh_scaling_factor = 50

######## Reference division counts to the minimum edge length (Even Number Recommended)
n = 10

for part_name in model.parts.keys():
    part = model.parts[part_name]
    edges = part.edges
    edge_lengths = []
    
    for edge in edges:
        edge_length = edge.getSize(printResults=False)
        edge_lengths.append(edge_length)
        
    for edge, edge_length in zip(edges, edge_lengths):
        num_divisions 
        = math.ceil((edge_length / min_edge_length) * n)
        
        ######## Mesh scaling example: Logarithm scaling
        # if edge_length / min_edge_length > Edge_scaling_factor:
            # num_divisions = math.ceil(math.log10(num_divisions)*Mesh_scaling_factor)
            # num_divisions = math.ceil((edge_length / min_edge_length)*n/4)

        ######## Odd counts are changed into even counts
        # if num_divisions % 2 != 0:
            # num_divisions += 1

        ######## If you want to set the specific edge division counts, TURN ON this code!
        # num_divisions = 50
        
        part.seedEdgeByNumber(edges=(edge,), number=num_divisions,
        constraint=FIXED)
    
    if part.cells:  # Solid Element Meshing
        part.setMeshControls(regions=part.cells, elemShape=HEX,
        technique=SWEEP, algorithm=MEDIAL_AXIS)
        elem_type = mesh.ElemType(elemCode=C3D20R, elemLibrary=STANDARD)
        part.setElementType(regions=(part.cells,), elemTypes=(elem_type,))
    else:  # Shell Element Meshing
        faces = part.faces
        part.setMeshControls(regions=faces, elemShape=QUAD,
        technique=FREE, algorithm=MEDIAL_AXIS)
        elem_type = mesh.ElemType(elemCode=S8R, elemLibrary=STANDARD)
        part.setElementType(regions=(faces,), elemTypes=(elem_type,))

######## Options for meshing types
########
######## SHELL elements
######## elemShape: QUAD, QUAD_DOMINATED, TRI
########
######## SOLID elements
######## elemShape: HEX, HEX_DOMINATED, TET, WEDGE
########
######## elemCode : STRI65, S4R, S8R, C3D20R
######## technique: FREE, STRUCTURED, SWEEP
######## algorithm: ADVANCING_FRONT, MEDIAL_AXIS

for part_name in model.parts.keys():
    part = model.parts[part_name]
    part.generateMesh()

\end{lstlisting}

\section{Connector Codes}
\label{AppendixA:Connector}

\begin{lstlisting}[
  language=Python,
  caption={Generating a pin joint between identical nodes that are on the selected edges},
  breaklines=true,
  breakatwhitespace=false,
  columns=fullflexible,
  keepspaces=true,
  breakindent=1em,
]
from abaqus import *
from abaqusConstants import *
from caeModules import *
import regionToolset
import math

model = mdb.models['Model-1']
assembly = model.rootAssembly

######## Round Function
def round_coordinates(node, precision=3):
    return tuple(round(coord, precision) for coord in node.coordinates)

######## Connector Creator
######## Here you can set the stretching and rotational stiffness between connected edges. If you want to add other mechanical constraints, such as plasticity, non-linearity, or locking, add codes in here.
def create_connector_section(edge_id, num_connected_nodes):
    Stretch_value = 1000000 / num_connected_nodes
    Bending_value = 0.001 / num_connected_nodes  
    elasticity = connectorBehavior.ConnectorElasticity(behavior=LINEAR, coupling=UNCOUPLED, components=(1, 2, 3, 4, 5, 6),table=((Stretch_value, Stretch_value, Stretch_value, Bending_value, Bending_value, Bending_value),))
    section_name = f'HingeConnectorSection_Edge{edge_id}' + '-' + manual_selection
    connector_section = model.ConnectorSection(name=section_name,assembledType=NONE,translationalType=CARTESIAN,rotationalType=ROTATION,behaviorOptions=[elasticity])
    return section_name

######## Manual Edge Selection
######## Please select the edges where you want to connect and generate the SET with the name of 'Selected Edges' (or whatever you want) in the Abaqus Interface
manual_selection = 'SelectedEdges'
selected_edges = assembly.sets[manual_selection].edges
nodes_in_selected_edges = []

for edge in selected_edges:
    nodes = edge.getNodes()
    for node in nodes:
        nodes_in_selected_edges.append(node)

node_pairs = set()
node_coords = {}
for node in nodes_in_selected_edges:
    rounded_coords = round_coordinates(node)
    if rounded_coords not in node_coords:
        node_coords[rounded_coords] = []
    node_coords[rounded_coords].append((node.instanceName, node.label))  

node_pairs = set()
for coords, node_group in node_coords.items():
    if len(node_group) > 1:
        for i in range(len(node_group)):
            for j in range(i + 1, len(node_group)):
                nodeA, nodeB = node_group[i], node_group[j]
                if nodeA != nodeB:
                    if (nodeA, nodeB) not in node_pairs and (nodeB, nodeA) not in node_pairs:
                        node_pairs.add((nodeA, nodeB))


used_node_pairs = set()
used_edge_list = []
connector_id = 0
edge_id = 0

for edge_id, edge in enumerate(selected_edges):
    edge_nodes = edge.getNodes()
    num_connected_nodes = 0
    edge_connected_pairs = set()
    for node in edge_nodes:
        node_rounded_coords = round_coordinates(node)
        for nodeA, nodeB in list(node_pairs):
            nodeA_rounded_coords = round_coordinates(assembly.instances[nodeA[0]].nodes.sequenceFromLabels([nodeA[1]])[0])
            if node_rounded_coords == nodeA_rounded_coords:
                if (nodeA, nodeB) not in used_node_pairs:
                    num_connected_nodes += 1
                    edge_connected_pairs.add((nodeA, nodeB))
                    
    if num_connected_nodes > 3:
        section_name = create_connector_section(edge_id, num_connected_nodes)

        for nodeA, nodeB in edge_connected_pairs:
            point1 = assembly.instances[nodeA[0]].nodes.sequenceFromLabels([nodeA[1]])[0]
            point2 = assembly.instances[nodeB[0]].nodes.sequenceFromLabels([nodeB[1]])[0]

            connector_wires = assembly.WirePolyLine(points=((point1, point2),), mergeType=SEPARATE, meshable=OFF)
            wire_name = 'Wire-1'
            feature_name = 'Wire' + manual_selection + '_' + str(connector_id)
            assembly.features.changeKey(fromName=wire_name, toName=feature_name)

            wire_set = assembly.Set(name='ConnectorSet' + '_' +  manual_selection + str(connector_id), edges=assembly.getFeatureEdges(connector_wires.name))

            assembly.SectionAssignment(region=wire_set, sectionName=section_name)
            
            connector_id += 1
            
        used_edge_list.append(edge_id)        
        used_node_pairs.update(edge_connected_pairs)

for edge_id in used_edge_list:
    edge = selected_edges[edge_id]
    edge_nodes = edge.getNodes()
    edge_node_labels = [node.label for node in edge_nodes]
    edge_node_coords = [round_coordinates(node) for node in edge_nodes]

    instance_name = None
    for instance in assembly.instances.values():
        for inst_edge in instance.edges:
            inst_edge_nodes = inst_edge.getNodes()
            inst_edge_node_coords = [round_coordinates(node) for node in inst_edge_nodes]

            if edge_node_coords == inst_edge_node_coords:
                instance_name = instance.name
                break
        if instance_name:
            break
    
    if instance_name:
        edge_set_name = f"EdgeSet_{edge_id}" + '_' + manual_selection
        
        assembly.Set(name=edge_set_name, nodes = assembly.instances[instance_name].nodes.sequenceFromLabels(edge_node_labels))

\end{lstlisting}

\section{Reset Codes}
\label{AppendixA:Reset}

\begin{lstlisting}[
  language=Python,
  caption={Reset generated Connectors, Connector Sections, Features, Section Assignments.},
  breaklines=true,
  breakatwhitespace=false,
  columns=fullflexible,
  keepspaces=true,
  breakindent=1em,
]
from abaqus import *
from abaqusConstants import *
from caeModules import *
import regionToolset
import math

model = mdb.models['Model-1']
assembly = model.rootAssembly

######## RESET Function
def clear_previous_features_and_sets():
    ######## Clear Connector Set
    for set_name in list(assembly.sets.keys()):
        if 'ConnectorSet' in set_name:
            del assembly.sets[set_name]
            
    for set_name in list(assembly.sets.keys()):
        if 'EdgeSet' in set_name:
            del assembly.sets[set_name]
    
    ######## Clear Feature Set
    for feat_name in list(assembly.features.keys()):
        if 'Wire' in feat_name:
            del assembly.features[feat_name]
    
    ######## Clear Connector Section Set
    sections_to_delete = [section_name for section_name in model.sections.keys() if 'ConnectorSection' in section_name]
    for section_name in sections_to_delete:
        del model.sections[section_name]
        
    ######## Clear Connector Section Assignments (Naming base)
    assignments_to_delete = [i for i, assignment in enumerate(assembly.sectionAssignments) 
                             if 'HingeConnectorSection' in assignment.sectionName]
    
    for index in sorted(assignments_to_delete, reverse=True):
        del assembly.sectionAssignments[index]
    
clear_previous_features_and_sets()

\end{lstlisting}
    \appendix
\setcounter{chapter}{1}
\setcounter{section}{0}
\renewcommand{\thesection}{\thechapter.\arabic{section}}

\chapter{Large-scale Prototyping Records}
\label{chap:AppendixB}

This appendix provides further details on large-scale prototype fabrications, including failures, weaknesses, improvements, and considerations.

\section{T-toroid: Deltoidal Hexecontahedron T-toroid}
\label{sec:AppendixBTtoroid}

For the large-scale T-toroid prototyping, two version trials were required.
This section provides the T-toroid prototyping records, which support the main contexts in Subsection~\ref{subsec:LargeTtoroidFabrication}.

\subsection{T-toroid Prototype: Version 1}

\begin{figure}[htbp]
    \centering
    \includegraphics[keepaspectratio,width=\linewidth, page=1]{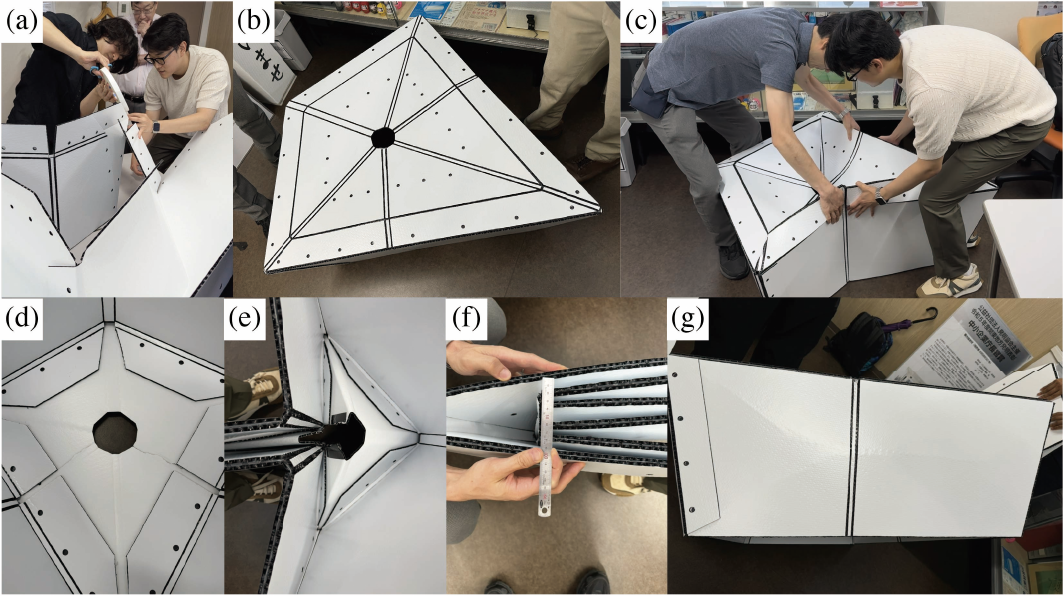}
    \caption{Deltoidal hexecontahedron T-toroid prototype Version 1: (a) Fabrication process, (b) T-toroid large-scale prototype Version 1, (c) Folding by 2 people, (d) Upper panel failure, (e) Temporary repair using duct tapes, (f) Thickness accommodation design failure, (g) Side panel local buckling by seat loading at seat-4 area.}
    \label{fig:TtoroidPrototype1}
\end{figure}

Figure~\ref{fig:TtoroidPrototype1} demonstrates the first trial of the deltoidal hexecontahedron T-toroid prototype.
Figure~\ref{fig:TtoroidPrototype1} (a) shows the fabrication process, in which the panels were connected using double-sided tape along the tab holes.
The prototype was fabricated in the deployed state (Figure~\ref{fig:TtoroidPrototype1} (b)) and then folded. 

As shown in Figure~\ref{fig:TtoroidPrototype1} (c), severe deformation occurred in the upper panel, leading to panel failure (Figure~\ref{fig:TtoroidPrototype1} (d)). 
This failure was mainly caused by the overly narrow holes in the upper panel, resulting in high stress (deformation) concentration.
Therefore, we decided to widen the upper panel hole in the next prototype.
The damaged regions were temporarily repaired using duct tape (Figure~\ref{fig:TtoroidPrototype1} (e)).

As shown in Figure~\ref{fig:TtoroidPrototype1} (f), the crease design also failed to accommodate the panel thickness in the folded state. 
This was due to a mis-calculation of the double-crease geometry, which did not account for the additional thickness introduced by the tabs and the adhesive layers. 
Therefore, the actual thickness in the flat state was measured (Figure~\ref{fig:TtoroidPrototype1} (f)) and incorporated into the next prototype.

In addition, seat loading at the seat-4 location caused local buckling of the side panels, as shown in Figure~\ref{fig:TtoroidPrototype1} (g).

\subsection{T-toroid Prototype: Version 2}

\begin{figure}[htbp]
    \centering
    \includegraphics[keepaspectratio,width=\linewidth, page=2]{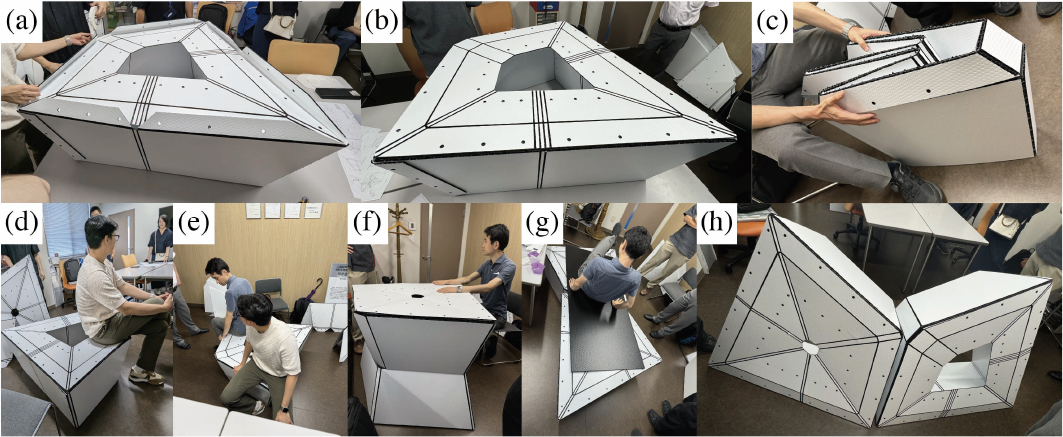}
    \caption{Deltoidal Hexecontahedron T-toroid prototype Version 2 (last): (a) Fabrication process, (b) T-toroid large-scale prototype Version 2, (c) Folded state. Application examples, such as (d) a stool, (e) a bench, (f) a desk, and (g) a bed. (h) The geometry design difference between versions 1 and 2.}
    \label{fig:TtoroidPrototype2}
\end{figure}

Figure~\ref{fig:TtoroidPrototype2} shows the final trial of the deltoidal hexecontahedron T-toroid prototype.
Figure~\ref{fig:TtoroidPrototype2} (a) presents the fabrication process, in which the panels were connected using double-sided tape along the tab holes. 
The prototype was fabricated in the deployed state (Figure~\ref{fig:TtoroidPrototype2} (b)).

Based on the issues identified in Version 1, the upper panel holes were widened, the double-crease geometry was revised to account for the total thickness in the flat state, and compliant zones were added to dissipate stress and accommodate thickness. 
As discussed in Chapter~\ref {chap:ttoroid}, prototype Version 2 exhibits snapping, multistability, and stable deployed states as intended.

Figure~\ref{fig:TtoroidPrototype2} (c) shows the folded state; although the stiffness of the creases and compliant zones slightly resists remaining flat-folded, the structure is geometrically flat-foldable. 
The prototype can support the weight of one or two people seated, as shown in Figure~\ref{fig:TtoroidPrototype2} (d--e).
Figures~\ref{fig:TtoroidPrototype2} (f--g) further demonstrate that assemblies of multiple blocks can function as a desk, table, or bed.
Figure~\ref{fig:TtoroidPrototype2} (h) shows the geometric differences of versions 1 and 2.

\section{CC-block: CC-boxes}
\label{sec:AppendixBCCbox}

\begin{table}[htbp]
    \centering
    \includegraphics[keepaspectratio,width=\linewidth, page=10]{figure/AppendixB.pdf}
    \caption{Geometric parameter changes during seven prototype trials.}
    \label{tab:CCboxGeoParam}
\end{table}

For the large-scale CC-block prototyping, the CC-column was successfully fabricated on the first attempt (Subsection~\ref{subsubsec:large_cccolumn}).
In contrast, seven version trials were required for the CC-box prototypes.
Table~\ref{tab:CCboxGeoParam} shows the geometry parameter changes during the seven trials.
Therefore, this section provides the CC-box prototyping records, which support the main contexts in Subsection~\ref{subsec:ccbox}.

\subsection{CC-box Prototype: Version 1}

\begin{figure}[htbp]
    \centering
    \includegraphics[keepaspectratio,width=\linewidth, page=3]{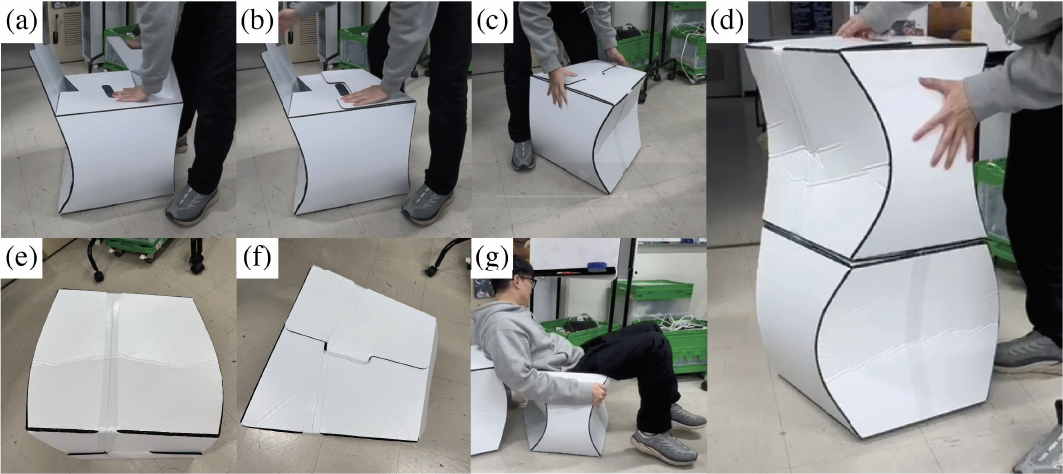}
    \caption{CC-box prototype Version 1: (a--c) Fabrication process of the seat parts. (d) The concave side panels of both CC-boxes undergo buckling during the snap-through process. (e) Buckling failures by snapping. The buckling is localized in the regions of highest curvature. (f) The seat part design and (g) failure during seating.}
    \label{fig:CCboxPrototype1}
\end{figure}

Figure~\ref{fig:CCboxPrototype1} shows the CC-box prototype Version 1.
All creases were fabricated by half-cutting, except for two creases on the concave panels, which were formed using duct tape.
The design parameters are shown in Table~\ref{tab:CCboxGeoParam}.
Figure~\ref{fig:CCboxPrototype1} (a--c) show the seat-part design and folding process. 
The prototypes exhibited snapping and were able to sustain their deployed states; however, local buckling failures occurred on the concave side panels, as shown in Figure~\ref{fig:CCboxPrototype1} (d) and (e).
The panel sheets were wrinkled near the mid-region, where the curvature is highest. 
This indicates that the profile curvature ratio ($c = 95\%$) was too large for the selected panel material (PDPPZ-100), and a lower profile curvature was required in subsequent prototypes.
Moreover, as seen in Figure~\ref{fig:CCboxPrototype1} (e) and (f), the duct tape overlapped with the seat part creases, introducing excessive constraint and wrinkling in the outer panel layers.
In addition, the seat design was insufficient to support the applied load; as shown in Figure~\ref{fig:CCboxPrototype1} (g), the prototype collapsed under the weight of a seated person.

\subsection{CC-box Prototype: Version 2}

\begin{figure}[htbp]
    \centering
    \includegraphics[keepaspectratio,width=\linewidth, page=4]{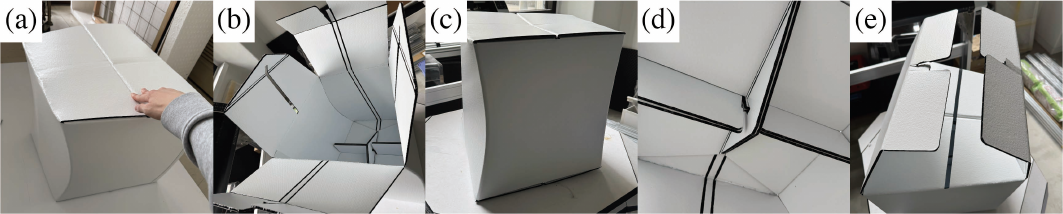}
    \caption{CC-box prototype Version 2: (a) Prototype design of Version 2. (b) Cutting pattern details, where the creases are placed on the inside. (c) The side panels pop out, and the blocks fail to maintain a stable shape. (d) Seat design details with cross ribs. (e) Failure of seat loading, where the rib intersections are highly stressed and torn.}
    \label{fig:CCboxPrototype2}
\end{figure}

Figure~\ref{fig:CCboxPrototype2} shows the CC-box prototype Version 2.
Here, we set the profile curvature as $c=99\%$.
To avoid the wrinkling problems on the outer surface layers, all mountain creases were replaced with valley creases, so that the thickness direction was oriented inward.
This can cause the bending stress of concave panels to be directed inward (thickness direction), thereby reducing bending stresses on the outer surface.
Accordingly, all creases were V-cuts, and the regions requiring 180-degree folding were implemented as double creases.
As shown in Figure~\ref{fig:CCboxPrototype2} (a), no creases are visible on the outer surface, while Figure~\ref{fig:CCboxPrototype2} (b) confirms that all creases are located on the inner side.

Consequently, a profile curvature ratio of $c=99\%$ was too low to achieve multistability.
As indicated by the simulation result in Figure~\ref{fig:results} (e), lower curvature reduces shape stability.
Additionally, the actual prototype was further reduced by hinge stiffness, reaction forces caused by contact in the single-line V-cut curved creases (Figure~\ref{fig:CCboxPrototype2} (b)), and play in the double-hinge regions, preventing the prototype from stabilizing in its designed stable shape.
As a result, the concave panels popped out, as shown in Figure~\ref{fig:CCboxPrototype2} (c).

Figure~\ref{fig:CCboxPrototype2} (d) shows the seat design.
To address the seating failure observed in Version 1, a cross-shaped rib was added.
However, this was still insufficient to support the seating load: stress concentrated at the central contact of the ribs, leading to tearing, as shown in Figure~\ref{fig:CCboxPrototype2} (e).

These results indicate that valley-crease-based designs make it more difficult to achieve shape stability and that higher profile curvature is required.
They also reveal the limitation of using a single material (PDPPZ-100), motivating the introduction of additional panel materials in subsequent designs.

\subsection{CC-box Prototype: Version 3}

\begin{figure}[htbp]
    \centering
    \includegraphics[keepaspectratio,width=\linewidth, page=1]{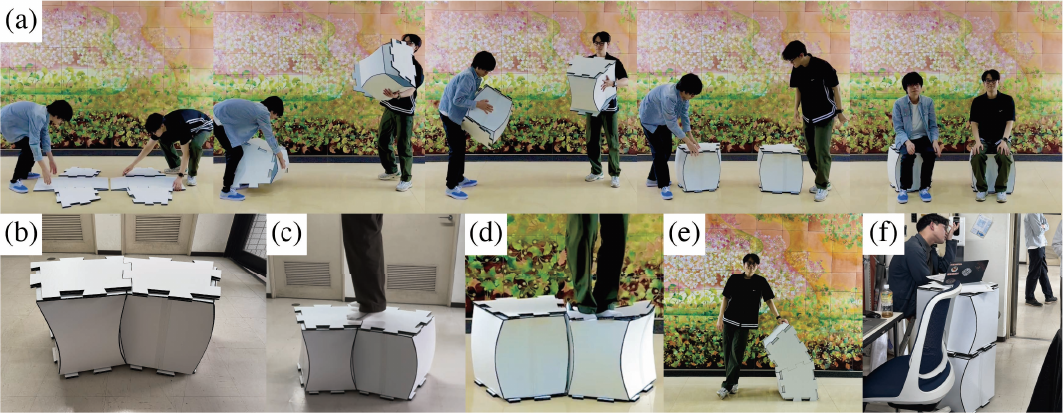}
    \caption{CC-box prototype Version 3: (a) Deployment motion of the prototype. Each prototype can support seat loading. (b) Horizontal connection of prototypes by puzzle-shaped seat assembly. (c) Assembled blocks can support a person standing or even jumping on them. (d) However, separated blocks cannot support a standing load. (e) arch configuration, and (f) actual use as the standing table.}
    \label{fig:CCboxPrototype3}
\end{figure}

Figure~\ref{fig:CCboxPrototype3} shows CC-box prototype Version 3.
The profile curvature ratio was set to $c=98.5\%$, and all creases were fabricated again as mountain folds by half-cutting from the outer surface, to minimize unexpected effects captured in Version 2, such as contact and play.

Figure~\ref{fig:CCboxPrototype3} (a) shows the deployment motion of Version 3.
The CC-box geometry was simplified by keeping only the cap parts and removing the tab elements.
The prototype exhibited snapping behavior and could sustain its deployed configuration; however, even under small external forces or slight geometric mismatch, the concave panels popped out, indicating that the profile curvature was still insufficient.

To prevent seating failures, seat panels made of a stiffer, thicker material (PGPPZ-300, \SI{10}{mm}-thick) were added. 
These panels were designed as puzzle-shaped, following the trapezoidal path, and were attached to the cap parts of the CC-box with double-sided tape.
The introduction of the seat panels was successful.
No failure occurred under sitting loads in the stable configuration, and each block could function as a stool.

Figure~\ref{fig:CCboxPrototype3} (b) shows horizontal assembly.
The puzzle-shaped seat panels enable tight interlocking, and the connected blocks exhibit higher load-bearing capacity than individual blocks.
As shown in Figure~\ref{fig:CCboxPrototype3} (c), the assembled blocks could support both standing loads and impact loads (jumping).
In contrast, as shown in Figure~\ref{fig:CCboxPrototype3} (d), a single block failed under standing load due to local buckling at the upper corners and popping out of concave panels.
It can be estimated that in horizontally assembled structures, contact between adjacent curved panels becomes tighter under load, allowing neighboring blocks to sustain stable shapes and suppress local buckling of each other, thereby increasing the overall load-bearing capacity.

As shown in Figure~\ref{fig:CCboxPrototype3} (e), multiple blocks can also be assembled into arch-like structures, and vertical assembly, as in Figure~\ref{fig:CCboxPrototype3} (f), allows the system to function as a desk or standing table.

Although Version 3 had not yet achieved sufficient shape stability, the introduction of the seat panels significantly improved the load-bearing performance.
However, practical issues remained: the sharp edges of the puzzle-shaped seat panels posed a risk of damaging clothing or skin, and the connections were too tight for easy assembly and disassembly.
These issues indicate the need for further refinement, such as edge filleting and connection tuning, in the next prototype.

\subsection{CC-box Prototype: Version 4}

\begin{figure}[htbp]
    \centering
    \includegraphics[keepaspectratio,width=\linewidth, page=6]{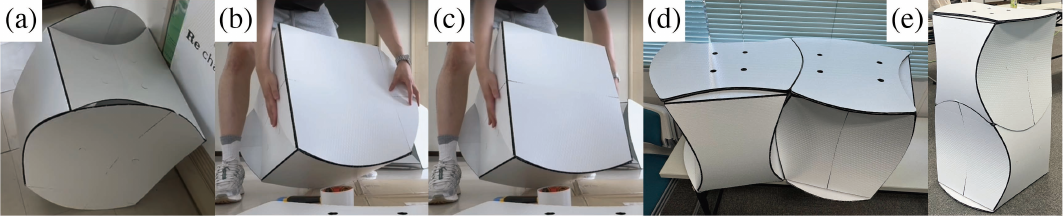}
    \caption{CC-box prototype Version 4: (a) The block snaps and maintains a stable shape when the tabs are not folded. (b--c) Due to the hinge stiffness of the tabs, the concave panels pop out when the tabs are folded. (d) Horizontal and (e) vertical assemblies.}
    \label{fig:CCboxPrototype4}
\end{figure}

Figure~\ref{fig:CCboxPrototype4} presents CC-box prototype Version 4.
We increased the profile curvature by setting $c=97\%$.
Based on the analysis results of the cap and tab parameters (Figure~\ref{fig:results} (i)), the tab parts were reintroduced to improve shape stability.
Figure~\ref{fig:CCboxPrototype4} (a) shows the CC-box body, in which the tab parts were cut as elastica shapes following the profile curvature.
Before folding the tabs, the CC-box exhibited clear snapping behavior and maintained a stable deployed configuration.

However, when the tab parts were folded (Figure~\ref{fig:CCboxPrototype4} (b--c)), the concave panels popped out, and the CC-box could no longer sustain its stable configuration.
This instability was pronounced in the widest concave panel, which was expected to have the lowest stability based on the horizontal slenderness analysis result (Figure~\ref{fig:results} (f)).
This behavior arises because, in the physical prototypes, the hinge stiffness of the tab creases generates moments that drive the concave panels outward.
As a result, the tabs reduced rather than increased the shape stability, in contrast to the analysis result of Figure~\ref{fig:results} (i).

Figures~\ref{fig:CCboxPrototype4} (d) and (e) show the horizontal and vertical assemblies, respectively.
The seat parts were designed in an hourglass shape using elastica curves, and the edge tips were filleted (the detail will be explained in the next section).
The two blocks matched geometrically well, and the tabs enabled a tighter and finer vertical assembly (Figure~\ref{fig:CCboxPrototype4} (e)).
For manufacturability, we tried using two layers of the PDPPZ-100 for the seat panels in this version.
However, this design was insufficient to support the seating load.

Excluding tab parts, Version 4 exhibits better shape stability than the previous versions, although manual testing suggests that a higher energy barrier is still desirable.
We therefore decided that, to improve stability, the tab parts should be removed and that the seat panels should again be fabricated from PGPPZ-300 to ensure sufficient load-bearing capacity.
These design decisions are incorporated into the subsequent prototype.

\subsection{CC-box Prototype: Version 5}

\begin{figure}[htbp]
    \centering
    \includegraphics[keepaspectratio,width=\linewidth, page=7]{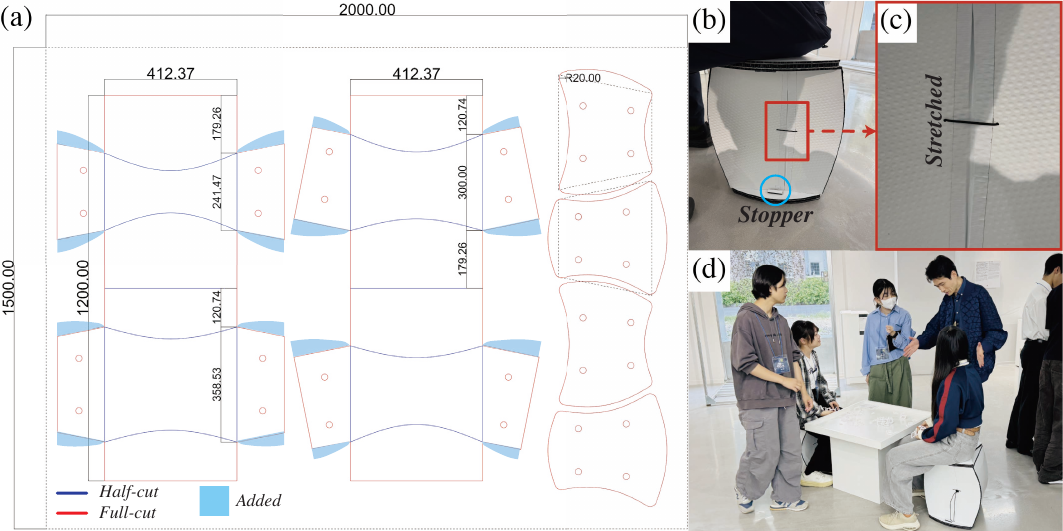}
    \caption{CC-box prototype Version 5: (a) Cutting patterns, (b) Rubber bands and stoppers are installed to ensure the shape during seating. (c) Delamination and stretching of the duct-tape crease regions under seat loading. (d) Version 5 is actually used as stools in an exhibition.}
    \label{fig:CCboxPrototype5}
\end{figure}

Figure~\ref{fig:largeccbox} (a, c) and Figure~\ref{fig:CCboxPrototype5} show the CC-box prototype Version 5, which was designed by incorporating the overall improvements identified in the previous prototypes.
In this version, the tab parts were removed, and the profile curvature ratio was set to $c=97\%$.
Based on the analysis of the vertical slenderness parameter ($VSL=b/a_1$) in Figure~\ref{fig:results} (e), the VSL was increased from $1$ to $4/3~(=1.\dot{3})$ by setting $a_1=300$, to improve the shape stability.

Figure~\ref{fig:CCboxPrototype5} (a) shows the cutting patterns of the CC-box body and the seat parts of Version 5.
All creases of the body were fabricated by half-cuts, except for one side concave panel, which was connected by a duct tape hinge.
The circular holes on the caps and seat parts were used for alignment and assembly.
The seat parts were designed in an hourglass shape using elastica curves with the same curvature ratio of $97\%$ along each trajectory edge of length $b$, and their edge tips were filleted with a radius of \SI{20}{mm}.
The CC-box body was fabricated from PDPPZ-100, while the seat parts were made from PGPPZ-300.

As shown in Figure~\ref{fig:largeccbox} (a), Version 5 exhibited clear snapping behavior, stable configurations of both blocks, and geometrically well-matched assemblies in both horizontal and vertical.
It also supported seating loads sufficiently (Figure~\ref{fig:largeccbox} (c)).
For improving and ensuring the load-bearing performance, we added \SI{5}{mm}-thick panel pieces on the blue region of Figure~\ref{fig:CCboxPrototype5} (a), so that the seating load is transferred directly to the curved panels.
Therefore, we consider the CC-box prototyping to have been successful from Version 5, and this version was used as stools for the exhibition.
To ensure shape stability in actual use, stoppers were added to prevent excessive deformation of the concave panels, and rubber bands were used to block the concave panel from popping out, as shown in Figure~\ref{fig:CCboxPrototype5} (b).

During seating tests, delamination and stretching of the duct-tape hinge were observed (Figure~\ref{fig:CCboxPrototype5} (c)).
Although this did not cause significant problems during use, due to the stoppers and rubber-band constraints, it should be improved in future prototypes to enhance durability and reusability.
Nevertheless, the CC-boxes Version 5 were successfully used as stools for an eight-week exhibition (Figure\ref{fig:CCboxPrototype5} (d)) without any structural failures.

\subsection{CC-box Prototype: Version 6}

\begin{figure}[htbp]
    \centering
    \includegraphics[keepaspectratio,width=\linewidth, page=2]{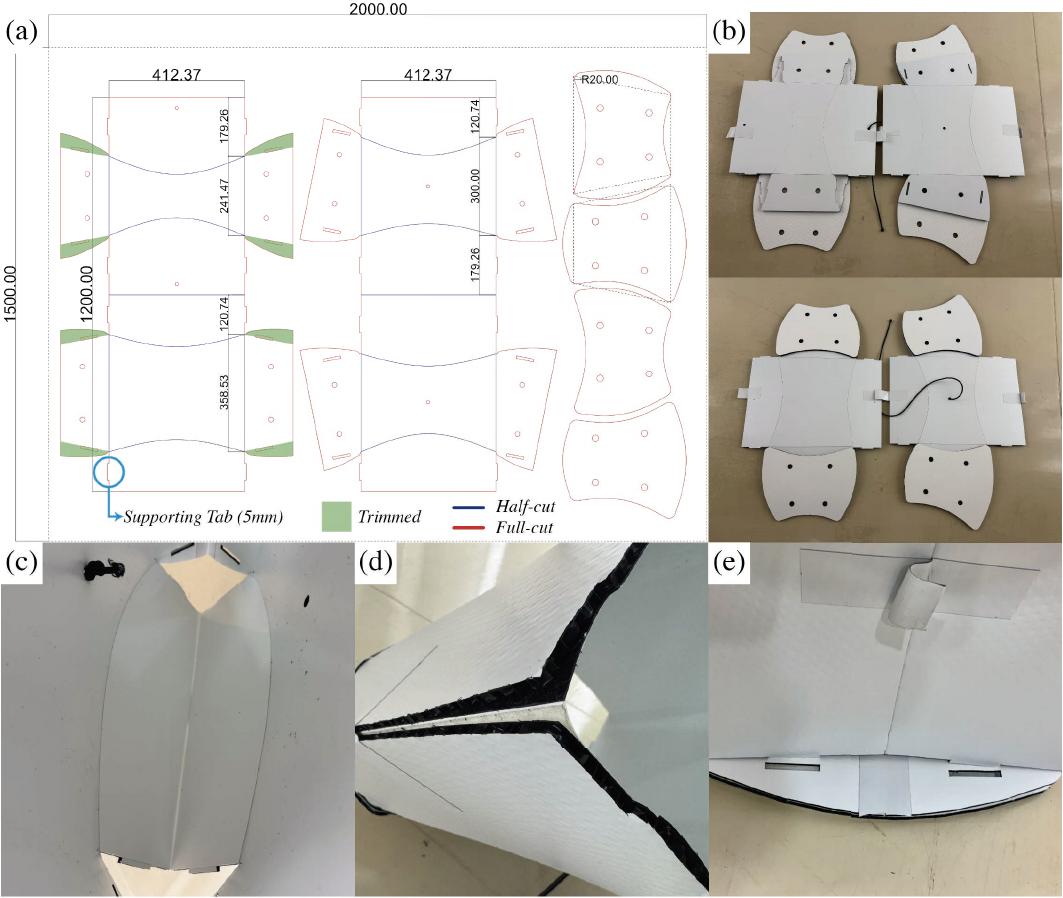}
    \caption{CC-box prototype Version 6: (a) Cutting patterns and (b) flat states of CC-boxes, back side (Top) and front side (Bottom), (c) Attached polypropylene-sheet adhesive layer for hinge, and (d) delamination issue in the edge tip, (e) Geometric mismatch by curvature relaxation.}
    \label{fig:CCboxPrototype6}
\end{figure}

Figure~\ref{fig:CCboxPrototype6} shows CC-box prototype Version 6.
Figure~\ref{fig:CCboxPrototype6} (a) shows the cutting patterns, and (b) shows the flat state of fabricated CC-boxes from front (top) and back (bottom) sides.
This version is based on the same geometric parameters as Version 5, with additional holes (\SI{5}{mm}--radius) introduced at the center of the side panels to allow the use of a tension cable (\SI{2.5}{mm}--radius rubber band).
The supporting pieces and stoppers were also integrated into the design. 
Supporting tabs were added on the edges of the concave panels; in the deployed state, these tabs are inserted into rectangular holes in the cap parts, where they function as stoppers. 
In addition, instead of duct-tape hinges, we used Polypropylene-sheet adhesive layers to form the hinges by attaching them to inner surfaces (Figure~\ref{fig:CCboxPrototype6} (c)).

Using PP-sheet hinges produced clearer snapping behavior than in Version 5, and sufficient load-bearing capacity under seating without the play issue caused by duct tape stretching. 
However, because of the repeated folding and deploying, delamination occurred in the tip of the linear creases (Figure~\ref{fig:CCboxPrototype6} (d)).

We then inserted a rubber band into the holes in the middle of the panels to pass through two prototypes horizontally (Figure~\ref{fig:CCboxPrototype6} (b) and Figure~\ref{fig:largeccbox} (d)). 
Here, we found that the green regions in Figure~\ref{fig:CCboxPrototype6} (a) interfered with the horizontal assembly, so these areas were trimmed eventually. 
After this modification, Version 6 could exhibit the intended assemblies and reconfiguration motion with a tension cable, as shown in Figure~\ref{fig:largeccbox} (e). 
While this modification removes the stopper function, the seating load was still transferred directly to the curved panels through the supporting tabs, with no significant change in load-bearing performance.

The supporting tabs and rectangular holes were designed to match the stable shape geometry of CC-boxes.
However, the observed stable shapes deviate slightly from the design, with the concave panels folded inward a bit, as shown in Figure~\ref{fig:CCboxPrototype6} (e).
This discrepancy is attributed to the free boundary conditions at the top and bottom edges of the concave panels.
Near the curved creases, the panels retain their designed curvature, but toward the linear crease, geometric constraints are relaxed, and the curvature gradually diminishes.
As a result, the linear creases become active, leading to a relaxed curvature state and a stable configuration that is slightly inset compared to the intended geometry.
This free-end induced curvature relaxation is more likely to occur in CC-boxes with relatively stiff panels and low vertical slenderness.
This issue likely also occurred in Version 5, as shown in Figure~\ref{fig:largeccbox} (c). 
Although the prototypes visually seem to match well with the designed geometry, this indicates that the curvature relaxation leads to minor geometric mismatch. 
Considering this mismatch, applying tension to ensure a secure interlock can be seen as an effective strategy.
Therefore, CC-columns, which have lower panel stiffness (\SI{3}{mm}-thick PCPPZ-050) and higher vertical slenderness, showed this issue much less.

\subsection{CC-box Prototype: Version 7}

\begin{figure}[htbp]
    \centering
    \includegraphics[keepaspectratio,width=\linewidth, page=3]{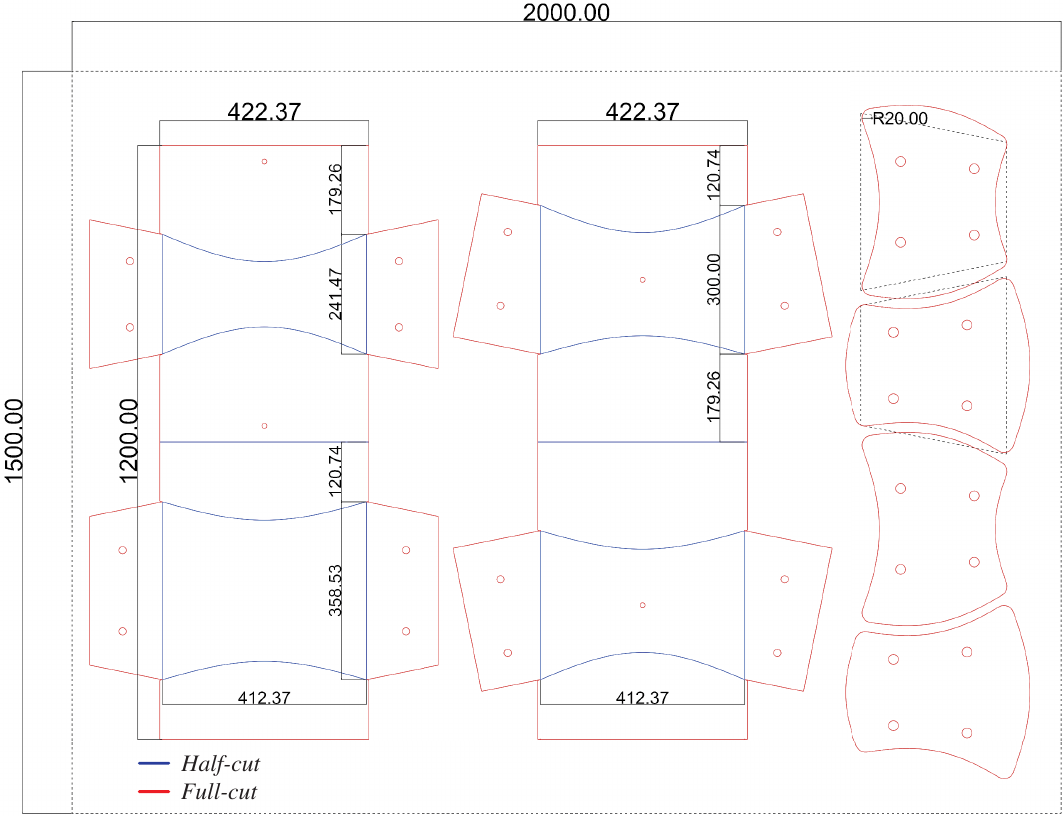}
    \caption{Cutting patterns of CC-box prototype Version 7.}
    \label{fig:CCboxPrototype7}
\end{figure}

Figure~\ref{fig:CCboxPrototype7} shows the cutting patterns of Version 7.
Version 7 has the same geometric parameters as Version 6 and incorporates improvements, including the removal of rectangular hole regions and the extension of supporting tabs along the concave panel edges to ensure effective seating load transfer and to prevent inward folding of the linear creases.
Due to time constraints, fabrication of Version 7 is left as future work.
However, we believe that the CC-box design is close to the final stage.

\end{document}